\documentclass[onecolumn,aps,nofootinbib,superscriptaddress,12pt]{revtex4-2}
\usepackage{amssymb,mathtools, graphicx, enumerate, subfig}
\usepackage{color}
\usepackage{siunitx}

\newcommand{\eq}[1]{Eq.\,\eqref{eq:#1}}
\newcommand{\fig}[1]{Fig.\,\ref{fig:#1}}
\newcommand{\mum}[1]{\SI{#1}{\micro\meter}}

\newcommand{\chie}{\chi_{\rm e}}
\newcommand{\chib}{\chi_{b}}
\newcommand{\chimax}{\chi_{\rm max}}
\newcommand{\LeL}{L_{\rm e\ell}}

\newcommand{\Arc}{Arcturus\,}
\newcommand{\TPW}{TPW\,}
\newcommand{\Cor}{CoReLS\,}
\newcommand{\OPAL}{OPAL\,}
\newcommand{\ELInp}{ELI-NP\,}
\newcommand{\ELIhapl}{ELI-L3\,}

\begin{document}
\title{Luminosity for laser-electron colliders}
\author{B. Manuel Hegelich}
\email{Authors alphabetical}
\affiliation{Center for High Energy Density Science, University of Texas, Austin, Texas, 78712}
\affiliation{Tau Systems, Inc., Austin, Texas, 78701}
\author{Calin I. Hojbota}
\affiliation{Center for Relativistic Laser Science, Institute for Basic Science (IBS), Gwangju 61005, Republic of Korea}
\author{Lance A. Labun}
\affiliation{Center for High Energy Density Science, University of Texas, Austin, Texas, 78712}
\affiliation{Tau Systems, Inc., Austin, Texas, 78701}
\author{Ou Z. Labun}
\affiliation{Center for High Energy Density Science, University of Texas, Austin, Texas, 78712}
\affiliation{Tau Systems, Inc., Austin, Texas, 78701}
\author{Dung D. Phan}
\affiliation{Center for High Energy Density Science, University of Texas, Austin, Texas, 78712}

\begin{abstract}
High intensity laser facilities are expanding their scope from laser and particle-acceleration test beds to user facilities and nuclear physics experiments.  A basic goal is to confirm long-standing predictions of strong-field quantum electrodynamics, but with the advent of high-repetition rate laser experiments producing GeV-scale electrons and photons, novel searches for new high-energy particle physics also become possible.  The common figure of merit for these facilities is the invariant $\chi\simeq 2\gamma_e|\vec E_{\rm laser}|/E_c$ describing the electric field strength in the electron rest frame relative to the ``critical'' field strength of quantum electrodynamics where the vacuum decays into electron-positron pairs.  However, simply achieving large $\chi$ is insufficient; discovery or validation requires statistics to distinguish physics from fluctuations.  The number of events delivered by the facility is therefore equally important.  In high-energy physics, luminosity quantifies the rate at which colliders provide events and data.  We adapt the definition of luminosity to high-intensity laser-electron collisions to quantify and thus optimize the rate at which laser facilities can deliver strong-field QED and potentially new physics events.  Modeling the pulsed laser field and electron bunch, we find that luminosity is maximized for laser focal spot size equal or slightly larger than the diameter of the dense core of the electron bunch.  Several examples show that luminosity can be maximized for parameters different from those maximizing the peak value of $\chi$ in the collision.  The definition of luminosity for electron-laser collisions is straightforwardly extended to photon-laser collisions and lepton beam-beam collisions.
\end{abstract}

\maketitle

\section{Introduction}

High-intensity lasers have long been promised as a new avenue to generating high-energy particle beams, studying nuclear physics and exploring the novel regime of particle dynamics in ultra-strong fields \cite{albert2016applications, national2018opportunities, albert20212020,falcone2020workshop, mp32022workshop}.    However, until recently high-intensity laser experiments were dominated by a few large-scale facilities focused on fusion science and smaller university-based experiments pushing the envelope on the laser technology and small-scale experiments in these areas of promise.
High-intensity laser facilities \cite{hooker2008commissioning, rus2011outline, sung20174, kuhn2017eli, le2017design,papadopoulos2019first,  cerchez2019arcturus, danson2019petawatt, doria2020overview, abramowicz2021conceptual,nees2021zettawatt, gan2021shanghai}) and organizations of existing facilities (LaserNetUS \cite{lasernet}, ELI ERIC \cite{elieric}, LaserLab Europe \cite{laserlab}) now aim to provide more consistent access for users and enhance opportunities to study fundamental science \cite{danson2019petawatt}.   

In many of these facilities, a basic experimental setup is the collision of a high-energy electron beam with an ultra-high intensity laser field.  The high-energy electron beam may be provided by a conventional accelerator (as in LUXE and FACET-II) or by another laser beam line in an ``all-optical'' setup.  After the pioneering E-144 experiment at SLAC \cite{Bula:1996st,Bamber:1999zt}, electron beam-laser collision experiments have  dramatically accelerated in the last ten years to study electromagnetic radiation in the strong-field regime and develop novel particle and photon beam sources \cite{schwoerer2006thomson, ta2012all, chen2013mev,powers2014quasi,sarri2014ultrahigh,khrennikov2015tunable, yu2016ultrahigh,yan2017high, cole2018experimental, poder2018experimental}.  Moreover, recent and planned  \cite{abramowicz2019letter} experiments are reaching a regime of electrodynamics where classical radiation reaction and quantized radiation processes are important \cite{di2012extremely, blackburn2020review, gonoskov2022charged, fedotov2022advances}.   CoReLS, ELI-NP, ZEUS, OPAL and LUXE all plan to conduct experiments accessing the regime of strong-field QED (sfQED) where the electric field seen in the rest frame of a high-energy counter-propagating electron is greater than the `critical' electric field strength
\begin{align}\label{eq:Ecdefn}
F_c=m_e^2c^3/e\hbar
\end{align}
near which the field rapidly converts its energy into electron-positron pairs \cite{schwinger1951gauge,labun2009vacuum}.  This condition is invariantly characterized by 
\begin{align}\label{eq:chibasic}
\chi=\frac{\hbar}{c^5}\frac{|p_\mu eF^{\mu\nu}|}{m_e^3}>1.
\end{align}
Photon emission in the strong-field regime is in some sense the simplest process, but is not the only process of interest in strong fields: high-energy photons can convert to electron-positron pairs (the Breit-Wheeler process), photons may split, and hypothetical particles such as axions or emergent scalars can be produced.  As a baseline, we consider photon emission first. 

The rate of photon emission is an increasing function of $\chi$, and it has therefore been assumed that the facility should maximize the $\chi$ achieved in the experiment.  The above facilities are designed with one laser accelerating electrons to sufficiently high-energy counter-propagating to a second laser so as to achieve $\chi\gtrsim 1$.  In this setup, a naive estimate of the $\chi$ achieved by electrons is obtained by combining the nominal electron energy with the peak laser electric field,
\begin{align}\label{eq:chibdefn}
\chib\equiv \frac{\hbar}{c^4}\frac{|p_\mu^{(0)} eF^{\mu\nu}_{\rm peak}|}{m_e^3}\simeq \frac{2E_e}{m_e}{|e\vec E_{\rm peak}|}{m_e^2}=\frac{2E_e}{m_e}\sqrt{\frac{I_{\rm peak}}{I_c}}
\end{align}
where $I_c=\epsilon_0c E_c^2=1.\times 10^{29}$\,W/cm$^2$ is the intensity of a linearly polarized wave corresponding to the critical electric field.  We refer to \eq{chibdefn} as the `beam' $\chi$ since it uses only nominal parameters of the electron and laser before collision.  As an estimate, it might be slightly improved by using the root-mean-square electric field as in Refs. \cite{Bula:1996st,Bamber:1999zt} rather than the peak field, but \eq{chibdefn} is commonly seen in the laser experiment literature and differs from the r.m.s. estimate only by a factor of $\sqrt{2}$.  Using $\chi_b$ suggests that maximizing the probability of an event requires simultaneously maximizing electron energy and laser intensity.  

The question should then be posed if maximizing the (peak) intensity and electron energy enhances the signal, i.e. ability of the experiment to measure and to quantitatively verify the theory of strong-field radiation processes.  Much previous work has pointed out how various features of the laser fields or electron bunch are imprinted on observables such as the energy or angular distribution of radiation or scattered electrons \cite{harvey2009signatures,di2009strong, Heinzl:2009nd, di2010quantum, Seipt:2010ya, seipt2011nonlinear, harvey2011symmetry, Seipt:2012tn,thomas2012strong,blackburn2014quantum,neitz2014electron,vranic2014all, blackburn2015measuring, heinzl2015detecting, lobet2017generation, ridgers2017signatures, blackburn2020radiation, holkundkar2020complete, tamburini2021efficient,golub2022non}.  Despite several rounds of a facility design, we do not know that the highest optically-achievable intensity, generally achieved by the shortest possible laser pulse and strongest possible focusing, always leads to larger or more reliable signal.  

In general, many other parameters also affect the significance of the signal in an experiment and therefore the physics that it can explore and test.  Clearly, the magnitude of the signal is proportional the number of photons produced which in turn scales with the number of electrons interacting with the high-intensity fields.  Thus a trade-off arises: a smaller focus increases the intensity but decreases the area available for electrons to interact with, decreasing the number of electrons interacting with the highest field intensity and/or the probability that the electron beam overlaps with the laser in a given shot.  Other parameters such as pulse length and pulse shape affect the signal in more subtle ways by changing the dynamics of the laser-electron collision.  Beyond tuning the design parameters of the experiment, the stability of the laser system in pulse energy, pointing, focal spot size/quality, pulse profile, determine how many shots provide relevant data by creating the expected laser field conditions at the collision.  The stability combined with the shot frequency determines the ability of the facility to collect statistics and thereby quantitatively validate theory predictions or discover new effects.  Such statistical effects will be addressed in later work.  

These are several reasons $\chi_b$ alone is an incomplete indicator of the physics capability of the facility.  Our aim is to establish a metric to evaluate the performance of laser facilities in the laser-electron collision geometry.  We draw guidance from high-energy particle physics experiment, for which there are two key performance metrics: the beam center-of-mass energy, which determines what physics is accessible in a given 2-particle collision, and luminosity, which characterizes how effectively/efficiently the machine collides particles and therefore accumulates measurement data toward the discovery of new physics.  A leading-order estimate of the luminosity combines the cross sectional areas of the colliding beams with the frequency of beam crossing; in more detail, the electromagnetic field of one beam perturbs particle trajectories in the other beam correcting the probability of collision.  The luminosity is now widely used and can estimate how long a given facility must run in order to be sensitive to rare events, such as the production of new high-mass particles.  In its role in laser facility design, $\chi_b$ is analogous to the beam center-of-mass energy.  We need the analog of the luminosity to both quantify--and optimize--the efficiency in colliding electrons with a focused laser pulse as well as estimating facility requirements in searching for rarer strong-field processes.

\section{Theory background}

\subsection{Event probabilities}

Collisions of electrons with high-intensity laser pulses differ from the bunch crossings in high-energy particle colliders in several important way.  (We discuss similarities and applications to proposed high-luminosity linear accelerators, which may also access the sfQED regime, at the end of this section.)  For the goal of defining a performance metric analogous to luminosity, the most important considerations are: 1) that the laser fields are extended in space and time, in contrast to the point-like particles of the bunches, 2) the fields are strong enough for the electron dynamics in the field to be accelerated and differing significantly from the inertial trajectory, and 3) probabilities of events are expressed as (quasi-local) rates rather than cross sections.  The impact of (1) and (2) are best understood after a review of relevant properties of the event probabilities.

For a high-energy electron $E_e\gg m_e$ in a focussed, high-intensity laser field, event probabilities are well-approximated by results for a constant ``crossed'' electromagnetic field, which is defined by vanishing Lorentz invariants $\mathcal{S}=\frac{1}{4}F^{\mu\nu}F_{\mu\nu}=\frac{1}{2}(\vec{B}^2-\vec{E}^2)=0$ and $\mathcal{P}=\frac{1}{8}F^{\mu\nu}\epsilon_{\mu\nu\kappa\lambda}F^{\kappa\lambda}=-\vec{E}\cdot \vec{B}=0$.  The physics reasoning is essentially the same as for the equivalent photon approximation.  In practice, we require $\chi$ is much larger than the two field invariants normalized to the critical QED electric field strength, $\chi^2\gg\mathcal{S,P}/F_c^2$ with $F_c$ defined in \eq{Ecdefn}.
This is generally true for high-energy electrons due to the $p/m$ factor in $\chi$ \cite{Ritus:1985}.  The invariants $\mathcal{S,P}$ are suppressed by the typically small divergence angle in weakly focussed beams.  Even for the most strongly focussed beams, $\mathcal{|S|,|P|}\lesssim |\vec E_\ell|^2/2\pi$.  For a constant field and unpolarized scattering, we need only examine the 3 invariants $\chi,\mathcal{S,P}$; others constructed from combinations of the electron 4-momentum and field tensor are not independent, all powers of the field tensor higher than $F^2$ vanish in the plane wave limit, and higher orders of the form $pF^{2n}p$ for $n>1$ are suppressed by even powers of the laser electric field over the critical field once the tensor structure is simplified for nearly plane-wave fields.

For electrons in electromagnetic fields, the primary event is always emission of a photon: pair-production, cascade development, radiation reaction and other higher order processes must begin with emission of a photon by at least one electron since there is no other interaction at leading order.  We can use this work to estimate the probability of more exotic processes such as axion production, but rare events are less useful for defining baseline performance.  

Constant field event probabilities suffice as an approximation because the wavelength of the laser is much greater than the electron de Broglie wavelength, $\lambda_\ell\gg 1/p_e$, and the energy of emitted photons is also much greater than the laser photon energy ($E_\gamma\gg\omega_\ell$).  More rigorously, semi-classical analysis of the photon emission amplitude shows that the dominant contributions come from a spatial region $\sim \lambda_\ell/a_0$ \cite{Ritus:1985} where $a_0=\chi/k\cdot p\simeq |e\vec E|/\omega_\ell m_e$ is generally greater than 1.  Separate analysis by Refs. \cite{baier1989quantum,baier2005concept} suggests a formation length
\begin{align}\label{eq:formlength}
\ell_f\simeq \frac{E_e}{m_e}\frac{\lambda_e}{\chi_e}\left(1+\frac{\chi_e(E_e-E_\gamma)}{E_\gamma}\right)^{1/3},
\end{align}
which is notably only weakly dependent on the photon energy $E_\gamma$, except at very low energy $E_\gamma/E_e\ll\chi_e$.  

To achieve $\chi\gtrsim 1$, most facilities will work in the $a_0\gtrsim 1$ regime because the frequency distribution of radiation broadens in comparison to the $a_0\lesssim 1$ regime and hence higher energy photons can be expected \cite{Esarey:1993zz,Ride:1995zz,harvey2009signatures}.  Corrections to the constant crossed field event probabilities for the $a_0\sim 1$ case have been suggested in Refs. \cite{di2019improved,ilderton2019extended} and can be added for facilities operating in that regime.

For the event probability, we have no experimental measurements of the total event rate.  Provided the separation of timescales between the quantized event dynamics \eq{formlength} and the classical dynamics, the total probability for something to happen during the electron laser collision is obtained by integrating the rate for the classical electron distribution function and classical trajectory,
\begin{align}\label{eq:Pinclusive}
P[e\to X]&=\int \frac{d^3pd^3x}{(2\pi)^3}f_{\rm e}(\vec x,\vec p,t)\Gamma(E_e,\chi(\vec x,\vec p))dt 
\end{align}
where $\Gamma(\chi)$ is the locally-defined rate, here for an electron to transition to anything (represented by '$X$').  The dependence on the laser fields is embedded in the spatial dependence of $\chi$. 

At order $\alpha=e^2/4\pi\simeq 1/137$ single-photon emission is the only process allowed, and photon production is a precursor to many of the other processes of interest, with a few exceptions \cite{bai2022new}.  Therefore, together with the smallness of $\alpha$, calculating the rate $\Gamma$ to order $\alpha$ suffices.  This calculation is equivalent to the imaginary part of the 1-loop self-energy.  The result is known as an integral over an Airy function \cite{Ritus:1985}
\begin{align}\label{eq:Gamma1loop}
\Gamma_{\rm tot}^{\rm 1-loop}(E_e,\chi)\simeq \frac{\alpha m^2}{3E_e}\int_0^\infty du\frac{5u^2+7u+5}{(1+u)^3(u/\chi)^{2/3}}(-1)\mathrm{Ai}'\!\Big((u/\chi)^{2/3}\Big)
\end{align}
where $-\mathrm{Ai}'(z):=-d\mathrm{Ai}(z)/dz>0$ is the derivative of the Airy function with respect to its argument.  A function only of $\alpha,\chi$ and the electron energy $E_e$, the Lorentz transformation properties of $\Gamma$ are entirely in the $1/E_e$ prefactor, which ensures that it transforms as a decay rate.  In the wavefunction methods commonly used, this is also equal to the rate of decay of the electron state \cite{tamburini2021efficient} (though it is not the sum of a geometric series of loop diagrams).  Despite the complicated appearance, the function is quite simple in form, well-described by power laws above and below the transitional regime $\chi\simeq 1$ \cite{Ritus:1985}:
\begin{align}\label{eq:Gamma1loopasymptotics}\Gamma_{\rm tot}^{\rm 1-loop}(E_e,\chi)& \simeq \frac{5\alpha}{2\sqrt{3}}\frac{m_e^2}{E_e}\chi\left(1-\frac{8\sqrt{3}}{15}\chi+...\right) &\chi \ll 1 \\
\Gamma_{\rm tot}^{\rm 1-loop}(E_e,\chi)&\simeq \frac{14\Gamma(2/3)3^{2/3}\alpha}{27}\frac{m_e^2}{E_e}\chi^{2/3}\left(1-\frac{45}{28\Gamma(2/3)}\chi^{-2/3}+...\right) &\chi\gg 1
\end{align}
where $\Gamma(z)$ is the Euler gamma function.  

The 1-loop result suffices here thanks to the smallness of the QED coupling, though in the strong field regime $\chi\gtrsim 1$ where radiation reaction is an order 1 correction to the classical trajectory \cite{Hadad:2010mt}, we still need to synthesize consistently classical radiation reaction with the quantum emission rates \cite{ilderton2013radiation1, ilderton2013radiation2, neitz2014electron, heinzl2021classical,edwards2021resummation, torgrimsson2021resummation}.  Achieving $\chi\sim \alpha^{-3/2}$ is hypothesized to require a resummation in $\alpha\chi$ \cite{fedotov2017conjecture,ilderton2019note, mironov2020resummation}, but facilities remain far from this regime.  Thus, in summary we expect \eq{Gamma1loop} to suffice for the present purposes and foreseeable facilities, particularly once classical dynamics between emission events are accounted for as accurately as possible.

In our study, the nominal $\chi$ are mostly less than 1, with a few reaching $\chi\lesssim 10$.  In this transitional regime $\chi\sim 1$, classical radiation reaction is an important (large but next-to-leading-order) correction to the dynamics \cite{heintzmann1972acceleration, piazza2008exact, Hadad:2010mt,ekman2021reduction} and the probability of photon emission becomes significant.  For the laser systems modeled below, limiting to $\chi\lesssim 10$ regime means that generally much less than one hard-emission event will occur per electron per collision, and therefore modeling the recoil is not important to computing the total number of events in each crossing of the electron bunch and laser pulse.

\subsection{Electron-laser collision dynamics}

The difficulty in determining the classical dynamics is not the theory but the inputs to the theory.
Neglecting radiation, the trajectory of a particle in a pulsed laser field is determined by solving the Lorentz force equation,
\begin{align}\label{eq:LFeqn}
\frac{dp^{\mu}}{d\tau}=qF^{\mu\nu}u_\nu
\end{align}
where $p^\mu=mu^\mu$.  Comparing \eq{LFeqn} to \eq{chibasic}, we immediately notice that $\chi$ is the magnitude of the Lorentz force on an electron, normalized to the electron mass, as well as (electron charge times) the center-of-mass energy in the collision between an electron and a Compton-volume of laser field
\begin{align}
\chi=\frac{|F^{\mu}_{\rm LF}|}{m_e^2}=\sqrt{e^2p_e\cdot P_\ell},\\
P_\ell^\mu = T^{\mu\nu}u_\nu
\end{align}
and hence also the leading-order acceleration of an electron in natural units \cite{rafelski2013critical}.  Solving the Lorentz force for arbitrary electron initial conditions requires complete knowledge of the field strength in space and time, which is technically difficult to achieve in an experiment.  The most common approach is to use generic models based around the approximately plane-wave character of a laser field.  An analytic solution for the momentum and trajectory is known in case of a plane wave \cite{heintzmann1972acceleration, piazza2008exact, Hadad:2010mt}, but no analytic solutions are known for even the simplest models of focused laser fields (bounds on the maximum scattering angle however have recently been obtained \cite{holkundkar2020complete}).

To highlight differences from more realistic models of the laser fields, we review properties of the plane wave limit in which \eq{Gamma1loop} is obtained.  A plane wave field is characterized by a single lightlike 4-vector.  We utilize the lightcone basis of 4-vectors $\{ n^\mu, \bar n^\mu, \epsilon_\perp^{\mu}\}$, which has the properties that $n\cdot \bar n=2$ and $n^2=\bar n^2=0=n\cdot \epsilon_\perp=\bar n\cdot\epsilon_\perp$.  Taking the wave to be propagating in the $+\hat z$ direction, a plane wave of arbitrary spectral composition is a function of only $x_-=\bar n\cdot x=t-z$.  This property implies that the minus lightcone momentum of the electron $p_-=\bar n\cdot p=E-p_z$ is conserved, and the Hamilton-Jacobi equation can be integrated directly \cite{landau1989classical} or the Hamiltonian solved algebraically with the help of additional constants of motion \cite{beers1972algebraic,heinzl2017exact}.  This lightcone symmetry also implies various algebraic identities obeyed by the field tensor, in particular $(F^2)^{\mu\nu}\propto \bar n^\mu\bar n^\nu$, which leads to the simple result $\chi=(p_-/m_e)a_0\omega_\ell/m_e$ for all time.  Thus in a plane wave field and neglecting radiation losses, $\chi$, which is in general a function of electron momentum $\vec p_e$ and position $\vec x$ and time $t$ through the electromagnetic field tensor, is a constant of motion 
\begin{align}\label{eq:chiplanewave}
\chi(\vec p,\vec x,t)=\frac{p_-}{m_e}\frac{a_0\omega_\ell}{m_e} \quad \mathrm{(planewave)}
\end{align}
and completely characterizes the interaction of the electron with the classical field.  This discussion reveals a second condition to be satisfied in order to use \eq{Gamma1loop} as a reasonable approximation to a local emission rate: $p_-$ must evolve slowly, on a timescale much greater than the coherence time for the emission process.

Away from the plane wave limit, $p_-$ is not conserved.  For example, the gaussian beam model reviewed in the appendix contains a factor depending on spatial distance from the focus and hence the longitudinal coordinate $z$ alone (not in the lightcone combination with $t$).  Finite transverse size of the beam and finite temporal duration of the pulse mean that electron trajectories need not pass within $\sim 1/e$ of the laser's peak field strength in either space or time.  As a rough estimate, consider that the magnetic field strengths in the focused laser will exceed $10^5$ T meaning that even a 5 GeV electron can be deflected transversely by 2 microns over 10 microns of longitudinal propagation--enough to miss the central $1/e$ radius of an f/1 focused laser pulse.

Classical radiation losses also cause $p_-$ to evolve in time, decreasing from cycle to cycle in the plane wave limit \cite{Hadad:2010mt}.  Momentum loss due to classical radiation is described by the Lorentz-Abraham-Dirac equation \cite{dirac1938classical}, which is equivalent at the same fixed order of precision to the Landau-Lifshitz equation \cite{landau1989classical,spohn2000critical}.  The Landau-Lifshitz equation of motion is
\begin{align}\label{eq:LLeom}
\frac{dp^\mu}{d\tau}=qF^{\mu\nu}u_\nu+\tau_0\left(u^\alpha\partial_\alpha qF^{\mu\nu} u_\nu+qF^{\mu\nu}qF_{\nu\alpha}u^\alpha - u^\mu (uqF)^2\right)
\end{align}
where the time constant is
\begin{align}
\tau_0=\frac{2e^2}{3m_ec^3}\simeq 6.24\times 10^{-24}\mathrm{s}.
\end{align}
The first term in \eq{LLeom} is the Lorentz force, and as the QED coupling is made arbitrarily small $e^2\to 0$, the time constant also vanishes $\tau_0\to 0$, the radiation-loss correction vanishes. The correction to the trajectory turns on smoothly as the electron energy and/or field strength increase toward the $\chi\gtrsim 1$ regime.  The primary results below utilize \eq{LLeom} to determine particle trajectories, and for completeness we briefly compare the outcome of Lorentz-force only dynamics to the Landau-Lifshitz dynamics.

Together these realities mean that $\chi_b$, constructed from the initial electron energy (equivalently initial $p_-$) and the peak laser field strength, need not be attained at all during the electron-laser crossing.  On the other hand, considering that the $\chi$ is equivalent to the magnitude of the acceleration, in a focussed beam $\chi$ could in principle dynamically exceed $\chi_b$.
Therefore, the beam estimate $\chi_b$ cannot be taken for granted as an accurate indicator of either the ``typical'' or the largest $\chi$ achieved during the collision.  In fact, hadron colliders have a similar issue: the center-of-mass energy of most events is not the beam center-of-mass energy but is instead the parton-level center-of-mass energy, which in all but a few cases is less than the beam center-of-mass energy.  

To obtain more information about how many strong-field QED events a facility may produce, we investigate the distribution of $\chi$ achieved by electrons in a bunch counter-propagating to a laser.  Since $\chi$ varies with time for each electron in the bunch but we want a distribution only in $\chi$ to characterize the entire collision dynamic, we select the maximum $\chi$ for each electron:
\begin{align}
\chimax \equiv \max_\tau \chi\left(x^\mu(\tau)\right)
\end{align}
where $x^\mu(\tau)$ is the trajectory of the electron parameterized by its proper time $\tau$.  In the next section, we obtain example distributions $dN/d\chimax$, showing how many electrons achieve given values of $\chi$ for varying facility parameters.

While the time-averaged $\chi$ for each particle might also be an interesting measure, the distribution of $\chi$ achieved dynamically provides only some visual intuition for our other quantitative results and therefore we do not pursue or compare the various options for defining it.  Because $dN/d\chimax$ does not invoke the probability of a QED event, it does not summarize the efficiency of the facility in producing QED events.  For that we need the laser-electron collider analog of particle colliders' luminosity.

\subsection{Luminosity for particle colliders}

Conventional high-energy particle collider experiments consist of two counter-propagating bunched beams of particles.  The beams are stored in two adjacent tubes of a circular ring and magnetically focused at an interaction point, where instrumentation is prepared to measure the properties and distributions of particles created in the collisions.  The number of collision events is proportional to the number of particles in each beam (bunch), the frequency with which they cross and the cross section for a collision.  The event rate is obtained from the standard statistical definition for 2-body collisions
\begin{align}
\frac{dN}{dt}=\int d^3xd^3p_1d^3p_2 f_1(\vec x,\vec p_1,t)f_2(\vec x,\vec p_2,t)v_{12}\sigma(E_{12}),
\end{align}
where $v_{12}=|\vec v_1-\vec v_2|$ is relative velocity and $E_{12}=\sqrt{E_1E_2-\vec p_1\cdot\vec p_2}$ is the center-of-mass energy, the conventional input for the cross section $\sigma$.  For monoenergetic beams, the momentum integrals are evaluated by $\delta$-functions and the cross section factors out, leaving the overlap integral in space, \cite{herr2006concept,grafstrom2015luminosity}
\begin{align}
\frac{dN}{dt}=\sigma(E_{cm})v_{12}\int d^3x n_1(\vec x,t)n_2(\vec x,t)\equiv \sigma L.
\end{align}
The latter equality defines the luminosity, which clearly depends only on the beam distributions in space and time.

Most facilities run particle beams that consist of temporally distinct bunches.  In this case, we can separate the timescale of the bunch crossing (e.g. $\sim 1$ ns for the LHC) from the bunch separation ($\sim 25$ ns) to write \cite{grafstrom2015luminosity}
\begin{align}
L=2c \nu_c\int  n_1(\vec x,t)n_2(\vec x,t) d^3xdt
\end{align}
where $\nu_c$ is the frequency of bunch crossing and $v_{12}\simeq 2c$ for the relevant case of highly relativistic particles.  In the baseline example that both bunches collide head-on (3-momenta antiparallel), are axially symmetric, and are described by the same gaussian distributions in the transverse and longitudinal directions, the overlap integral is easily evaluated to
\begin{align}
L=\frac{
\nu_cN_1N_2}{4\pi \sigma_\perp^2}
\end{align}
where $N_i=\int n_i d^3x $ is the total number of particles in bunch $i=1,2$ and $\sigma_\perp$ is the width of the gaussian transverse distribution.  $2\sigma_\perp^2$ is generalized to dependence on 4 distinct widths in case the bunches are not the same shape.  It is straightforward to show that this result does not depend on the temporal/longitudinal shape of the bunches.

More important for our discussion are the corrections from more realistic conditions, including non-vanishing crossing angle, focusing of beams near the design collision point and electromagnetic fields in each bunch dynamically affecting particle trajectories.  These corrections are geometric and generally of order 1, and therefore the parametric dependence of $L$ on the bunch shape and crossing frequency are preserved.  We may absorb these corrections into a constant coefficient $F$ to write \cite{pdg2020}
\begin{align}\label{eq:Lwithprefac}
L=F\frac{\nu_cN_1N_2}{4\pi \sigma_\perp^2}.
\end{align}
Explicit examples of corrections included in $F$ can be found in the references \cite{grafstrom2015luminosity,pdg2020}.

The total number of events expected for a given process with cross section $\sigma$ is the time integral of $dN/dt$.  Since the cross section is independent of time, we see that the relevant quantity for determining the number of events is the integrated luminosity,
\begin{align}
N_{events}=\int \frac{dN}{dt}dt=\sigma\int Ldt,
\end{align}
which does not have a specialized notation.  The integrated luminosity has units of inverse area and is now frequently used to measure data accumulation and progress toward a rare low-cross section event.

\subsection{Luminosity for laser-electron colliders}

As described above, the luminosity of a particle collider measures the efficiency of the facility in delivering particles to collide.  It is independent of the center-of-mass energy of the 2-particle collisions and is equally important in quantifying how useful the facility is in studying new physics.  Due to the essential differences in the collision dynamics, we cannot directly use the particle physics definition of luminosity and must adapt a modified defintion.  Key features of the luminosity to reproduce are a) an estimate of the event rate, and b) isolating the characteristics of the colliding beams/particles from the probability of an event.
We define a normalized event rate, which we might also call `luminosity', as
\begin{align}\label{eq:Leldefn}
\LeL &=\frac{1}{\Gamma_{b}}\frac{dN}{dt} \\
\frac{dN}{dt}&=\nu_cP[e\to X]
\end{align}
where $dN/dt$ is the (macroscopic) event rate, $\nu_c$ is the repetition rate of the experiment (measured in Hz) and $P[e\to X]$ is given \eq{Pinclusive}, i.e. the probability for an electron to generate any observable event.  We write the event rate with the facility repetition rate, because the time separating collisions (milliseconds at best) is much greater than the duration of the collision (roughly the laser pulse duration, 200 fs or less), as for conventional collider facilities.  Since we are interested in how much time, on the hours or days scale, that the facility must run, the repetition rate is the more relevant timescale. 
Clearly $L_{\rm e\ell}$ depends on a large set of collision parameters, notably the electron spatial distribution, laser pulse field spatial and temporal distributions.  Similar to the dynamical corrections to the collider luminosity (recall \eq{Lwithprefac}), $\LeL$ depends on the electron energy through the electron dynamics and trajectories. Unlike collider luminosity having units of (area$\times$time)$^{-1}$ (i.e. inverse to the cross section and duration), $\LeL$ is dimensionless.

To demonstrate how \eq{Leldefn} parallels the definition of luminosity in conventional colliders, we consider conditions under which $\LeL$ can be evaluated analytically.  Using the fact that $\chi$ in a planewave \eq{chiplanewave} is a constant of motion, we can compute the luminosity analytically in the limit of a planewave laser pulse.  A counterpropagating monoenergetic ultrarelativistic electron beam can be modeled by a distribution function
\begin{align}
f_e(\vec x,\vec p,t)\simeq n(\vec x_\perp,x_+)(2\pi)^3\delta^2(p_\perp)\delta\left(p_z-\sqrt{E_b^2-m^2}\right)
\end{align}
where $E_b$ is the electron beam energy before interacting with the laser.  The spatiotemporal distribution depends only on $x_+$ since the bunch is moving in $-\hat z$ direction at $v_e\simeq 1$.  Inserting this and other definitions into \eq{Leldefn}, the integrations are straightforwardly carried out in lightcone coordinates, showing that the integration over the bunch decouples from the integration over the event rate.  Note that three of four components of the canonical momentum $\Pi^\mu=p^\mu-eA^\mu$ are not conserved, but only the energy appears in the event rate and is a function only of $x_-$, $\Pi^0=E_e(x_-)$.  We can exhibit the (in)dependence on the electron beam energy by separating the dimensionful prefactor $\alpha m^2/3E_e$ from the expression for the 1-loop rate \eq{Gamma1loop}, writing $\Gamma(E_e,\chi_e)=(\alpha m^2/3E_e)G(\chi_e)$, 
\begin{align}
\LeL&=\nu_c\left(\frac{\alpha m^2}{3E_b}G(\chi_b)\right)^{\!-1}\int n(x_\perp,x_+)d^2x_\perp dx_+\int \frac{\alpha m^2}{3E_e(x_-)}G\big(\chi_e(x_-)\big) \frac{dx_-}{2}\notag\\
\label{eq:LeLplanewave}
&=\nu_cN_b\frac{E_b}{G(\chi_b)}\int \frac{dx_-}{2}\frac{G\big(\chi_e(x_-)\big)}{E_e(x_-)}.
\end{align}
In the second line, we define $N_b$ as the number of electrons in the bunch.  Now since $\Gamma$ has a simple power-law dependence on $\chi_e$ and depends on the constant $p_-$ in the planewave limit, we write
\begin{align}
\chi_e(x_-)=\frac{p_-}{m}\frac{a_0\omega_\ell}{m}f(x_-)
\end{align}
isolating the $x_-$ dependence in the temporal profile function describing the laser pulse.  Thus, numerical constants and the peak laser field strength in $\Gamma$ cancel between the rate in the integrand and the normalization, with the result
\begin{align}
\LeL=\nu_cN_bE_b\times
\begin{cases}
\int\frac{dx_-}{2}\frac{f(x_-)}{E_e(x_-)} & \chi\ll 1 \\
\int\frac{dx_-}{2}\frac{\left(f(x_-)\right)^{2/3}}{E_e(x_-)} & \chi\gg 1
\end{cases},
\end{align}
where the 2/3 power in the second result comes from the $\chi$ scaling in \eq{Gamma1loopasymptotics} (with some obvious caveats about how that case can be applied).  The energy in the denominator can be computed from the analytic solution to the Landau-Lifshitz equation \cite{Hadad:2010mt}, which contains in the limit $e^2\to 0$ the solution for the Lorentz force.  To illuminate the parameter dependence, consider the case that radiation reaction can be neglected.  Then the energy can be written in terms of the lightcone momenta
\begin{align}
2E_e(x_-)&=\Pi_+(x_-)+p_-=p_+^{(0)}+\frac{(eA_\perp)^2(x_-)}{2p_-}+p_- \notag \\
&=2E_b\left(1+\frac{1}{2}\left(\frac{m}{E_b}\frac{a_0}{2}\right)^2\mathcal{F}(x_-)^2\right),
\end{align}
where $\mathcal{F}(x_-)$ is the profile function obtained for the (transverse) vector potential from the profile function $f(x_-)$ defined for the field strength.  Note that for $a_0\simeq 100-150$ the second term in parentheses is $\ll 1$ for $E_b\gtrsim 200$ MeV, but for $E_b\gg 200$ MeV radiation reaction corrections become relevant.  Thus for this case and for lower intensities $a_0< 100$, the luminosity manifestly depends only on the laser and electron bunch parameters,
\begin{align}\label{eq:LeLsmallchi}
\LeL&\simeq\frac{1}{2}\nu_cN_b\Delta\tau \int d(x_-/\Delta\tau)f(x_-)\qquad (\mathrm{planewave}, \chi\ll 1), 
\end{align}
where the laser pulse duration $\Delta\tau$ is factored out as the timescale characterizing the profile function $f(x_-)$.  

This expression is limited to the $\chi\ll 1$ in order to ensure that radiation reaction is negligible, in which case the luminosity depends only on what fraction of the running time electrons are crossing the laser $(\sim \Delta\tau/\Delta T$(between collisions)$=\Delta\tau \nu_c)$ and the number of electrons in each bunch.  This result is intuitive and shows the definition \eq{Leldefn} reproduces some basic properties of the collider luminosity.  It suggests we can write, paralleling \eq{Lwithprefac},
\begin{align}
\LeL&=\frac{1}{2}\nu_cN_b\Delta\tau F_{\mathrm e\ell} 
\end{align}
where the numerical factor $F_{\mathrm e\ell} $ contains the remaining physics that must in general be evaluated by simulation, including dynamics arising from focused laser fields, nontrivial electron trajectories and radiation reaction.  In case the electron beam is much smaller than the laser spot size and the laser weakly focused (e.g. a waist parameter $\gtrsim 10\mu$m compared to an electron beam 2$\mu$m wide), the planewave approximation may suffice and we have
\begin{align}
F_{\mathrm e\ell}=\int dx_-\frac{G\big(\chi_e(x_-)\big)}{G\big(\chi_b)\big)}\frac{E_b}{E_e(x_-)} \qquad (\mathrm{planewave})
\end{align}
The finite transverse extensions of the laser pulse and electron beam would introduce an additional geometric factor effectively modifying the number of electrons interacting with the laser. 

The time integral of the luminosity multiplied by the rate of a desired event (e.g. pair creation) yields an estimate of the number of desired events
\begin{align}
\Gamma_{\rm event} \int L dt\simeq N_{\rm events}.
\end{align}
Here $\Gamma_{\rm event}$ can be approximated by the rate at $\chi_b$ without affecting the precision of the equality, which is anyway an estimate.  This streamlines comparing the capabilities of different facilities.

These definitions are straightforwardly applied to beam-beam collisions, where beamstrahlung at $\chi\sim 1$ is either an important consideration for the lepton collider luminosity \cite{noble1987beamstrahlung, telnov1990problems, chen1992differential, yokoya1992beam, esberg2014strong, bambade2019international, yakimenko2019prospect} or a potential avenue to a photon-photon collider \cite{ginzburg1981production, telnov1995principles,gronberg2014photon,takahashi2019future, tamburini2021efficient}.  In that case, the classical electromagnetic field is generated by one highly relativistic electron bunch, and the colliding bunch provides the electrons that radiate while passing through the relativistically boosted, counter-propagating EM field.  This relationship is a priori symmetric: the second bunch also generates an EM field which is highly boosted relative to electrons in the first bunch.  However, bunch shape can be manipulated to reduce beamstrahlung \cite{yakimenko2019prospect} or enhance photon emission, optimizing one bunch as the source of strong field and the other as the source of electrons \cite{tamburini2021efficient}. In any case, the luminosity is computed from \eq{Leldefn} with the laser field replaced by a model of the electromagnetic field sourced by a specified bunch.  As the electromagnetic field of a bunch is highly boosted and high amplitude, the constant crossed field emission rate \eq{Gamma1loop} can again be used, with $\chi$ defined as in \eq{chibasic} with field tensor of the bunch.

The luminosity can also be applied to estimate the event rate for experiments with other kinematics, including electrons at rest or even moving colinearly to the laser.  Of course, the electrons being at rest or comoving does not maximize $\chi$ for given laser field strength and can result in the electrons being ejected from the focal region before the highest-intensity field arrives (if they are free), and therefore reduces the event rate by orders of magnitude.  More interesting results could be obtained, and luminosity used to estimate expectations, for electrons from high-Z atoms, which are freed closer to the peak field strength and see a sudden jump in net force and acceleration (see Refs. \cite{yandow2019direct,hegelich2020reconciling} and references therein).  We may consider the detailed exploration of these kinematics in future work.

\section{Comparing and optimizing electron-laser luminosity}

With the definitions of the previous section, we can quantify and compare the capabilities of different facilities, addressing such questions as the importance of repetition rate and laser spot size.  There are two general types of facility we might consider with different implications for the electron beam. At present, radio-frequency (RF) accelerators generate electron beams with much higher quality than laser wakefield accelerators  (LWFA).  For RF accelerators, the electron beam can have sub-0.1\% energy spread ($\Delta E/E\lesssim 10^{-3}$), sub-mm-mrad normalized emittance and 100s pC of charge (see e.g. the E-144 experiment \cite{Bamber:1999zt}).  For LWFA, electron beams have achieved at best $\sim 1\%$ energy spread \cite{wang2021free}, mrad beam divergence \cite{ke2021near}, 0.1 mm mrad emittance \cite{weingartner2012ultralow,plateau2012low}  and up to a few nanocoulombs of charge \cite{couperus2017demonstration,gotzfried2020physics}, though only in a few cases are multiple of these achieved in the same experiment \cite{wang2021free,aniculaesei2022}.  The transverse sizes of the beams are similar, 1-10 $\mu$m in each case.  While the length of the electron bunches is quite different, ps-scale for RF accelerators and 10-100 fs for LWFA, this parameter does not enter the luminosity.

Following the usual practice in RF particle colliders, we evaluate the luminosity under nominal collision conditions.  The center of the bunch is aligned with the axis of the laser, i.e. the impact parameter of the bunch-pulse collision is zero.  We have checked that electrons closest to the axis generally achieve the highest $\chie$, so zero impact parameter also generally achieves the highest average $\chi$.
The peak of the laser intensity and center of the electron bunch are coordinated to pass through the same point at the same time; we have checked that either positive or negative delay relative to this coincidence generally results in smaller values of $\chi$ for all electrons in the bunch.  For simplicity in this work, the laser propagation axis and the electron bunch velocity vector are anti-parallel, though in experiment the crossing angle may be offset slightly from 180 degrees, just as in RF particle colliders.  

To be systematic in this initial study, we fix the model of the laser field to be the gaussian beam with first order spatial and temporal corrections \cite{Quesnel:1998zz}, reviewed in the appendix.  We fix a relatively simple model of the electron bunch.  Although wakefield can achieve much higher electron energies, we consider a well-collimated, small energy-spread bunch because it is likely to be more useful for experiments aiming at quantitative discovery.  This suggests considering a moderate $\chi$ regime, roughly $0.01\ll\chi\lesssim 10$, so that radiation events are probable but current, simplified numerical models of well-separated emission events can be used in simulations. 

The counter-propagating electron bunch is chosen monoenergetic with energy $E_e=400$ MeV and uniformly distributed in a cylinder of radius 2$\mu$m and length 10$\mu$m.  This beam size is chosen to be small but reasonable for a laser wakefield generated high-energy bunch.  The spatial size is chosen for demonstration, being larger than the focal volume $\sim \pi w_0^2z_R$ of the most strongly focused lasers (f/1) but significantly smaller than the focal volume of more weakly focused lasers (f/5).   Electrons outside the typical focal spot radius only contribute additional scattering at small $\chi$ and contribute much less to the luminosity. Our choice thus amounts to importance sampling.  The size of the bunch relative to the focal volume determines in part the efficiency of the facility in generating events; while specific numbers are used here for focal spot sizes, the results can be generalized by rescaling with fixed beam-diameter-to-focal-spot ratio.

We define a set of characteristic laser systems, based on the nominal system parameters of facilities worldwide and listed in Table \ref{tab:facilities}.  Using the nominal pulse energy, duration and spot size, the peak intensity and $a_0$ are estimated with the temporal profile assumed to be gaussian.  The peak intensity computed for other profiles differs by a factor of order 1, much greater than the correction from $\mathcal{O}(\epsilon_k),\mathcal{O}(\epsilon_t)$ corrections to the fields.  We will demonstrate the impact of different temporal profiles among our results.

\begin{table}
\begin{tabular}{l|c c c c c c c c c c}
Facility & $E_l$[J] & $\Delta\tau$[fs] & $P_{\rm peak}$ [PW] & $w_0$[$\mu$m] & $\lambda_l$[$\mu$m] & $I_{\rm peak}$[W/cm$^2$] & $a_{\rm peak}$ & $\epsilon_w$ & $\epsilon_t$ & Rep.[Hz] \\\hline
\TPW f/1 & 150 	& 150 	& 1 	& 1.25 	& 1.054 & $2.76\times 10^{22}$ 	& 142.5 & 0.127 & 0.022 & $10^{-4}$\\\hline
\ELInp   & 244 	& 22.5 	& 10.8	& 2 	& 0.8 	& $1.17 \times 10^{23}$ & 234.6 & 0.064 & 0.119 & 0.02 	\\\hline
\Arc     & 7 	& 30 	& 0.23	& 1.1 	& 0.8 	& $8.26\times 10^{21}$ 	& 62.34 & 0.115 & 0.089 & 5 	\\\hline
\ELIhapl\footnote{Original specifications used here rather than best performance to date}
	 & 30 	& 28 	& 1.1	& 1.9 	& 1.054 & $1.28\times 10^{22}$	& 102.38 & 0.088 & 0.126 & 10	\\\hline
\Cor\footnote{CoReLS can now achieve $>10^{23}$ W/cm$^2$ with $>100$ J pulse \cite{yoon2021realization}}
 	 & 50 	& 30 	& 1.67	& 1.8 	& 0.8	& $2.22\times 10^{22}$ 	& 127.7 & 0.088 & 0.111 & 0.1 	\\\hline
\OPAL 	 & 600 	& 20 	& 30	& 1.25 	& 0.9 	& $8.28\times 10^{23}$ & 702.4 & 0.0839 & 0.156 & $10^{-4}$
\end{tabular}
\caption{The pulse duration is measured as intensity full-width half-max.  For peak intensity estimates, the temporal profile is assumed to be gaussian. \label{tab:facilities}}
\end{table}

The distributions and expectation values of quantities are evaluated by Monte Carlo simulation, typically with 250 sample particles, which we expect is much smaller than the real bunches used but provides sufficient numbers for reliable estimates of the measured distributions.  We make no simplifications or approximations in the numerical implementation of the particle dynamics \eq{LLeom}, the given prediction of the event rate \eq{Gamma1loop} or specified models of the laser and electron beam.  As noted above, the $\chi\lesssim 10$ regime does not require modeling the recoil, which would introduce an additional stochastic dynamic into the electron trajectories and thus require many more samples.

\subsection{Focal spot size}
As mentioned in the introduction, an obvious trade-off arises in the choice of the focal spot size of the high intensity laser: a smaller spot size increases the field strength (for fixed pulse energy) but decreases the cross sectional area and focal region volume generally decreasing the overlap with the electron beam.  Higher field strength increases the probability of photon emission by a single electron but fewer electrons limits the number of photons that can be emitted.  A likely optimum is to match the electron beam size to the focal spot size so that the majority of electrons are aligned with the region within $1/e$ of the maximum field strength.  However, we will find that not all facilities adhere to this expectation, for various reasons.


We vary the laser spot size keeping the laser pulse energy fixed, assuming most facilities will prefer to run at maximum (nominal) amplification to maximize performance.  Scanning pulse energy as a way to scan interaction strength at the collision however is highly encouraged as an experimental method to enhance quantitative understanding of the dynamics and outcomes of the laser-electron collisions.  

\begin{figure}
\includegraphics[width=0.48\textwidth]{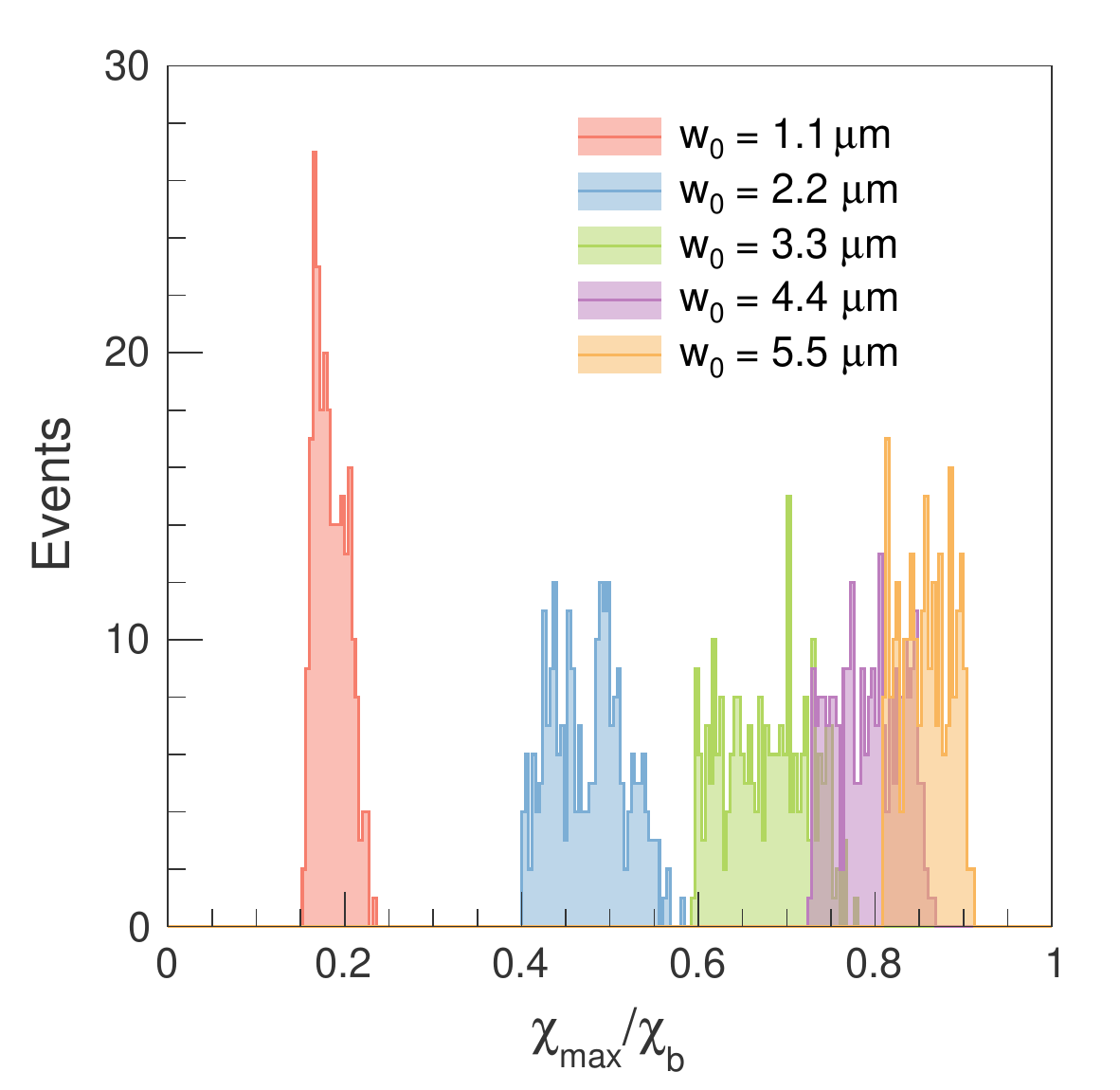}
\includegraphics[width=0.48\textwidth]{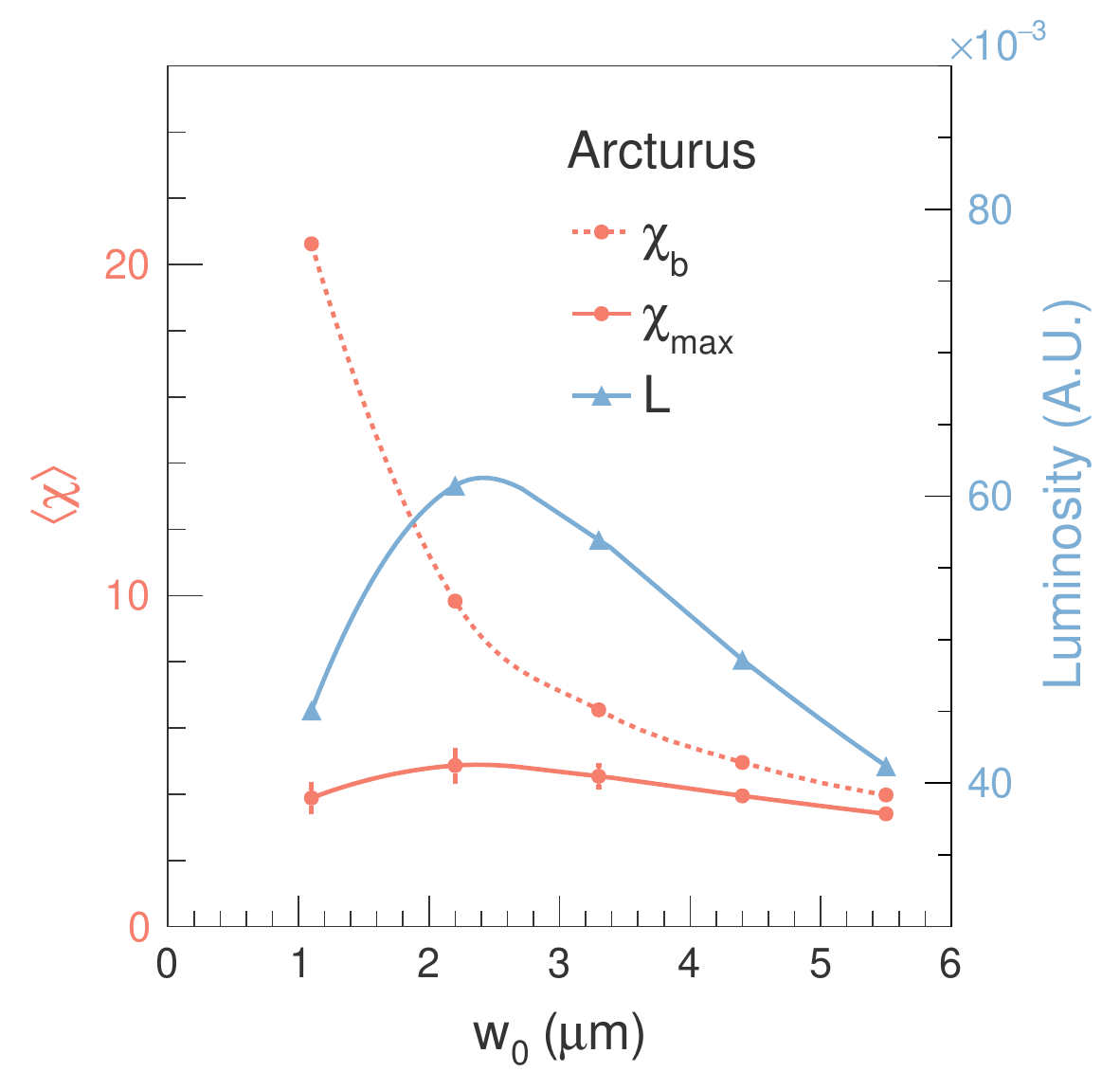}
\caption{\label{fig:arcturusw0}  The $\chimax$ distributions for various $w_0$ achieved for a 400 MeV electron beam counterpropagating to \Arc and corresponding luminosity as a function of $w_0$.  The maximum luminosity is achieved for the same spot size as the optimal distribution in $\chimax$, that is where the spot size roughly matches the electron beam radius of $\mum{2}$.}
\end{figure}

For several of the example facilities, looking at the distribution of $\chimax$ achieved is sufficient to determine the spot size that maximizes the event rate.  Arcturus, the $2\times 200$ TW laser at U. Dusseldorf \cite{cerchez2019arcturus}, a mid-size facility with a two-beam arrangement, is a good example showing that the $\chimax$ supports the optimum spot size found by calculating $\LeL$.  Figure \ref{fig:arcturusw0} shows that for $w_0=\mum{2.2}$ the $dN/d\chimax$ distribution has the highest average $\langle\chimax\rangle\simeq 5$ and a reasonably narrow distribution $\langle\Delta\chimax\rangle\simeq 0.5$.  Likewise the luminosity is maximized at $w_0=\mum{2.2}$. The strongest focusing $w_0=\mum{1.1}$ yields an average $\chimax$ similar to weaker focusing $w_0=\mum{4.4}$, but the larger focus results in a narrower distribution.  (In \fig{arcturusw0}(left), the scaling by $\chib$ also applies to the width of each distribution so that the factor four higher $\chib$ translates the roughly half as wide distribution for $w_0=\mum{1.1}$ into a distribution twice as wide in absolute terms as for $w_0=\mum{4.4}$.)  This outcome is explained by electron dynamics: the fields and gradients in the fields are strong enough at $w_0=\mum{1.1}$ to cause the electrons to diverge from the laser axis and thus miss the highest field region.   Consequently the larger focal spot sizes $w_0=\mum{3.3}$ and $w_0=\mum{4.4}$ also provide improved event rate over the strongest focusing.  For the same reason, larger focal spot sizes approach the beam estimate $\chib$. Thus, for \Arc, the luminosity partially tracks the average $\chimax$ but the narrower distribution in $\chimax$ means the luminosity at $w_0=\mum{4.4}$ is higher.  

As we have only sampled a discrete set of focal spots, we can only estimate the optimal focal spot size as in 2-\mum{3} range.  The same result is seen for \Cor (not shown): maximum luminosity is obtained for focal spots in the 2-\mum{3} range, corresponding to a similar pattern in the $\chimax$ distribution, but the smallest focal spot is similar to the beam size $w_0=\mum{1.8}$ so that the luminosity is only slightly lower than for the next larger focal spot tested, $w_0=\mum{3.6}$.  Note the contrast to the planewave estimate of the luminosity \eq{LeLplanewave}

\begin{figure}
\includegraphics[width=0.48\textwidth]{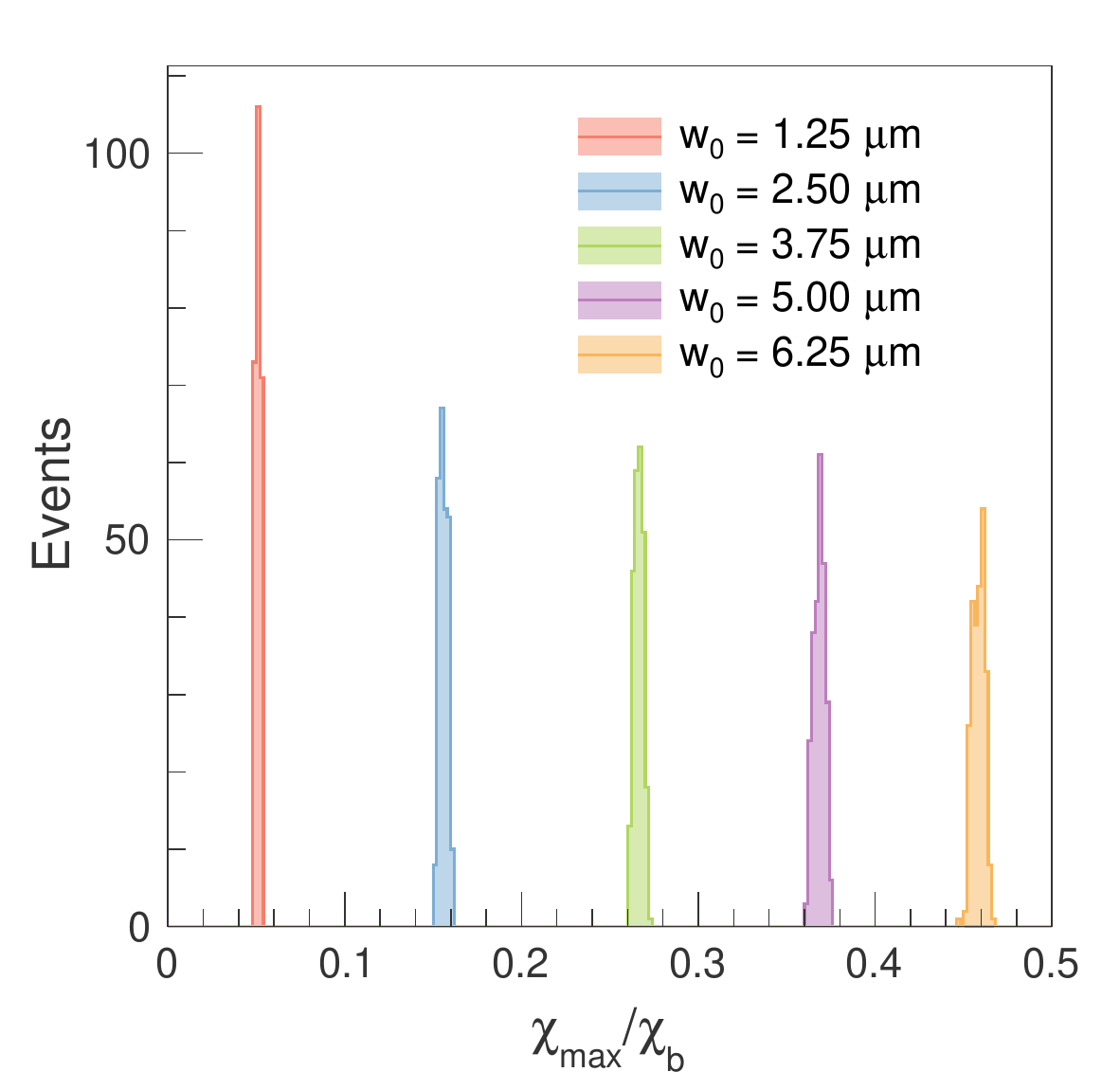}
\includegraphics[width=0.48\textwidth]{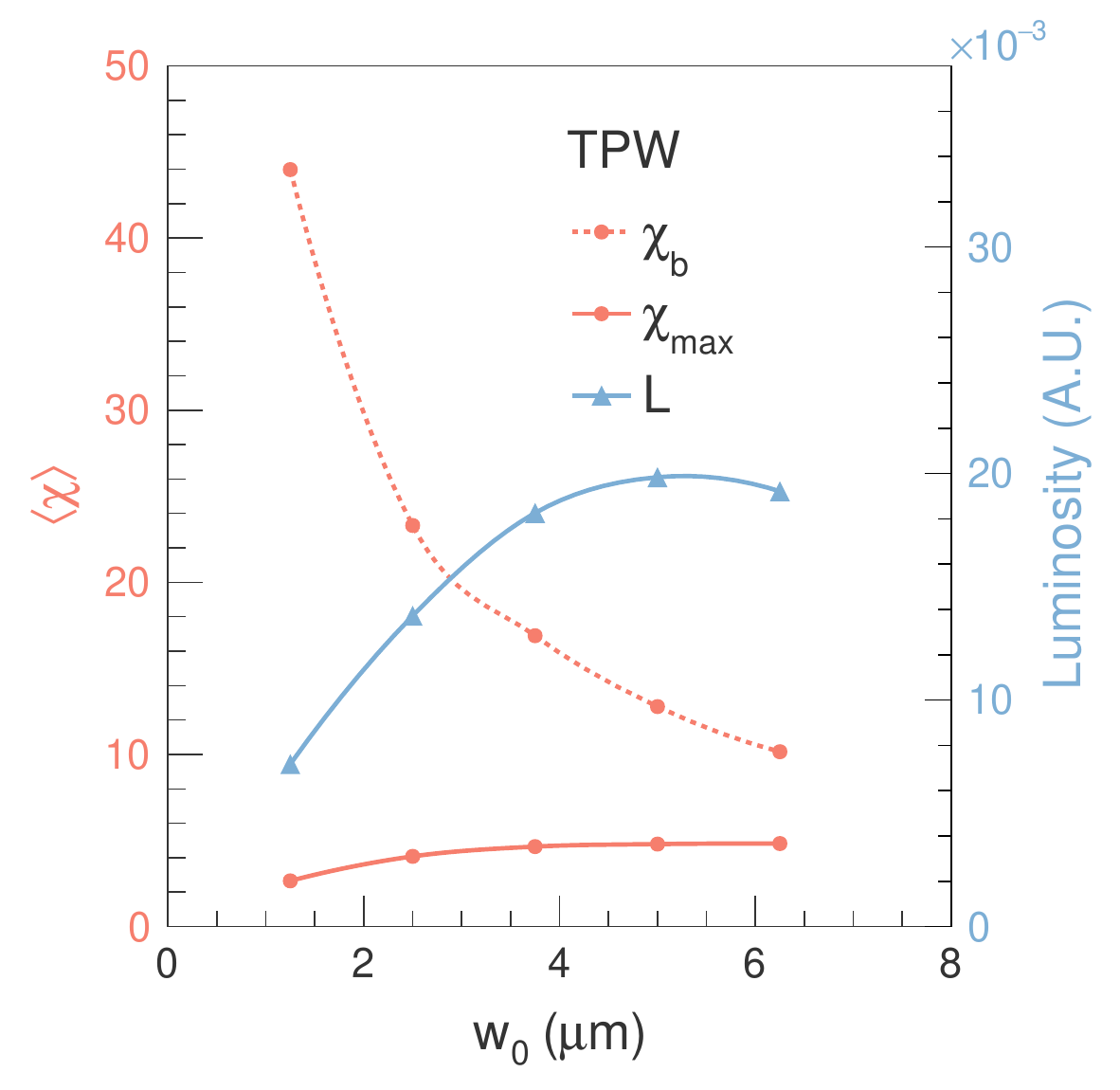}
\caption{\label{fig:TPWw0}  The $\chimax$ distributions for various $w_0$ achieved for a 400-MeV $\mum{2}$-radius electron beam counterpropagating to \TPW and corresponding luminosity as a function of $w_0$.}
\end{figure}

The \TPW facility exhibits a qualitatively different outcome, see \fig{TPWw0}.  The longer pulse length ($\sim 150$ fs) means that the acceleration of particles is generally maximized by a larger focal spot.  The larger focal spot increases the Rayleigh range $z_R\simeq \pi w_0^2/\lambda_l$.  Combined with the longer pulse length, a test particle has more distance and time in which to interact with the strongest $<1/e$ fields of the laser pulse \cite{hegelich2020reconciling}.  The luminosity is maximized for $w_0=\mum{5.0}$, and examination of the $\chimax$ distribution shows a preference for larger spot size but does not significantly distinguish between $w_0=\,3.75, 5.0$ or \mum{6.25}.

For the same reason, larger spot size increases the luminosity in case the initial electron bunch is free and not counter-propagating.  This is consistent with higher particle energies observed for f/3 focusing as compared to f/1 focusing \cite{hegelich2020reconciling}.

\begin{figure}
\includegraphics[width=0.48\textwidth]{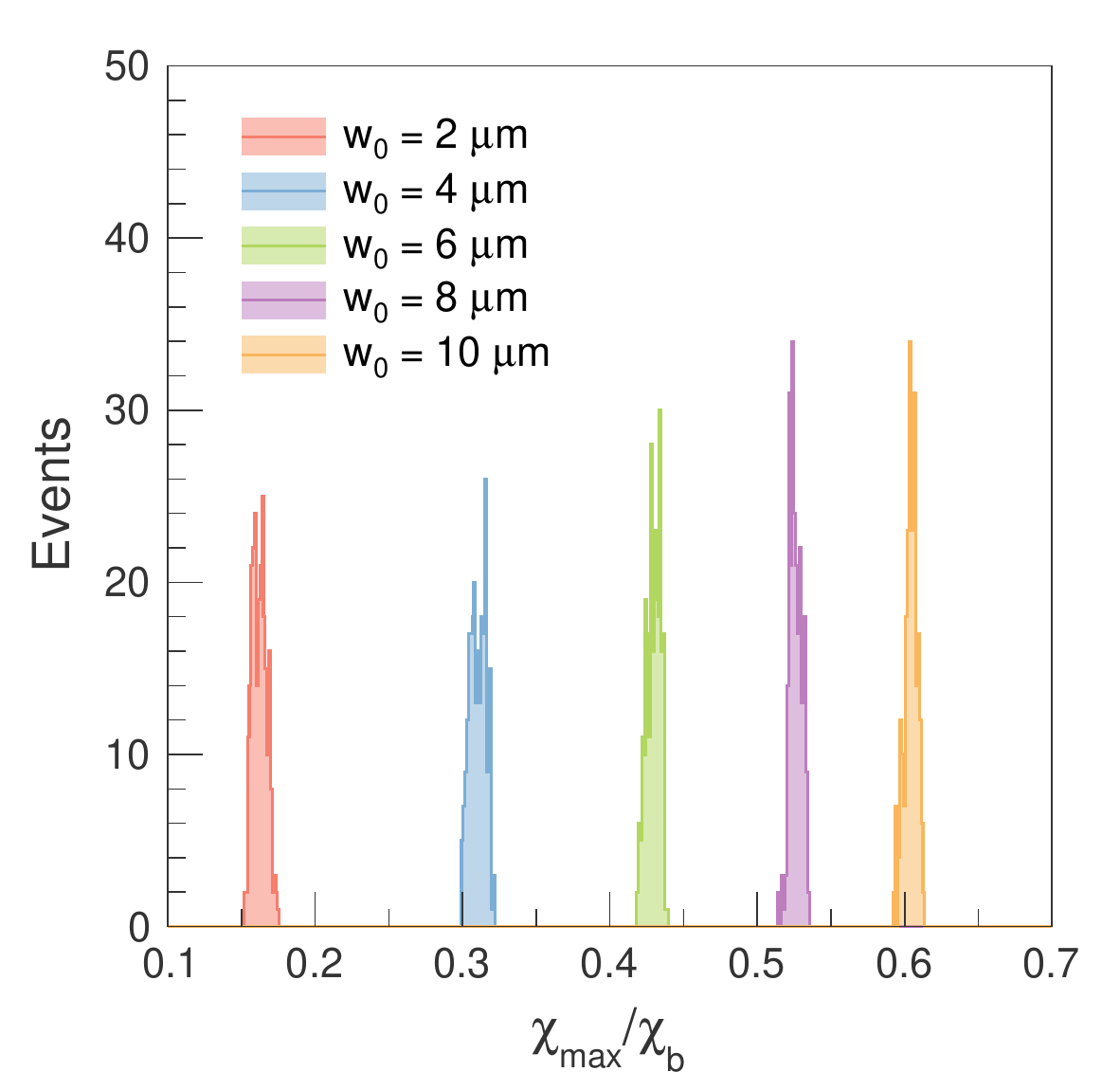}
\includegraphics[width=0.48\textwidth]{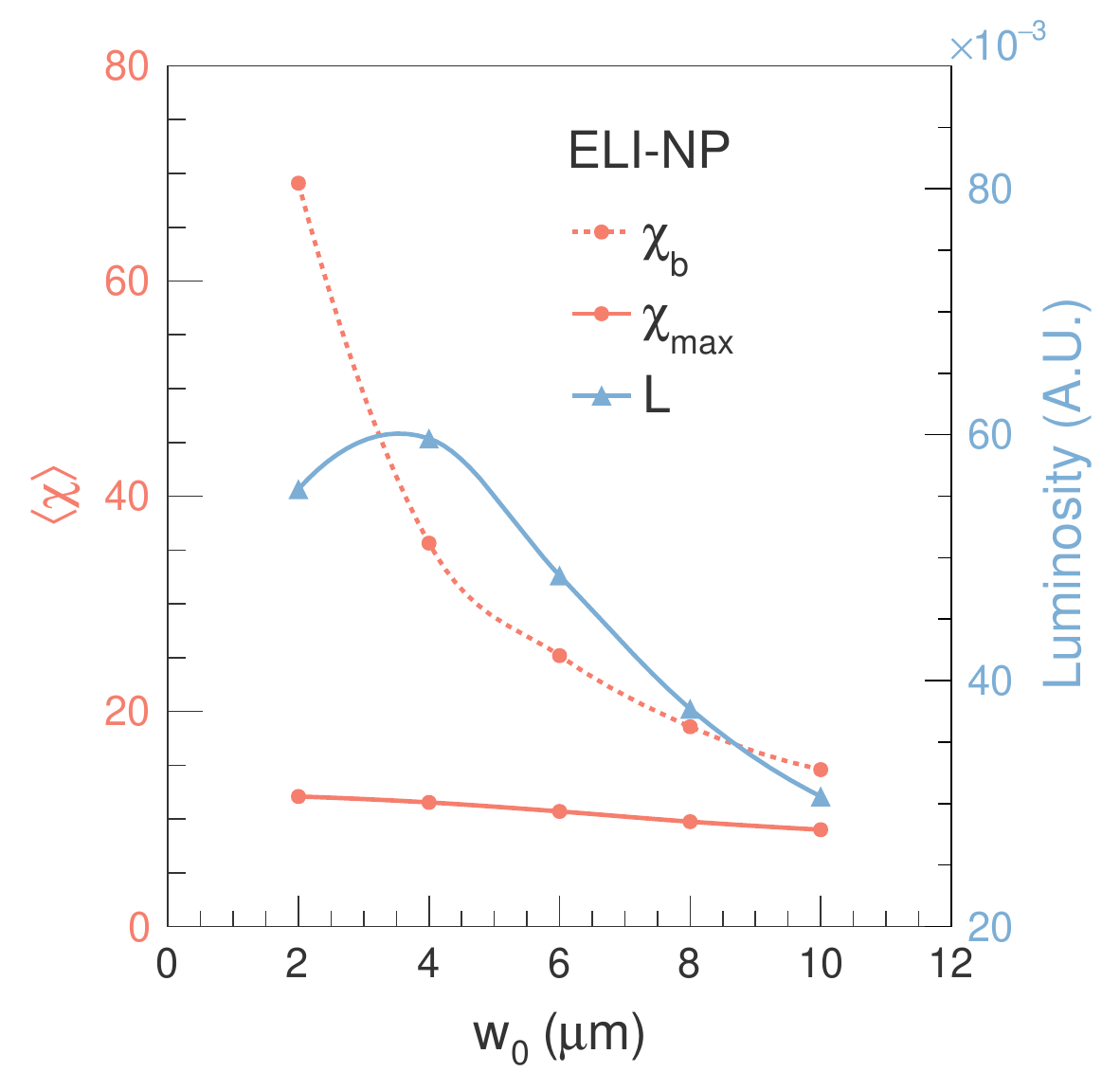}
\caption{\label{fig:ELIw0}  The $\chimax$ distributions for various $w_0$ achieved for a 400 MeV $\mum{2}$-radius electron beam counterpropagating to \ELInp and corresponding luminosity as a function of $w_0$.}
\end{figure}

For two of the facilities in our study however, \ELInp and \OPAL, the luminosity suggests an optimal spot size different from that suggested by the $\chimax$ distribution.  In \fig{ELIw0}, the $\chimax$ distribution is clearly maximized for the smallest spot size,  $\langle\chimax\rangle\simeq 12.1$ at $w_0=\mum{2.0}$ as compared to $\langle\chimax\rangle=11.5$ at $w_0=\mum{4.0}$.  A 10\% larger luminosity is obtained for $w_0=\mum{4.0}$, which is due to large focal volume allowing for longer interaction of the beam with the laser pulse.  Since the luminosity depends on the duration of the interaction, the scaling of the focal volume's length $z_R\propto w_0^2$ can compensate $\chi_b\propto w_0^{-1}$.

Finally, we note that for all facilities, the achieved $\chimax$ is less than $\chi_b$, frequently by a factor of 5-10 for the smallest focal spots.  This overestimation is due to dynamics: the electrons lose energy due to radiation reaction and can be diverted from the focal region.  In general, larger spot size is more efficient with respect to the expected $\chi_b$ \eq{chibdefn}.  For fixed energy, increasing the spot size decreases the peak field strength and therefore $\chi_b$.  Once the spot size is significantly larger than the beam, the electrons are not diverted enough from the beam axis to miss the highest field $\gtrsim 1/e$ region.  The $dN/d\chimax$ distributions show that the largest tested spot size maximizes $\chimax/\chi_b$.  For sufficiently large spot size, we expect that the achieved $\chimax$ approaches $\chi_b$, as the field strength also becomes low enough that radiation reaction ceases to be important.  An overview comparison of all six facilities' luminosities is discussed below in the context of the trade off between higher intensity and higher repetition rate.

\subsection{Temporal profile effects}
\begin{figure}
\includegraphics[width=0.48\textwidth]{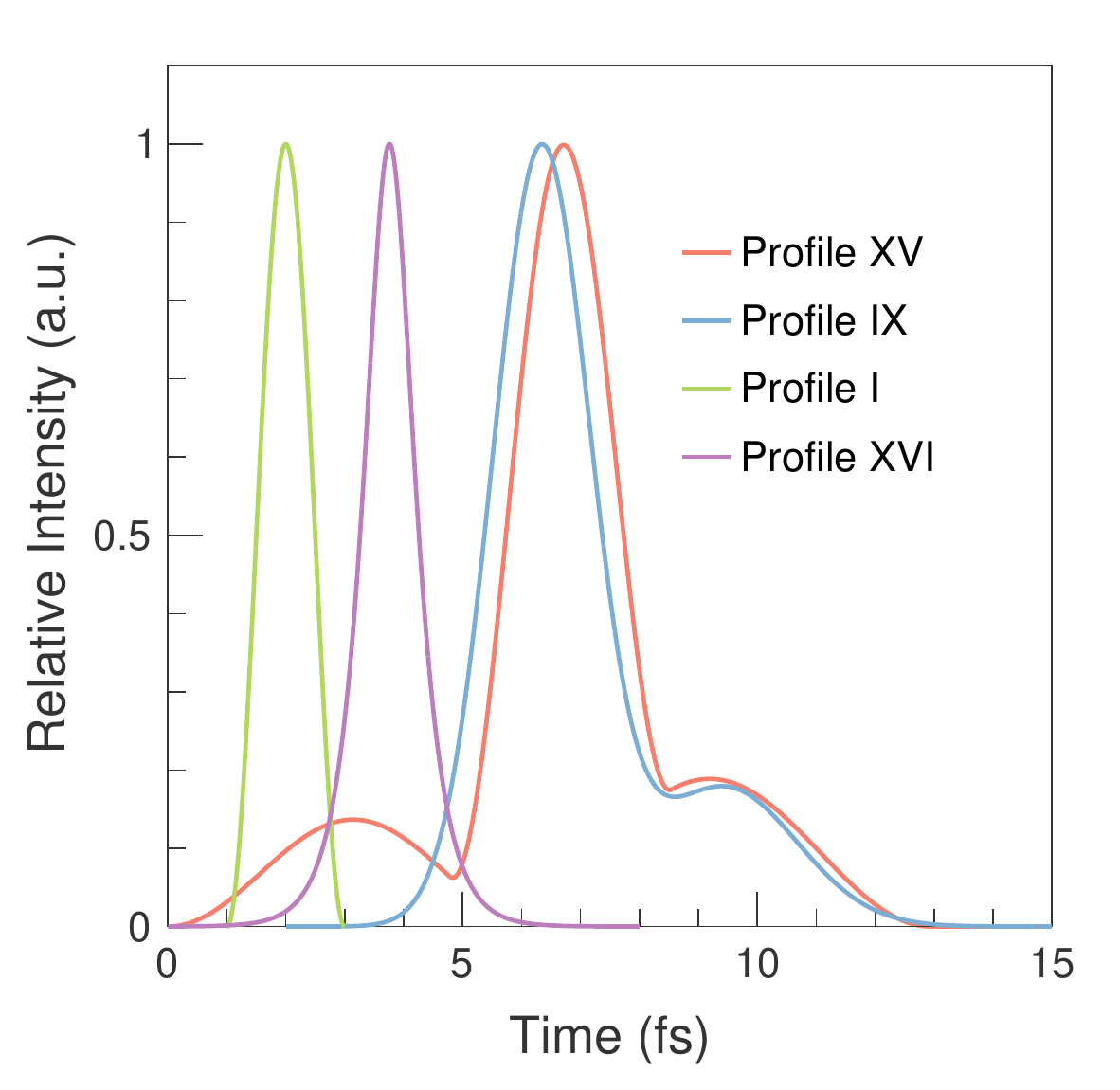}
\caption{\label{fig:pulseprofiles} Temporal profiles used in this study.
}
\end{figure}

The temporal profile of the laser pulse has long been recognized as an approach to modulating photon emission in strong-field QED \cite{Heinzl:2009nd,Seipt:2010ya,seipt2011nonlinear,Seipt:2012tn}.  Analytic calculations utilizing plane waves with different temporal profiles do not account for additional temporal modification due to focusing.  Single-particle and PIC simulations are thus required \cite{ghebregziabher2013spectral,holkundkar2015thomson,rykovanov2016controlling}.  No studies have systematically investigated the impact of pulse profile function--the problem being that some models can significantly over-estimate the energy shift of a scattered electron \cite{mcdonald1998comment,hegelich2020reconciling}.  In the present study, we use 4 profile functions: sin$^2$, gaussian, hyperbolic secant and a model of a prepulse consisting of a $\simeq 15\%$ secondary maximum $\simeq 1.5$ pulse durations ahead of the central maximum.  The prepulse-containing profile is described by the function
\begin{align}\label{eq:prepulseprofile}
f_{\rm prepulse}(\phi)=0.952f(\phi)+0.137f\left(0.6(\phi+2.93)\right)+0.189f\left(0.5(\phi-2.11)\right).
\end{align}
The relative magnitudes of the peaks and their temporal separation are based on profile measurements at one of the facilities we model.  Having found that the qualitative impact of the prepulse is the same for any input single-peak profile $f(\phi)$, we use here $f(\phi)\propto \sin^2(\phi)$ for demonstration.  These profile functions are plotted in \fig{pulseprofiles}.

\begin{figure}
\includegraphics[width=0.48\textwidth]{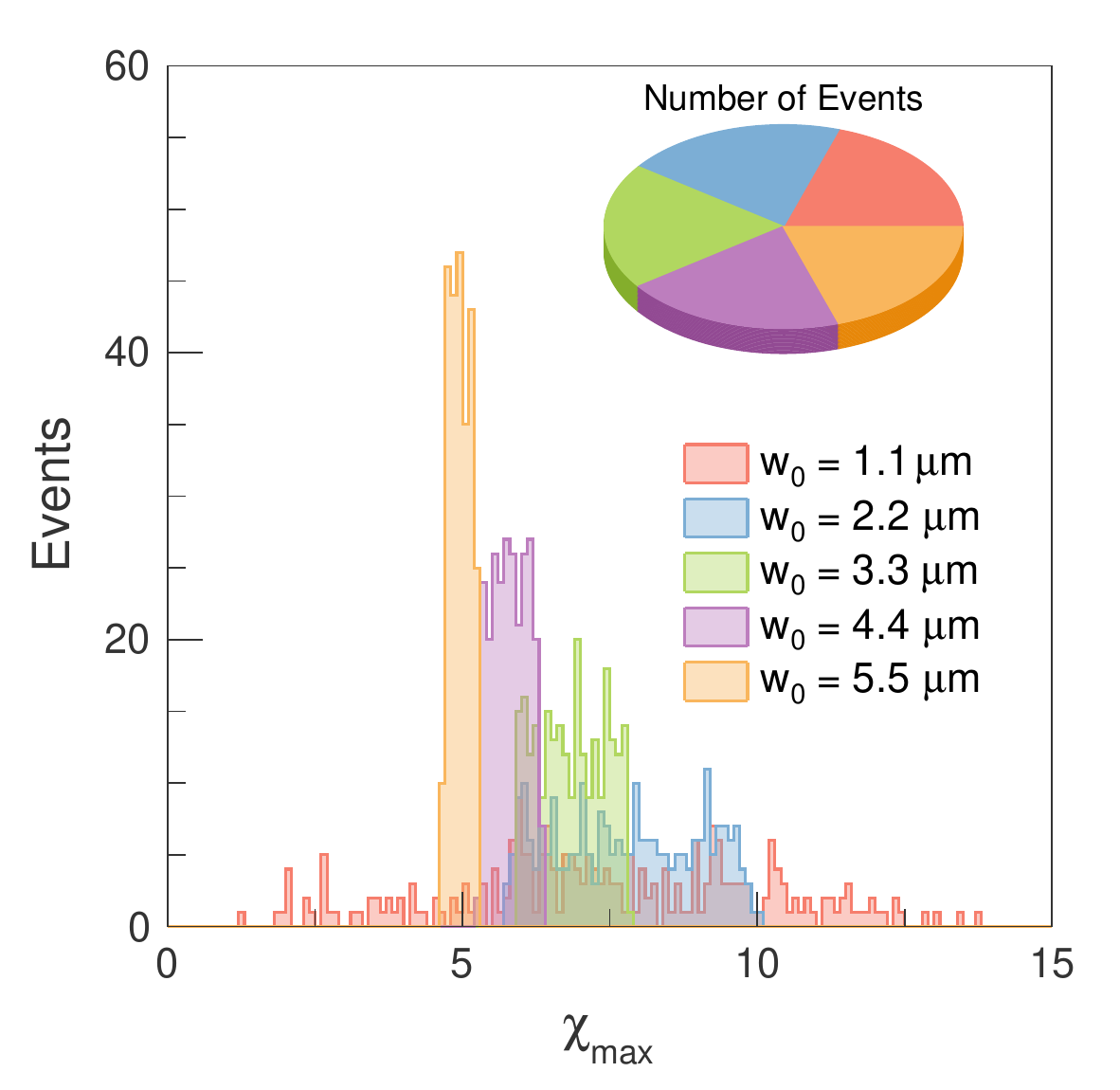}
\includegraphics[width=0.48\textwidth]{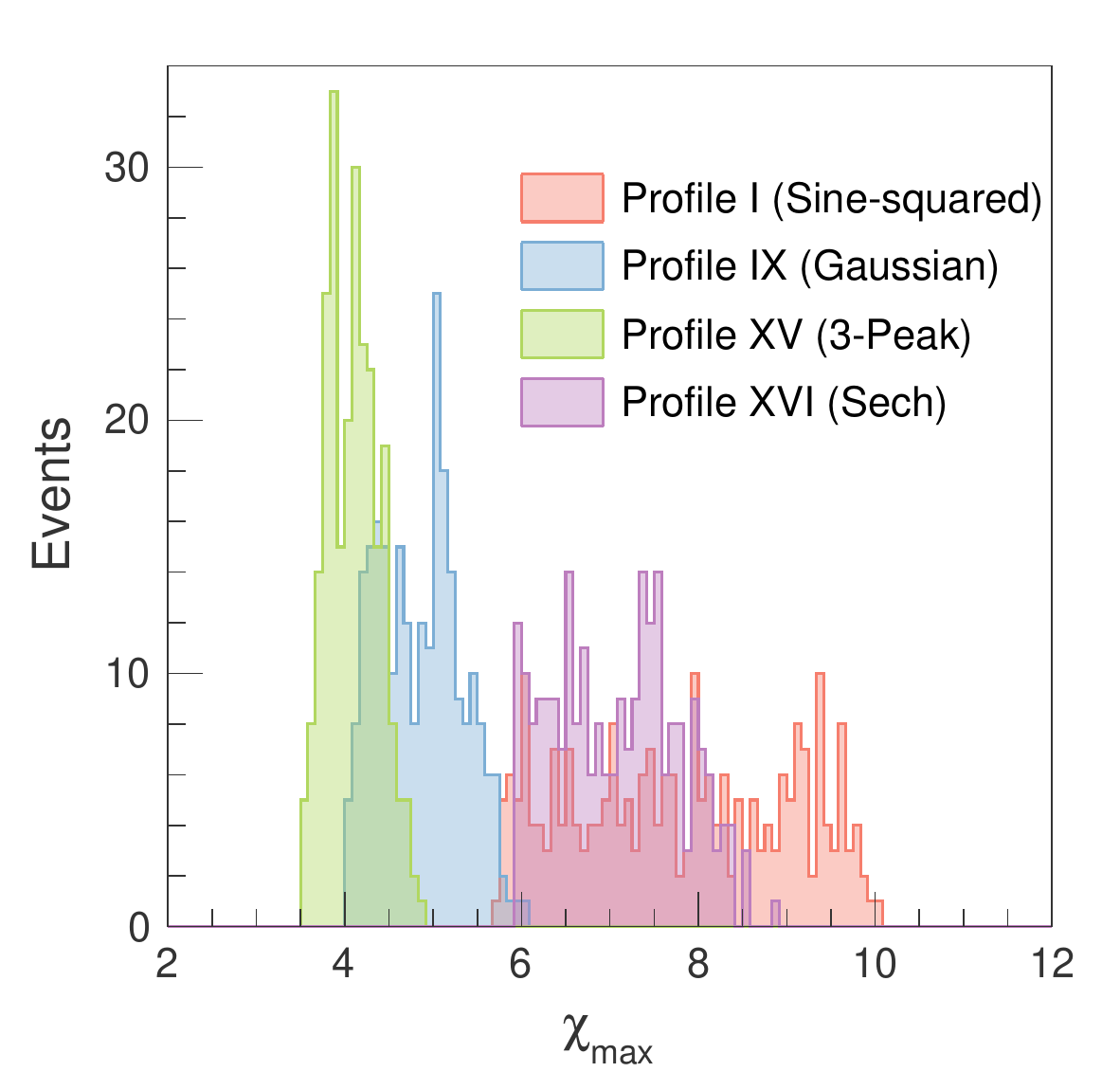}
\caption{\label{fig:Arcprofiles} Left: $dN/d\chimax$ for sin$^2$ profile and varying spot size, same electron beam as above (monoenergetic 400 MeV and radius \mum{2}).  Right: $dN/d\chimax$ for fixed spot size $w_0=\mum{2.2}$ and varying profiles with the same electron beam.
}
\end{figure}

As for vacuum laser acceleration, we find that numerically convenient pulse profiles, such as sin$^2$ or any discontinuous jump to set the initial field strength to zero, tend to overestimate the acceleration of the particles.  The sin$^2$ profile generally predicts higher $\chimax$ and a wider distribution in $\chimax$, effects consistent with the criticism that the sin$^2$ profile has unphysically large gradients \cite{mcdonald1998comment}.  Two other patterns in the results support this explanation: first, the effect is most pronounced for the shorter pulses (20-30 fs) and smaller focal spot sizes, and second the gaussian temporal profile tends to display the same enhancements relative to the hyperbolic secant.  These patterns are demonstrated with the \Arc facility in \fig{Arcprofiles}, but they hold generally to varying degree for the other facilities.

Non-ideal pulse shapes, common in short-pulse systems, can also affect the acceleration of particles entering the field.  For a model suggested by a fit to the CoReLS laser, secondary maxima of $\sim 10\%$ peak intensity dramatically decrease the energy gain when accelerating particles from rest \cite{hegelich2020reconciling}. One may expect the prepulse to be less important in the collision geometry, because the electrons have sufficient rigidity to continue through the prepulse to encounter the central maximum in intensity.  The prepulse also causes energy loss, which can be significant if either the magnitude of the prepulse is too large or the overall intensity is too high meaning that even small $\lesssim 0.1I_0$ prepulses contain very strong fields.  

\begin{figure}
\includegraphics[width=0.48\textwidth]{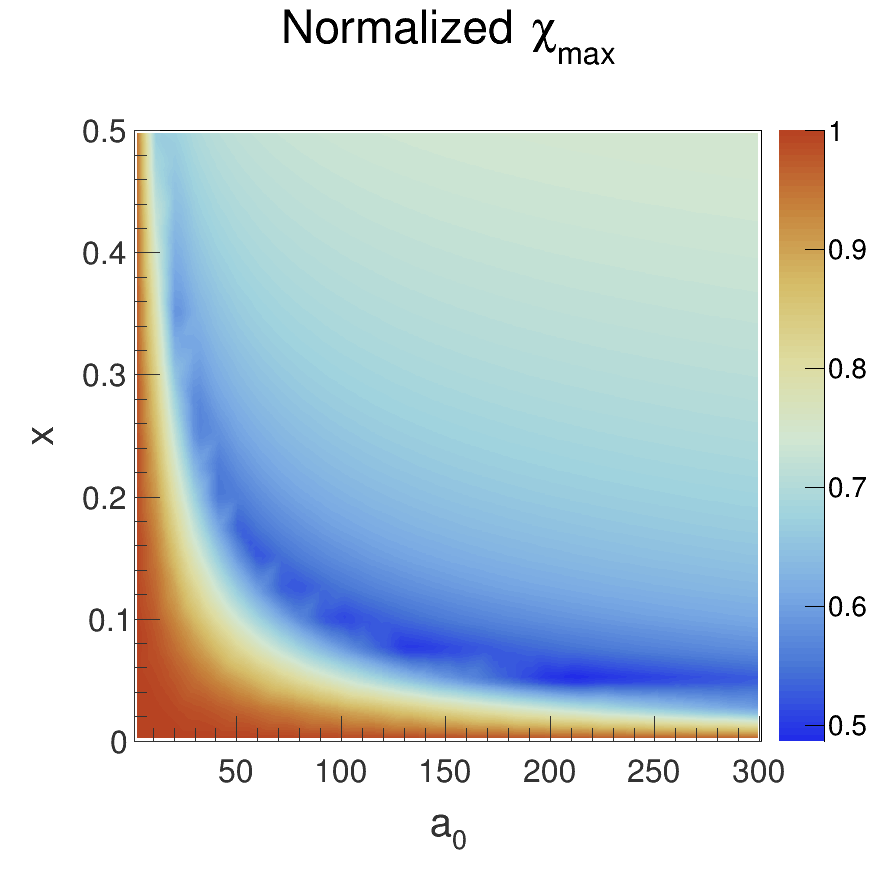}
\caption{\label{fig:1dprepulse} $\chimax$ achieved for a profile with a prepulse normalized to $\chimax$ without a prepulse, i.e. a single peak.  $x=|\vec E_{\rm max,prepulse}|/|\vec E_{\rm max,main}| $ is the ratio of the field strength achieved by the prepulse compared to main peak.
}
\end{figure}

A 1-d plane wave model builds intuition for the phenomenology of the energy-loss effect.  Taking the profile \eq{prepulseprofile} and varying the relative field strength of the prepulse (second term) compared to the central maximum as well as the overall peak intensity, we find from the analytic solution of Ref. \cite{Hadad:2010mt} that the prepulse decreases $\chimax$ by as much as 50\%.  However above a certain combination of $x$ and $a_0$, $\chimax$ is achieved in the prepulse rather than the main peak and consequently increases again slightly relative to $\chimax$ for the single-peak profile.  This transition in the maximum-achieved $\chi$ causes the nontrivial minimum curve in \fig{1dprepulse}.

A 1-dimensional model however does not account for the 3-dimensional divergence of trajectories around the region of peak intensity, which we might expect to reduce $\chimax$ even more.  Just as we have observed that the differing gradients in the profile functions have a significant impact on the achieved $\chimax$, so we find that the prepulse generally significantly decreases the achieved $\chimax$ by causing energy loss before they can encounter the central maximum and diverting the electrons from the beam path.  

Despite decreasing the $\chimax$ achieved in the collision, the prepulse generally increases the luminosity for all of the facilities modeled here.  As an integral of the event rate, the luminosity increases with the extended interaction time in the prepulse profile.  In the planewave limit, this scaling is apparent in \eq{LeLsmallchi}.  Spreading a fixed total laser energy over a longer pulse decreases $\chi$ along the trajectory by less than the linear increase in the duration of the electron-laser interaction.

\begin{figure}
\includegraphics[width=0.48\textwidth]{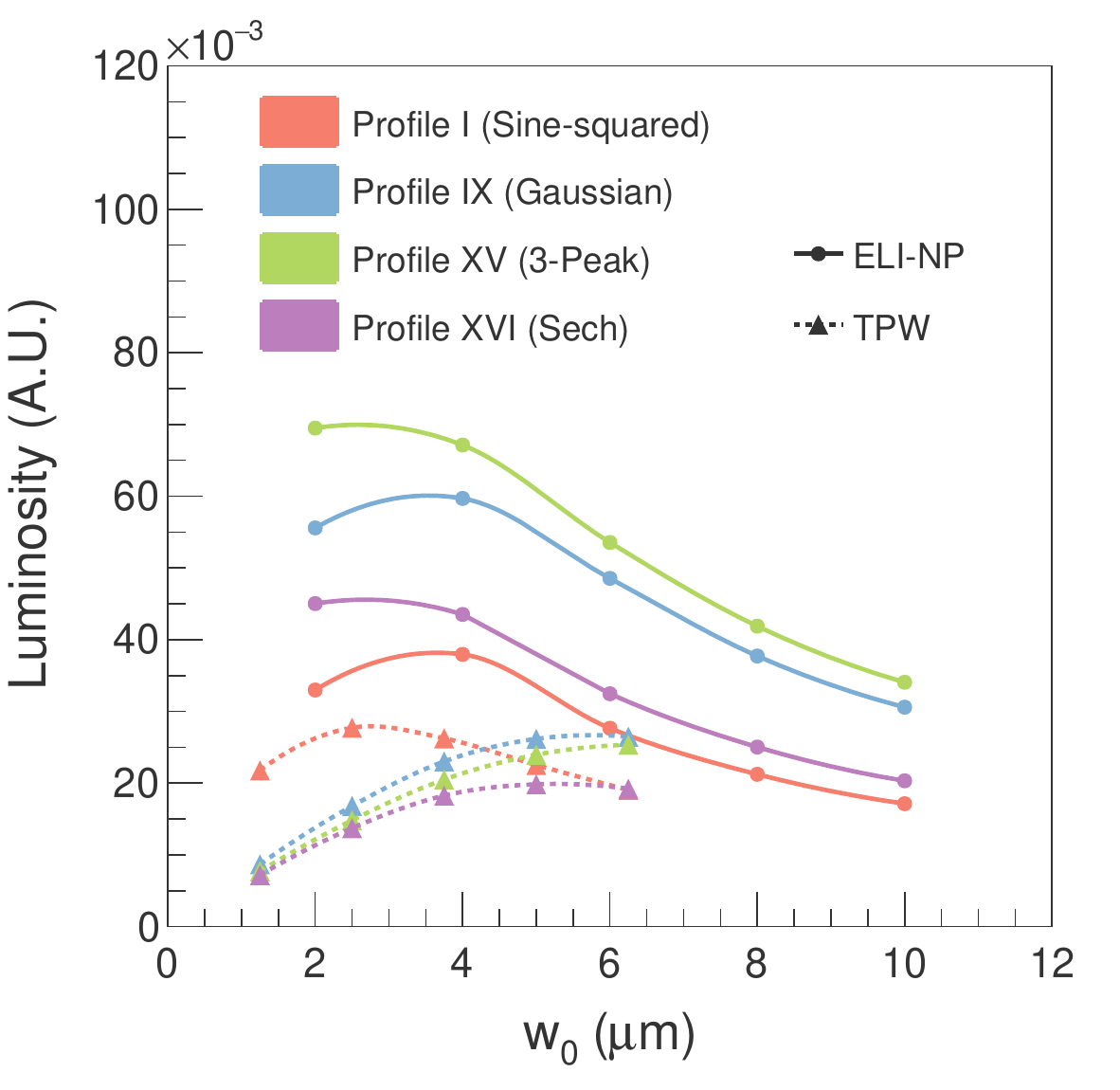}
\caption{\label{fig:ELITPWluminosity} Luminosity as a function of spot size for different profile functions for \TPW and \ELInp with the same electron beam as above (400 MeV, radius \mum{2}, length \mum{10})
}
\end{figure}

We have checked but do not show here results for profiles with an introduced jump discontinuity at the beginning of the pulse. As for previous studies of electron acceleration from rest, the discontinuity can be used as a simple threshold model of field ionization and thus the luminosity estimates the rate of strong-field radiation events in such experiments.  The presence of the discontinuity generally enhances the observed $\chimax$ and thus also the luminosity over the case of free electrons.

\subsection{Radiation reaction effect}

Here we briefly compare dynamics with radiation reaction, as modeled by the Landau-Lifshitz equation, to dynamics with radiation reaction unphysically suppressed.  To exclude radiation reaction effects, we simulate the electron dynamics with the Lorentz force only.  

\begin{figure}
\includegraphics[width=0.48\textwidth]{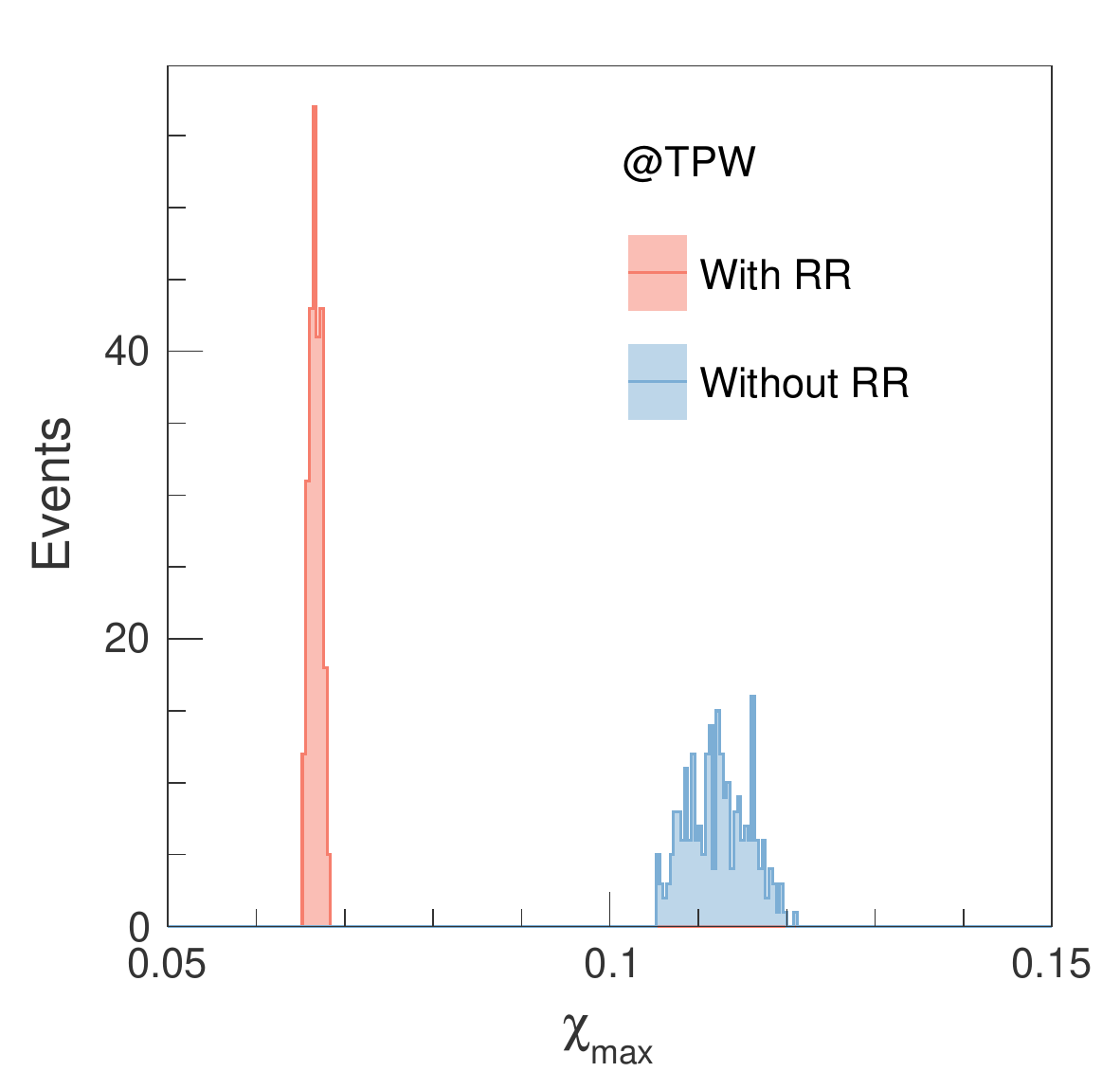}
\includegraphics[width=0.48\textwidth]{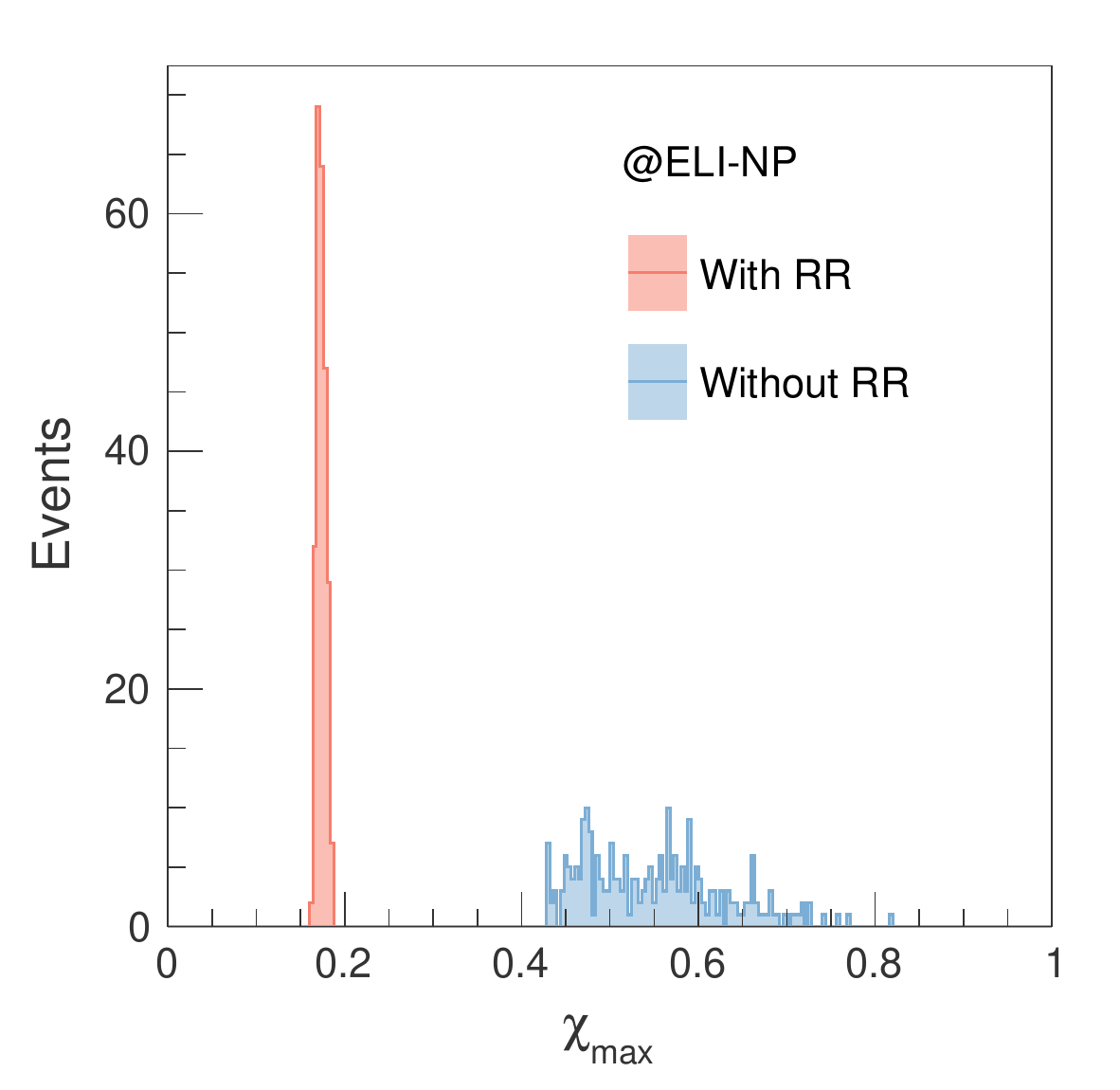}
\caption{\label{fig:chimaxRR} Left: The $dN/d\chimax$ distributions with radiation reaction (Landau-Lifshitz dynamics) and without radiation reaction (Lorentz force) for TPW $w_0=\mum{3.75}$ and the same 400-MeV \mum{2}-radius electron beam.  Right: the same comparison for ELI-NP $w_0=\mum{2.0}$ and the same electron beam. 
}
\end{figure}

As shown in Figure \ref{fig:chimaxRR}, radiation reaction generally decreases the mean $\chimax$ achieved.  The energy loss also compresses the distribution of $\chimax$, as higher energy electrons lose more energy.  The luminosity is consequently also decreased, due to two reinforcing effects.  First, radiation reaction decreases the energy of the colliding electron, and second, with smaller momentum the gradients in the laser field more strongly divert the electron trajectory from the high-intensity region.  These effects are greater for higher electron energy and negligible for electrons at rest or co-propagating, since $\chi$ values and the radiation reaction correction are strongly suppressed in the latter case.  The reduction in $\chimax$ and luminosity is greater for smaller focal spots, consistent with the importance of the field gradients in diverting electron trajectories from the high-intensity region.

\subsection{Trading intensity for repetition rate}

\begin{figure}
\includegraphics[width=0.48\textwidth]{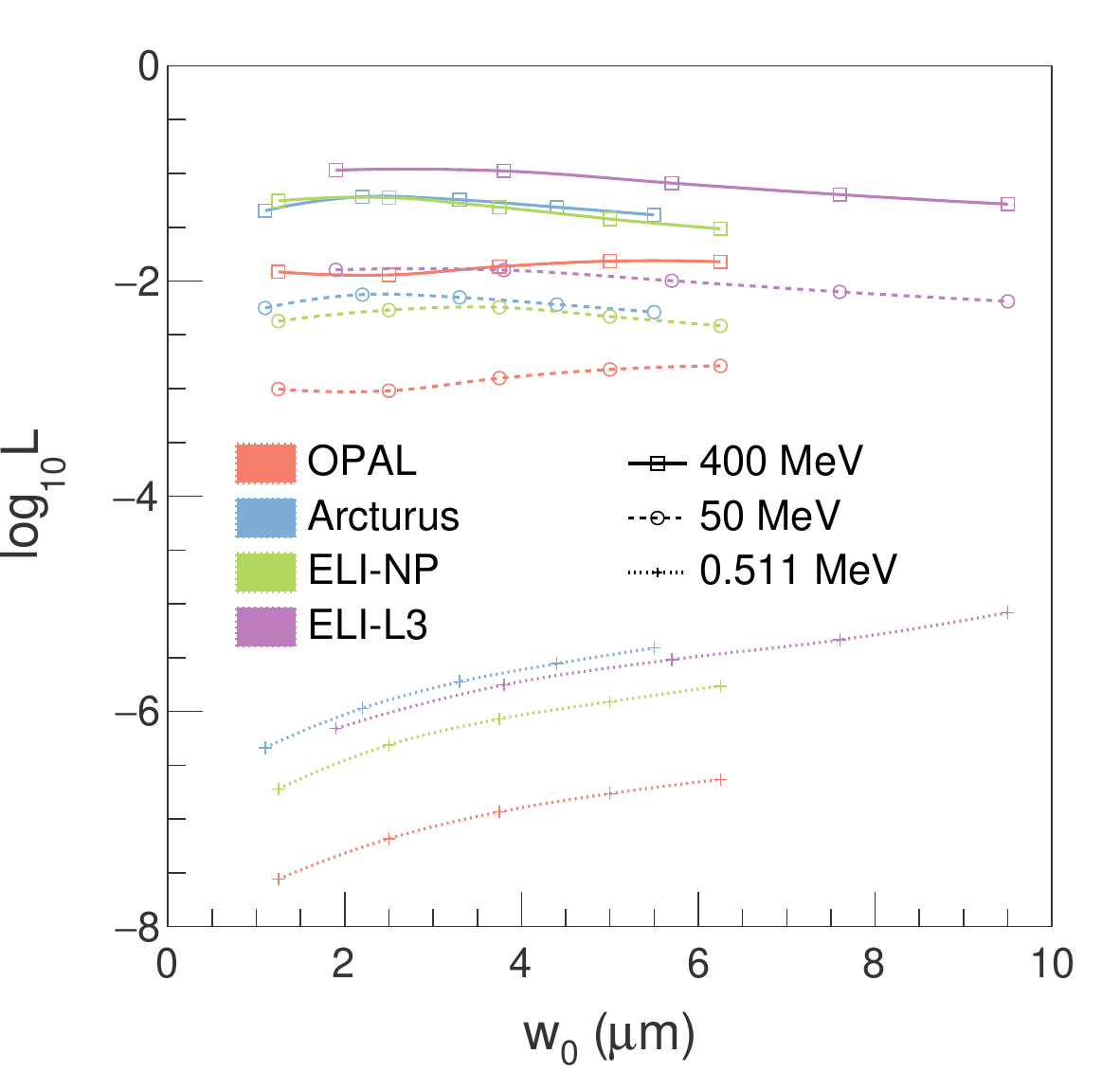}
\caption{\label{fig:reprate}  Luminosity comparison for \ELInp, \ELIhapl, \OPAL and \Arc. 
}
\end{figure}

Lastly we address an important design question.  Can a lower-intensity laser operating at higher repetition rate produce an equal or greater number of events than a higher-intensity laser at lower repetition rate?  In the absence of a fixed scaling law describing a trade-off between laser pulse energy and repetition rate, we select two pairs of facilities to compare.  In \fig{reprate}, \ELInp is compared to \ELIhapl, and \Arc is compared to \OPAL.  

In both cases, we find that the high repetition rate laser achieves higher luminosity.  Note that large differences in repetition rate are necessary: \ELIhapl has $600\times$ higher repetition rate with only $10\times$ smaller peak intensity in order to achieve a factor $\sim 4\times$ increase in luminosity.  More interestingly, \Arc achieves higher luminosity than \OPAL with $100\times$ smaller peak intensity and a repetition rate of 5 Hz ($300\times$ higher).  The $50\%$ longer pulse duration of \Arc enables a longer interaction also marginally increasing the luminosity.

Since the luminosity \eq{Leldefn} is linear in the repetition rate, the extent to which each facility falls short of their design repetition rate is the extent to which the luminosity is decreased.  Differing de facto running capabilities will therefore determine whether the advantage of the high rep-rate laser remains or is decreased.  While these results only show that higher repetition rate-lower intensity lasers can yield as many or more events, they do not completely determine the outcome of a cost-benefit analysis.  The relative cost of the two design approaches will depend on many other characteristics of the facilities, most notably the efficiency of the laser systems.

\section{Generalizing to radiation efficiency}

Having established the definition \eq{Leldefn} and its similarities to lepton and hadron collider luminosity, we now generalize slightly.  Since we are interested in how the particle dynamics, as determined by choices on the laser design side, affect the number of observed events, we define a per-particle luminosity such that
\begin{align}
\LeL\equiv N_e \langle\eta_{\rm e\ell}\rangle,
\end{align}
where $N_e$ is the number of electrons in the bunch.
The per-particle quantity $\eta_{\rm e\ell}$ is a measure of the efficiency of the laser pulse in causing the particle to emit a photon, and can be written simply
\begin{align}
\eta_{\rm e\ell}=\frac{1}{\Gamma_b}\nu_c\int \Gamma(E_e,\chi(\vec x,\vec p))dt,
\end{align}  
where the integral represents the total probability for a given particle to emit.  Implicit in the emission probability is dependence on the particle's initial conditions.

Since the electron bunch is necessarily distributed in space and time, the values of $\eta_{\rm e\ell}$ for randomly initialized individual particles are not useful so much as the statistics.  For example, for the purposes of more precise quantitative measurements of strong-field radiation event probabilities, it may be useful to have the majority of electrons in the bunch undergo similar interactions with the laser in the sense of having similar total probabilities to radiate.  In other words, we may like to engineer a narrow distribution in $\eta_{\rm e\ell}$.  To define the width of the distribution in $\eta_{\rm e\ell}$, we may formally write
\begin{align}\label{eq:Lperparticle}
\langle\eta_{\rm e\ell}\rangle&=\frac{1}{N_e}\int \frac{dN}{d\eta}\eta d\eta  \\
\langle\eta_{\rm e\ell}^2\rangle&=\frac{1}{N_e}\int \frac{dN}{d\eta}\eta^2 d\eta,\\
\label{eq:DeltaL}
\langle\Delta\eta_{\rm e\ell}\rangle&=\left(\langle\eta_{\rm e\ell}^2\rangle-\langle\eta_{\rm e\ell}\rangle^2\right)^{1/2}
\end{align}
with the integrals on the right hand side calculated straightforwardly from the Monte Carlo simulations above.

\begin{figure}
\includegraphics[width=0.48\textwidth]{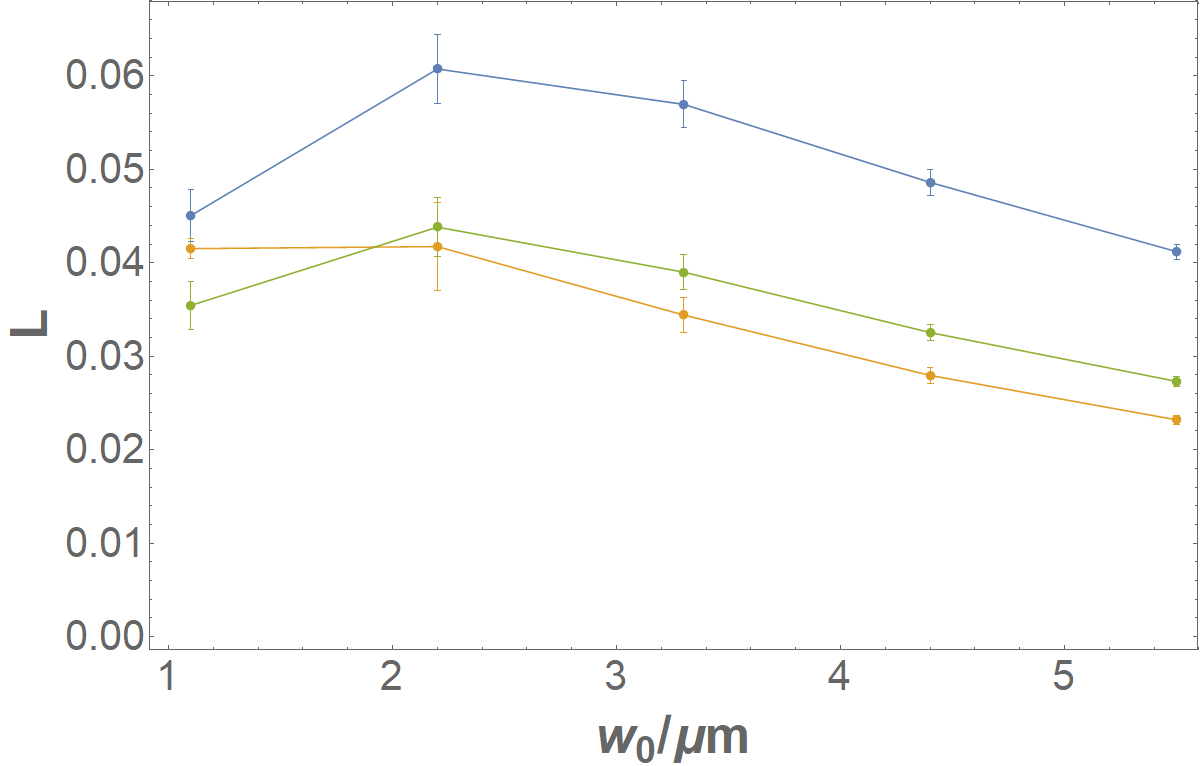}
\caption{\label{fig:etadeta}  Comparison of the radiation efficiency \eq{Lperparticle} and its width \eq{DeltaL} for the 200-TW \Arc laser for different temporal profile functions.  Gauss and sech profile points are horizontally offset slightly for clarity.
}
\end{figure}

Depending on the facility and the pulse profile, different choices of spot size provide a good compromise of both high total luminosity $\LeL$ and small $\langle\Delta\eta_{\rm e\ell}\rangle$.  The distribution width, measured by $\langle\Delta\eta_{\rm e\ell}\rangle/\LeL$, is generally smaller for high-energy collision geometry than for acceleration-from-rest.

\section{Conclusions}

We have defined the luminosity for electron-laser collisions as a general metric to streamline and quantify the design of experiments.  It builds on similar, complementary quantitative studies of sensitivity of specific processes to features of the laser field or electron beam and proposals for features that can serve as signals in an experiment.  We anticipate extending this optimization study to higher $\chi$ once a satisfactory equation of motion and more consistent theory of dynamics is available in the $\chi\gtrsim 10$ regime where the classical equations of motion suggest the dynamics are nonperturbative \cite{Hadad:2010mt,heinzl2021classical} and QED emission probabilities suggest a large fraction of momentum is lost in individual high-energy photon emission events \cite{tamburini2021efficient}.

We have also introduced the luminosity spread, which correlates directly with number and energy spread of expected events.  The luminosity spread should be useful to the development of applications and advanced experiments with laser-driven photon sources \cite{bai2022new}.

An important optimization we have not studied here is the division of laser energy for all-optical facilities.  However to undertake this optimization, a reliable estimate of the output electron beam energy as a function of input laser energy is required.  A 1-dimensional model \cite{esarey2009physics} maximizes $\chi_b$ with a larger fraction $\sim 0.75$ of energy given to the wakefield accelerator, whereas a 3-dimensional model \cite{lu2007generating} maximizes $\chi_b$ at half the energy in each beam line.  However, neither of these models performs well in comparison to published performance of laser wakefield acceleration experiments.  Based on general findings here, we recommend the following rule-of-thumb: optimize laser wakefield process for the maximal electron beam energy, because increasing the rigidity of the electrons relative to the field strength reduces transverse diversion of the electrons.  Radiation losses scale with $\chi$ and are thus unavoidable in seeking a maximal $\chi$.  However due to the absence of quantitative predictors of wakefield-generated electron beam energy, we leave more quantitative study, optimizing $\chimax$ and the luminosity for future work.

In this initial study, a number of simplifications have been made in modeling the laser and electron beams.  Since we assume the collective and coherent dynamics in the electron beam is negligible, a simple Monte Carlo approach to event probabilities suffices.  However for sufficiently dense bunches, or the possibility of high multiplicity electron-positron pair production at higher $\chi$,  may call for self-consistent simulations of the bunch dynamics.  Note that the luminosity can be applied to high-multiplicity particle production, including both ``seeded'' and  ``avalanche-type'' cascades \cite{fedotov2010limitations, mironov2014collapse}, just as it is in high-energy hadron collisions: the event is the cascade, so the luminosity quantifies the contribution of the beam geometries to the rate at which cascades are created.
A model of a OAP-focused field and initial studies of electron dynamics have recently been published \cite{vais2018direct}, but the change to electron trajectories has yet to be quantified. It may be worth repeating our work with this and potential measurement-derived models of the laser fields.  

Ref. \cite{vais2018direct} also investigates to some extent the effect of the secondary maximum in intensity due to the spatial Airy distribution.  In practice, the Airy distribution has two consequences for our estimates.  The first is that the secondary maximum is a ring in the transverse distribution that contains $\sim 10\%$ of the energy, effectively decreasing the peak intensity and hence the overall scale of the intensity for a given total pulse energy.  The second is relevant to another simplification, namely that in practice most experiments will cross the electron and laser beams at a nonzero angle $\sim 10^\circ-30^\circ$ in order to prevent damage to optics.  The nonzero crossing angle will slightly reduce the impact of secondary maxima in the temporal profile, because the beam approaches the focus off-axis where the magnitude of the fields is reduced.  The presence of the Airy disc second maximum however effectively re-introduces a low-intensity pre-pulse on the electrons' trajectories before the main pulse, with a consequent decrease in the acceleration seen by the electrons \cite{vais2018direct}.


Lastly, we observe that the luminosity defined here for strong-field QED events in electron laser collisions is trivially extended to dedicated photon-laser collision experiments such as possible with a gamma factory \cite{budker2022gamma, karbstein2022birefringence, blackburn2018nonlinear}.  The event rate \eq{Gamma1loop} obtained for an electron propagating in a constant crossed field is replaced by the rate for photon conversion into electron-positron pairs, obtained from the imaginary part of the photon self energy and available from Ref. \cite{Ritus:1985} and studied in a variety of more recent work \cite{meuren2015polarization, meuren2016semiclassical}.  Monte Carlo simulations are no longer necessary because the photon dynamics are negligibly affected by the high-intensity laser field; the event rate can be straightforwardly integrated over the overlap of the photon bunch and laser pulse, and then integrated over the photon distribution in energy and impact parameter.  The result is identical in form to the conventional collider luminosity \eq{Lwithprefac}.

\begin{appendix}
\section{Gaussian beam model}
For completeness, we give here the expressions for the model of the laser field.  If we assume the intensity in the focal plane is gaussian distributed, the solution to Maxwell's equations reproduces the gaussian beam as an expansion in powers of \cite{Quesnel:1998zz}
\begin{align}
\epsilon_w = \frac{1}{k_0 w_0}.
\end{align}
Even for the most strongly focused systems, with waist size similar to the wavelength $w_0\simeq \lambda_l$, $\epsilon_w$ is at most 0.13.  Second order corrections $\epsilon_w^2\lesssim 0.017$ are therefore of the same order as shot-to-shot variance in pulse energy and other likely corrections to wavefronts  \cite{pariente2016space,tiwari2019beam}.  As coefficients of the higher order corrections appear to grow less than factorially \cite{salamin2007fields}, the series appears reasonably convergent.  No new field components (which may qualitatively change trajectories) appear at higer orders at the expansion. Together these facts justify truncating after $\mathcal{O}(\epsilon_w)$.

Finite pulse duration affects the focusing both analytically in the momentum-space solution to Maxwell's equations \cite{Quesnel:1998zz} and experimentally due to variations in the amplification spectrum \cite{pariente2016space}.  We define the corresponding expansion parameter as
\begin{align}
\epsilon_t=\frac{\lambda_l}{c\Delta\tau}.
\end{align}

The resulting set of fields is, for linear polarization,
\begin{align}
\vec E^{(0)} &=\vec e E_0\frac{w_0}{w(z)}e^{-r^2/w(z)^2}\sin\!\Big(\Phi^{(0)}(t,z)\Big) \\
\vec B^{(0)} &=c^{-1}\hat k\times\vec E^{(0)} \\
\vec E^{(1)} &=\hat k 2\epsilon_w E_0 (\vec x\cdot\vec e)\frac{w_0}{w(z)}e^{-r^2/w(z)^2}\cos\!\Big(\Phi^{(1)}(t,z)\Big)\\
\vec B^{(1)} &=c^{-1} 2\epsilon_w E_0 (\hat k\times\vec x)\frac{w_0}{w(z)}e^{-r^2/w(z)^2}\cos\!\Big(\Phi^{(1)}(t,z)\Big)
\end{align}
where $\vec e$ is a polarization unit vector orthogonal to the wavevector $\vec k=\hat k\omega_0/c$, $w_0$ is the waist radius, $E_0$ is the peak electric field,
\begin{align}
w(z)&=w_0\sqrt{1+z^2/z_R^2}
\end{align}
is the beam radius as a function of distance from the focal plane and
\begin{align}
\Phi^{(0)}(t,z)=\omega_0(t-z/c)+\tan^{-1}(z/z_R)-\frac{zr^2}{z_Rw(z)^2}-\phi_0 \\
\Phi^{(1)}(t,z)=\Phi^{(0)}(t,z)+\tan^{-1}(z/z_R)
\end{align}
is the phase.  The $\mathcal{O}(\epsilon_t)$ fields differ for each profile function $f(t-z/c)$ and are derived by expanding the Fourier transform with respect to $t$,
\begin{align}
\vec E(t,x,y,z)&\simeq \vec e\tilde{E}(\omega_0,x,y,z)e^{i\omega_0(t-z/c)}f(t-z/c)+\vec e\frac{\partial\tilde{E}}{\partial\omega}(\omega_0,x,y,z)e^{i\omega_0(t-z/c)}f'(t-z/c)+...\\
&\equiv \vec E^{(0)}(t,x,y,z)+\vec E^{(1)}(t,x,y,z)+...
\end{align}
since the derivative on the profile function $f'$ bring in powers of $\epsilon_t$.

When a given $\epsilon_w$ or $\epsilon_t$ exceeds 0.06 for a given laser system, we include the corresponding correction to the fields in the simulation.

\end{appendix}

\begin{acknowledgments}
This work is supported by Air Force Office of Scientific Research (FA9550-14-1-0045), the Los Alamos National Lab Office of Experimental Sciences (RFP475852) and Tau Systems, Inc.  L.L. and O.Z.L. thank the Center for Relativistic Laser Science at the Gwangju Institute for Science and Technology for hospitality and Lynn Labun for continuing support.
\end{acknowledgments}

\bibliography{sf}

\begin{thebibliography}{127}%
\makeatletter
\providecommand \@ifxundefined [1]{%
 \@ifx{#1\undefined}
}%
\providecommand \@ifnum [1]{%
 \ifnum #1\expandafter \@firstoftwo
 \else \expandafter \@secondoftwo
 \fi
}%
\providecommand \@ifx [1]{%
 \ifx #1\expandafter \@firstoftwo
 \else \expandafter \@secondoftwo
 \fi
}%
\providecommand \natexlab [1]{#1}%
\providecommand \enquote  [1]{``#1''}%
\providecommand \bibnamefont  [1]{#1}%
\providecommand \bibfnamefont [1]{#1}%
\providecommand \citenamefont [1]{#1}%
\providecommand \href@noop [0]{\@secondoftwo}%
\providecommand \href [0]{\begingroup \@sanitize@url \@href}%
\providecommand \@href[1]{\@@startlink{#1}\@@href}%
\providecommand \@@href[1]{\endgroup#1\@@endlink}%
\providecommand \@sanitize@url [0]{\catcode `\\12\catcode `\$12\catcode
  `\&12\catcode `\#12\catcode `\^12\catcode `\_12\catcode `\%12\relax}%
\providecommand \@@startlink[1]{}%
\providecommand \@@endlink[0]{}%
\providecommand \url  [0]{\begingroup\@sanitize@url \@url }%
\providecommand \@url [1]{\endgroup\@href {#1}{\urlprefix }}%
\providecommand \urlprefix  [0]{URL }%
\providecommand \Eprint [0]{\href }%
\providecommand \doibase [0]{https://doi.org/}%
\providecommand \selectlanguage [0]{\@gobble}%
\providecommand \bibinfo  [0]{\@secondoftwo}%
\providecommand \bibfield  [0]{\@secondoftwo}%
\providecommand \translation [1]{[#1]}%
\providecommand \BibitemOpen [0]{}%
\providecommand \bibitemStop [0]{}%
\providecommand \bibitemNoStop [0]{.\EOS\space}%
\providecommand \EOS [0]{\spacefactor3000\relax}%
\providecommand \BibitemShut  [1]{\csname bibitem#1\endcsname}%
\let\auto@bib@innerbib\@empty
\bibitem [{\citenamefont {Albert}\ and\ \citenamefont
  {Thomas}(2016)}]{albert2016applications}%
  \BibitemOpen
  \bibfield  {author} {\bibinfo {author} {\bibfnamefont {F.}~\bibnamefont
  {Albert}}\ and\ \bibinfo {author} {\bibfnamefont {A.~G.}\ \bibnamefont
  {Thomas}},\ }\bibfield  {title} {\bibinfo {title} {Applications of laser
  wakefield accelerator-based light sources},\ }\href@noop {} {\bibfield
  {journal} {\bibinfo  {journal} {Plasma Physics and Controlled Fusion}\
  }\textbf {\bibinfo {volume} {58}},\ \bibinfo {pages} {103001} (\bibinfo
  {year} {2016})}\BibitemShut {NoStop}%
\bibitem [{nat(2018)}]{national2018opportunities}%
  \BibitemOpen
  \href@noop {} {\emph {\bibinfo {title} {Opportunities in Intense Ultrafast
  Lasers: Reaching for the Brightest Light}}}\ (\bibinfo  {publisher} {National
  Academies Press},\ \bibinfo {year} {2018})\BibitemShut {NoStop}%
\bibitem [{\citenamefont {Albert}\ \emph {et~al.}(2021)\citenamefont {Albert},
  \citenamefont {Couprie}, \citenamefont {Debus}, \citenamefont {Downer},
  \citenamefont {Faure}, \citenamefont {Flacco}, \citenamefont {Gizzi},
  \citenamefont {Grismayer}, \citenamefont {Huebl}, \citenamefont {Joshi} \emph
  {et~al.}}]{albert20212020}%
  \BibitemOpen
  \bibfield  {author} {\bibinfo {author} {\bibfnamefont {F.}~\bibnamefont
  {Albert}}, \bibinfo {author} {\bibfnamefont {M.}~\bibnamefont {Couprie}},
  \bibinfo {author} {\bibfnamefont {A.}~\bibnamefont {Debus}}, \bibinfo
  {author} {\bibfnamefont {M.~C.}\ \bibnamefont {Downer}}, \bibinfo {author}
  {\bibfnamefont {J.}~\bibnamefont {Faure}}, \bibinfo {author} {\bibfnamefont
  {A.}~\bibnamefont {Flacco}}, \bibinfo {author} {\bibfnamefont {L.~A.}\
  \bibnamefont {Gizzi}}, \bibinfo {author} {\bibfnamefont {T.}~\bibnamefont
  {Grismayer}}, \bibinfo {author} {\bibfnamefont {A.}~\bibnamefont {Huebl}},
  \bibinfo {author} {\bibfnamefont {C.}~\bibnamefont {Joshi}}, \emph {et~al.},\
  }\bibfield  {title} {\bibinfo {title} {2020 roadmap on plasma accelerators},\
  }\href@noop {} {\bibfield  {journal} {\bibinfo  {journal} {New Journal of
  Physics}\ }\textbf {\bibinfo {volume} {23}},\ \bibinfo {pages} {031101}
  (\bibinfo {year} {2021})}\BibitemShut {NoStop}%
\bibitem [{\citenamefont {Falcone}\ \emph {et~al.}(2020)\citenamefont
  {Falcone}, \citenamefont {Albert}, \citenamefont {Beg}, \citenamefont
  {Glenzer}, \citenamefont {Ditmire}, \citenamefont {Spinka},\ and\
  \citenamefont {Zuegel}}]{falcone2020workshop}%
  \BibitemOpen
  \bibfield  {author} {\bibinfo {author} {\bibfnamefont {R.}~\bibnamefont
  {Falcone}}, \bibinfo {author} {\bibfnamefont {F.}~\bibnamefont {Albert}},
  \bibinfo {author} {\bibfnamefont {F.}~\bibnamefont {Beg}}, \bibinfo {author}
  {\bibfnamefont {S.}~\bibnamefont {Glenzer}}, \bibinfo {author} {\bibfnamefont
  {T.}~\bibnamefont {Ditmire}}, \bibinfo {author} {\bibfnamefont
  {T.}~\bibnamefont {Spinka}},\ and\ \bibinfo {author} {\bibfnamefont
  {J.}~\bibnamefont {Zuegel}},\ }\bibfield  {title} {\bibinfo {title} {Workshop
  report: brightest light initiative (march 27-29 2019, osa headquarters,
  washington, dc)},\ }\href@noop {} {\bibfield  {journal} {\bibinfo  {journal}
  {arXiv preprint arXiv:2002.09712}\ } (\bibinfo {year} {2020})}\BibitemShut
  {NoStop}%
\bibitem [{\citenamefont {Di~Piazza}\ \emph {et~al.}(2022)\citenamefont
  {Di~Piazza}, \citenamefont {Willingale},\ and\ \citenamefont
  {Zuegel}}]{mp32022workshop}%
  \BibitemOpen
  \bibfield  {author} {\bibinfo {author} {\bibfnamefont {A.}~\bibnamefont
  {Di~Piazza}}, \bibinfo {author} {\bibfnamefont {L.}~\bibnamefont
  {Willingale}},\ and\ \bibinfo {author} {\bibfnamefont {J.}~\bibnamefont
  {Zuegel}},\ }\bibfield  {title} {\bibinfo {title} {Multi-petawatt physics
  prioritization (mp3) workshop report},\ }\href@noop {} {\bibfield  {journal}
  {\bibinfo  {journal} {arXiv preprint arXiv:2211.13187}\ } (\bibinfo {year}
  {2022})}\BibitemShut {NoStop}%
\bibitem [{\citenamefont {Hooker}\ \emph {et~al.}(2008)\citenamefont {Hooker},
  \citenamefont {Blake}, \citenamefont {Chekhlov}, \citenamefont {Clarke},
  \citenamefont {Collier}, \citenamefont {Divall}, \citenamefont {Ertel},
  \citenamefont {Foster}, \citenamefont {Hawkes}, \citenamefont {Holligan}
  \emph {et~al.}}]{hooker2008commissioning}%
  \BibitemOpen
  \bibfield  {author} {\bibinfo {author} {\bibfnamefont {C.}~\bibnamefont
  {Hooker}}, \bibinfo {author} {\bibfnamefont {S.}~\bibnamefont {Blake}},
  \bibinfo {author} {\bibfnamefont {O.}~\bibnamefont {Chekhlov}}, \bibinfo
  {author} {\bibfnamefont {R.}~\bibnamefont {Clarke}}, \bibinfo {author}
  {\bibfnamefont {J.}~\bibnamefont {Collier}}, \bibinfo {author} {\bibfnamefont
  {E.}~\bibnamefont {Divall}}, \bibinfo {author} {\bibfnamefont
  {K.}~\bibnamefont {Ertel}}, \bibinfo {author} {\bibfnamefont
  {P.}~\bibnamefont {Foster}}, \bibinfo {author} {\bibfnamefont
  {S.}~\bibnamefont {Hawkes}}, \bibinfo {author} {\bibfnamefont
  {P.}~\bibnamefont {Holligan}}, \emph {et~al.},\ }\bibfield  {title} {\bibinfo
  {title} {Commissioning the astra gemini petawatt ti: sapphire laser system},\
  }in\ \href@noop {} {\emph {\bibinfo {booktitle} {Conference on Lasers and
  Electro-Optics}}}\ (\bibinfo {organization} {Optica Publishing Group},\
  \bibinfo {year} {2008})\ p.\ \bibinfo {pages} {JThB2}\BibitemShut {NoStop}%
\bibitem [{\citenamefont {Rus}\ \emph {et~al.}(2011)\citenamefont {Rus},
  \citenamefont {Batysta}, \citenamefont {{\v{C}}{\'a}p}, \citenamefont
  {Divok{\`y}}, \citenamefont {Fibrich}, \citenamefont {Griffiths},
  \citenamefont {Haley}, \citenamefont {Havl{\'\i}{\v{c}}ek}, \citenamefont
  {Hlav{\'a}c}, \citenamefont {H{\v{r}}eb{\'\i}{\v{c}}ek} \emph
  {et~al.}}]{rus2011outline}%
  \BibitemOpen
  \bibfield  {author} {\bibinfo {author} {\bibfnamefont {B.}~\bibnamefont
  {Rus}}, \bibinfo {author} {\bibfnamefont {F.}~\bibnamefont {Batysta}},
  \bibinfo {author} {\bibfnamefont {J.}~\bibnamefont {{\v{C}}{\'a}p}}, \bibinfo
  {author} {\bibfnamefont {M.}~\bibnamefont {Divok{\`y}}}, \bibinfo {author}
  {\bibfnamefont {M.}~\bibnamefont {Fibrich}}, \bibinfo {author} {\bibfnamefont
  {M.}~\bibnamefont {Griffiths}}, \bibinfo {author} {\bibfnamefont
  {R.}~\bibnamefont {Haley}}, \bibinfo {author} {\bibfnamefont
  {T.}~\bibnamefont {Havl{\'\i}{\v{c}}ek}}, \bibinfo {author} {\bibfnamefont
  {M.}~\bibnamefont {Hlav{\'a}c}}, \bibinfo {author} {\bibfnamefont
  {J.}~\bibnamefont {H{\v{r}}eb{\'\i}{\v{c}}ek}}, \emph {et~al.},\ }\bibfield
  {title} {\bibinfo {title} {Outline of the eli-beamlines facility},\ }in\
  \href@noop {} {\emph {\bibinfo {booktitle} {Diode-Pumped High Energy and High
  Power Lasers; ELI: Ultrarelativistic Laser-Matter Interactions and Petawatt
  Photonics; and HiPER: the European Pathway to Laser Energy}}},\ Vol.\
  \bibinfo {volume} {8080}\ (\bibinfo {organization} {SPIE},\ \bibinfo {year}
  {2011})\ pp.\ \bibinfo {pages} {163--172}\BibitemShut {NoStop}%
\bibitem [{\citenamefont {Sung}\ \emph {et~al.}(2017)\citenamefont {Sung},
  \citenamefont {Lee}, \citenamefont {Yoo}, \citenamefont {Yoon}, \citenamefont
  {Lee}, \citenamefont {Yang}, \citenamefont {Son}, \citenamefont {Jang},
  \citenamefont {Lee},\ and\ \citenamefont {Nam}}]{sung20174}%
  \BibitemOpen
  \bibfield  {author} {\bibinfo {author} {\bibfnamefont {J.~H.}\ \bibnamefont
  {Sung}}, \bibinfo {author} {\bibfnamefont {H.~W.}\ \bibnamefont {Lee}},
  \bibinfo {author} {\bibfnamefont {J.~Y.}\ \bibnamefont {Yoo}}, \bibinfo
  {author} {\bibfnamefont {J.~W.}\ \bibnamefont {Yoon}}, \bibinfo {author}
  {\bibfnamefont {C.~W.}\ \bibnamefont {Lee}}, \bibinfo {author} {\bibfnamefont
  {J.~M.}\ \bibnamefont {Yang}}, \bibinfo {author} {\bibfnamefont {Y.~J.}\
  \bibnamefont {Son}}, \bibinfo {author} {\bibfnamefont {Y.~H.}\ \bibnamefont
  {Jang}}, \bibinfo {author} {\bibfnamefont {S.~K.}\ \bibnamefont {Lee}},\ and\
  \bibinfo {author} {\bibfnamefont {C.~H.}\ \bibnamefont {Nam}},\ }\bibfield
  {title} {\bibinfo {title} {4.2 pw, 20 fs ti: sapphire laser at 0.1 hz},\
  }\href@noop {} {\bibfield  {journal} {\bibinfo  {journal} {Optics letters}\
  }\textbf {\bibinfo {volume} {42}},\ \bibinfo {pages} {2058} (\bibinfo {year}
  {2017})}\BibitemShut {NoStop}%
\bibitem [{\citenamefont {K{\"u}hn}\ \emph {et~al.}(2017)\citenamefont
  {K{\"u}hn}, \citenamefont {Dumergue}, \citenamefont {Kahaly}, \citenamefont
  {Mondal}, \citenamefont {F{\"u}le}, \citenamefont {Csizmadia}, \citenamefont
  {Farkas}, \citenamefont {Major}, \citenamefont {V{\'a}rallyay}, \citenamefont
  {Cormier} \emph {et~al.}}]{kuhn2017eli}%
  \BibitemOpen
  \bibfield  {author} {\bibinfo {author} {\bibfnamefont {S.}~\bibnamefont
  {K{\"u}hn}}, \bibinfo {author} {\bibfnamefont {M.}~\bibnamefont {Dumergue}},
  \bibinfo {author} {\bibfnamefont {S.}~\bibnamefont {Kahaly}}, \bibinfo
  {author} {\bibfnamefont {S.}~\bibnamefont {Mondal}}, \bibinfo {author}
  {\bibfnamefont {M.}~\bibnamefont {F{\"u}le}}, \bibinfo {author}
  {\bibfnamefont {T.}~\bibnamefont {Csizmadia}}, \bibinfo {author}
  {\bibfnamefont {B.}~\bibnamefont {Farkas}}, \bibinfo {author} {\bibfnamefont
  {B.}~\bibnamefont {Major}}, \bibinfo {author} {\bibfnamefont
  {Z.}~\bibnamefont {V{\'a}rallyay}}, \bibinfo {author} {\bibfnamefont
  {E.}~\bibnamefont {Cormier}}, \emph {et~al.},\ }\bibfield  {title} {\bibinfo
  {title} {The eli-alps facility: the next generation of attosecond sources},\
  }\href@noop {} {\bibfield  {journal} {\bibinfo  {journal} {Journal of Physics
  B: Atomic, Molecular and Optical Physics}\ }\textbf {\bibinfo {volume}
  {50}},\ \bibinfo {pages} {132002} (\bibinfo {year} {2017})}\BibitemShut
  {NoStop}%
\bibitem [{\citenamefont {Le~Garrec}(2017)}]{le2017design}%
  \BibitemOpen
  \bibfield  {author} {\bibinfo {author} {\bibfnamefont {B.}~\bibnamefont
  {Le~Garrec}},\ }\bibfield  {title} {\bibinfo {title} {Design update and
  recent results of the apollon 10 pw facility},\ }in\ \href@noop {} {\emph
  {\bibinfo {booktitle} {CLEO: Science and Innovations}}}\ (\bibinfo
  {organization} {Optica Publishing Group},\ \bibinfo {year} {2017})\ pp.\
  \bibinfo {pages} {SF1K--3}\BibitemShut {NoStop}%
\bibitem [{\citenamefont {Papadopoulos}\ \emph {et~al.}(2019)\citenamefont
  {Papadopoulos}, \citenamefont {Zou}, \citenamefont {Le~Blanc}, \citenamefont
  {Ranc}, \citenamefont {Druon}, \citenamefont {Martin}, \citenamefont
  {Fr{\'e}neaux}, \citenamefont {Beluze}, \citenamefont {Lebas}, \citenamefont
  {Chabanis} \emph {et~al.}}]{papadopoulos2019first}%
  \BibitemOpen
  \bibfield  {author} {\bibinfo {author} {\bibfnamefont {D.}~\bibnamefont
  {Papadopoulos}}, \bibinfo {author} {\bibfnamefont {J.}~\bibnamefont {Zou}},
  \bibinfo {author} {\bibfnamefont {C.}~\bibnamefont {Le~Blanc}}, \bibinfo
  {author} {\bibfnamefont {L.}~\bibnamefont {Ranc}}, \bibinfo {author}
  {\bibfnamefont {F.}~\bibnamefont {Druon}}, \bibinfo {author} {\bibfnamefont
  {L.}~\bibnamefont {Martin}}, \bibinfo {author} {\bibfnamefont
  {A.}~\bibnamefont {Fr{\'e}neaux}}, \bibinfo {author} {\bibfnamefont
  {A.}~\bibnamefont {Beluze}}, \bibinfo {author} {\bibfnamefont
  {N.}~\bibnamefont {Lebas}}, \bibinfo {author} {\bibfnamefont
  {M.}~\bibnamefont {Chabanis}}, \emph {et~al.},\ }\bibfield  {title} {\bibinfo
  {title} {First commissioning results of the apollon laser on the 1 pw beam
  line},\ }in\ \href@noop {} {\emph {\bibinfo {booktitle} {CLEO: Science and
  Innovations}}}\ (\bibinfo {organization} {Optical Society of America},\
  \bibinfo {year} {2019})\ pp.\ \bibinfo {pages} {STu3E--4}\BibitemShut
  {NoStop}%
\bibitem [{\citenamefont {Cerchez}\ \emph {et~al.}(2019)\citenamefont
  {Cerchez}, \citenamefont {Prasad}, \citenamefont {Aurand}, \citenamefont
  {Giesecke}, \citenamefont {Spickermann}, \citenamefont {Brauckmann},
  \citenamefont {Aktan}, \citenamefont {Swantusch}, \citenamefont {Toncian},
  \citenamefont {Toncian} \emph {et~al.}}]{cerchez2019arcturus}%
  \BibitemOpen
  \bibfield  {author} {\bibinfo {author} {\bibfnamefont {M.}~\bibnamefont
  {Cerchez}}, \bibinfo {author} {\bibfnamefont {R.}~\bibnamefont {Prasad}},
  \bibinfo {author} {\bibfnamefont {B.}~\bibnamefont {Aurand}}, \bibinfo
  {author} {\bibfnamefont {A.}~\bibnamefont {Giesecke}}, \bibinfo {author}
  {\bibfnamefont {S.}~\bibnamefont {Spickermann}}, \bibinfo {author}
  {\bibfnamefont {S.}~\bibnamefont {Brauckmann}}, \bibinfo {author}
  {\bibfnamefont {E.}~\bibnamefont {Aktan}}, \bibinfo {author} {\bibfnamefont
  {M.}~\bibnamefont {Swantusch}}, \bibinfo {author} {\bibfnamefont
  {M.}~\bibnamefont {Toncian}}, \bibinfo {author} {\bibfnamefont
  {T.}~\bibnamefont {Toncian}}, \emph {et~al.},\ }\bibfield  {title} {\bibinfo
  {title} {Arcturus laser: a versatile high-contrast, high-power multi-beam
  laser system},\ }\href@noop {} {\bibfield  {journal} {\bibinfo  {journal}
  {High Power Laser Science and Engineering}\ }\textbf {\bibinfo {volume} {7}}
  (\bibinfo {year} {2019})}\BibitemShut {NoStop}%
\bibitem [{\citenamefont {Danson}\ \emph {et~al.}(2019)\citenamefont {Danson},
  \citenamefont {Haefner}, \citenamefont {Bromage}, \citenamefont {Butcher},
  \citenamefont {Chanteloup}, \citenamefont {Chowdhury}, \citenamefont
  {Galvanauskas}, \citenamefont {Gizzi}, \citenamefont {Hein}, \citenamefont
  {Hillier} \emph {et~al.}}]{danson2019petawatt}%
  \BibitemOpen
  \bibfield  {author} {\bibinfo {author} {\bibfnamefont {C.~N.}\ \bibnamefont
  {Danson}}, \bibinfo {author} {\bibfnamefont {C.}~\bibnamefont {Haefner}},
  \bibinfo {author} {\bibfnamefont {J.}~\bibnamefont {Bromage}}, \bibinfo
  {author} {\bibfnamefont {T.}~\bibnamefont {Butcher}}, \bibinfo {author}
  {\bibfnamefont {J.-C.~F.}\ \bibnamefont {Chanteloup}}, \bibinfo {author}
  {\bibfnamefont {E.~A.}\ \bibnamefont {Chowdhury}}, \bibinfo {author}
  {\bibfnamefont {A.}~\bibnamefont {Galvanauskas}}, \bibinfo {author}
  {\bibfnamefont {L.~A.}\ \bibnamefont {Gizzi}}, \bibinfo {author}
  {\bibfnamefont {J.}~\bibnamefont {Hein}}, \bibinfo {author} {\bibfnamefont
  {D.~I.}\ \bibnamefont {Hillier}}, \emph {et~al.},\ }\bibfield  {title}
  {\bibinfo {title} {Petawatt and exawatt class lasers worldwide},\ }\href@noop
  {} {\bibfield  {journal} {\bibinfo  {journal} {High Power Laser Science and
  Engineering}\ }\textbf {\bibinfo {volume} {7}} (\bibinfo {year}
  {2019})}\BibitemShut {NoStop}%
\bibitem [{\citenamefont {Doria}\ \emph {et~al.}(2020)\citenamefont {Doria},
  \citenamefont {Cernaianu}, \citenamefont {Ghenuche}, \citenamefont {Stutman},
  \citenamefont {Tanaka}, \citenamefont {Ticos},\ and\ \citenamefont
  {Ur}}]{doria2020overview}%
  \BibitemOpen
  \bibfield  {author} {\bibinfo {author} {\bibfnamefont {D.}~\bibnamefont
  {Doria}}, \bibinfo {author} {\bibfnamefont {M.}~\bibnamefont {Cernaianu}},
  \bibinfo {author} {\bibfnamefont {P.}~\bibnamefont {Ghenuche}}, \bibinfo
  {author} {\bibfnamefont {D.}~\bibnamefont {Stutman}}, \bibinfo {author}
  {\bibfnamefont {K.}~\bibnamefont {Tanaka}}, \bibinfo {author} {\bibfnamefont
  {C.}~\bibnamefont {Ticos}},\ and\ \bibinfo {author} {\bibfnamefont
  {C.}~\bibnamefont {Ur}},\ }\bibfield  {title} {\bibinfo {title} {Overview of
  eli-np status and laser commissioning experiments with 1 pw and 10 pw
  class-lasers},\ }\href@noop {} {\bibfield  {journal} {\bibinfo  {journal}
  {Journal of Instrumentation}\ }\textbf {\bibinfo {volume} {15}}\bibinfo
  {number} { (09)},\ \bibinfo {pages} {C09053}}\BibitemShut {NoStop}%
\bibitem [{\citenamefont {Abramowicz}\ \emph {et~al.}(2021)\citenamefont
  {Abramowicz}, \citenamefont {Acosta}, \citenamefont {Altarelli},
  \citenamefont {Assmann}, \citenamefont {Bai}, \citenamefont {Behnke},
  \citenamefont {Benhammou}, \citenamefont {Blackburn}, \citenamefont
  {Boogert}, \citenamefont {Borysov} \emph
  {et~al.}}]{abramowicz2021conceptual}%
  \BibitemOpen
\bibfield  {number} {  }\bibfield  {author} {\bibinfo {author} {\bibfnamefont
  {H.}~\bibnamefont {Abramowicz}}, \bibinfo {author} {\bibfnamefont
  {U.}~\bibnamefont {Acosta}}, \bibinfo {author} {\bibfnamefont
  {M.}~\bibnamefont {Altarelli}}, \bibinfo {author} {\bibfnamefont
  {R.}~\bibnamefont {Assmann}}, \bibinfo {author} {\bibfnamefont
  {Z.}~\bibnamefont {Bai}}, \bibinfo {author} {\bibfnamefont {T.}~\bibnamefont
  {Behnke}}, \bibinfo {author} {\bibfnamefont {Y.}~\bibnamefont {Benhammou}},
  \bibinfo {author} {\bibfnamefont {T.}~\bibnamefont {Blackburn}}, \bibinfo
  {author} {\bibfnamefont {S.}~\bibnamefont {Boogert}}, \bibinfo {author}
  {\bibfnamefont {O.}~\bibnamefont {Borysov}}, \emph {et~al.},\ }\bibfield
  {title} {\bibinfo {title} {Conceptual design report for the luxe
  experiment},\ }\href@noop {} {\bibfield  {journal} {\bibinfo  {journal} {The
  European Physical Journal Special Topics}\ }\textbf {\bibinfo {volume}
  {230}},\ \bibinfo {pages} {2445} (\bibinfo {year} {2021})}\BibitemShut
  {NoStop}%
\bibitem [{\citenamefont {Nees}\ \emph {et~al.}(2021)\citenamefont {Nees},
  \citenamefont {Maksimchuk}, \citenamefont {Kalinchenko}, \citenamefont {Hou},
  \citenamefont {Ma}, \citenamefont {Campbell}, \citenamefont {McKelvey},
  \citenamefont {Willingale}, \citenamefont {Jovanovic}, \citenamefont {Kuranz}
  \emph {et~al.}}]{nees2021zettawatt}%
  \BibitemOpen
  \bibfield  {author} {\bibinfo {author} {\bibfnamefont {J.}~\bibnamefont
  {Nees}}, \bibinfo {author} {\bibfnamefont {A.}~\bibnamefont {Maksimchuk}},
  \bibinfo {author} {\bibfnamefont {G.}~\bibnamefont {Kalinchenko}}, \bibinfo
  {author} {\bibfnamefont {B.}~\bibnamefont {Hou}}, \bibinfo {author}
  {\bibfnamefont {Y.}~\bibnamefont {Ma}}, \bibinfo {author} {\bibfnamefont
  {P.}~\bibnamefont {Campbell}}, \bibinfo {author} {\bibfnamefont
  {A.}~\bibnamefont {McKelvey}}, \bibinfo {author} {\bibfnamefont
  {L.}~\bibnamefont {Willingale}}, \bibinfo {author} {\bibfnamefont
  {I.}~\bibnamefont {Jovanovic}}, \bibinfo {author} {\bibfnamefont
  {C.}~\bibnamefont {Kuranz}}, \emph {et~al.},\ }\bibfield  {title} {\bibinfo
  {title} {Zettawatt equivalent ultrashort pulse laser system: An nsf mid-scale
  facility for laser-driven science in the qed regime},\ }in\ \href@noop {}
  {\emph {\bibinfo {booktitle} {2021 Conference on Lasers and Electro-Optics
  (CLEO)}}}\ (\bibinfo {organization} {IEEE},\ \bibinfo {year} {2021})\ pp.\
  \bibinfo {pages} {1--2}\BibitemShut {NoStop}%
\bibitem [{\citenamefont {Gan}\ \emph {et~al.}(2021)\citenamefont {Gan},
  \citenamefont {Yu}, \citenamefont {Wang}, \citenamefont {Liu}, \citenamefont
  {Xu}, \citenamefont {Li}, \citenamefont {Li}, \citenamefont {Yu},
  \citenamefont {Wang}, \citenamefont {Liu} \emph {et~al.}}]{gan2021shanghai}%
  \BibitemOpen
  \bibfield  {author} {\bibinfo {author} {\bibfnamefont {Z.}~\bibnamefont
  {Gan}}, \bibinfo {author} {\bibfnamefont {L.}~\bibnamefont {Yu}}, \bibinfo
  {author} {\bibfnamefont {C.}~\bibnamefont {Wang}}, \bibinfo {author}
  {\bibfnamefont {Y.}~\bibnamefont {Liu}}, \bibinfo {author} {\bibfnamefont
  {Y.}~\bibnamefont {Xu}}, \bibinfo {author} {\bibfnamefont {W.}~\bibnamefont
  {Li}}, \bibinfo {author} {\bibfnamefont {S.}~\bibnamefont {Li}}, \bibinfo
  {author} {\bibfnamefont {L.}~\bibnamefont {Yu}}, \bibinfo {author}
  {\bibfnamefont {X.}~\bibnamefont {Wang}}, \bibinfo {author} {\bibfnamefont
  {X.}~\bibnamefont {Liu}}, \emph {et~al.},\ }\bibfield  {title} {\bibinfo
  {title} {The shanghai superintense ultrafast laser facility (sulf) project},\
  }\href@noop {} {\bibfield  {journal} {\bibinfo  {journal} {Progress in
  Ultrafast Intense Laser Science XVI}\ ,\ \bibinfo {pages} {199}} (\bibinfo
  {year} {2021})}\BibitemShut {NoStop}%
\bibitem [{las({\natexlab{a}})}]{lasernet}%
  \BibitemOpen
  \href@noop {} {\bibinfo {title} {Lasernetus}} ({\natexlab{a}}),\ \bibinfo
  {note} {https://lasernetus.org/}\BibitemShut {NoStop}%
\bibitem [{eli()}]{elieric}%
  \BibitemOpen
  \href@noop {} {\bibinfo {title} {Eli eric}},\ \bibinfo {note}
  {https://eli-laser.eu/organisation/eli-eric/}\BibitemShut {NoStop}%
\bibitem [{las({\natexlab{b}})}]{laserlab}%
  \BibitemOpen
  \href@noop {} {\bibinfo {title} {Laserlab europe}} ({\natexlab{b}}),\
  \bibinfo {note} {https://www.laserlab-europe.eu/}\BibitemShut {NoStop}%
\bibitem [{\citenamefont {Bula}\ \emph {et~al.}(1996)\citenamefont {Bula} \emph
  {et~al.}}]{Bula:1996st}%
  \BibitemOpen
  \bibfield  {author} {\bibinfo {author} {\bibfnamefont {C.}~\bibnamefont
  {Bula}} \emph {et~al.} (\bibinfo {collaboration} {E144}),\ }\bibfield
  {title} {\bibinfo {title} {{Observation of nonlinear effects in Compton
  scattering}},\ }\href {https://doi.org/10.1103/PhysRevLett.76.3116}
  {\bibfield  {journal} {\bibinfo  {journal} {Phys. Rev. Lett.}\ }\textbf
  {\bibinfo {volume} {76}},\ \bibinfo {pages} {3116} (\bibinfo {year}
  {1996})}\BibitemShut {NoStop}%
\bibitem [{\citenamefont {Bamber}\ \emph {et~al.}(1999)\citenamefont {Bamber}
  \emph {et~al.}}]{Bamber:1999zt}%
  \BibitemOpen
  \bibfield  {author} {\bibinfo {author} {\bibfnamefont {C.}~\bibnamefont
  {Bamber}} \emph {et~al.},\ }\bibfield  {title} {\bibinfo {title} {{Studies of
  nonlinear QED in collisions of 46.6-GeV electrons with intense laser
  pulses}},\ }\href {https://doi.org/10.1103/PhysRevD.60.092004} {\bibfield
  {journal} {\bibinfo  {journal} {Phys. Rev.}\ }\textbf {\bibinfo {volume}
  {D60}},\ \bibinfo {pages} {092004} (\bibinfo {year} {1999})}\BibitemShut
  {NoStop}%
\bibitem [{\citenamefont {Schwoerer}\ \emph {et~al.}(2006)\citenamefont
  {Schwoerer}, \citenamefont {Liesfeld}, \citenamefont {Schlenvoigt},
  \citenamefont {Amthor},\ and\ \citenamefont
  {Sauerbrey}}]{schwoerer2006thomson}%
  \BibitemOpen
  \bibfield  {author} {\bibinfo {author} {\bibfnamefont {H.}~\bibnamefont
  {Schwoerer}}, \bibinfo {author} {\bibfnamefont {B.}~\bibnamefont {Liesfeld}},
  \bibinfo {author} {\bibfnamefont {H.-P.}\ \bibnamefont {Schlenvoigt}},
  \bibinfo {author} {\bibfnamefont {K.-U.}\ \bibnamefont {Amthor}},\ and\
  \bibinfo {author} {\bibfnamefont {R.}~\bibnamefont {Sauerbrey}},\ }\bibfield
  {title} {\bibinfo {title} {Thomson-backscattered x rays from
  laser-accelerated electrons},\ }\href@noop {} {\bibfield  {journal} {\bibinfo
   {journal} {Physical Review Letters}\ }\textbf {\bibinfo {volume} {96}},\
  \bibinfo {pages} {014802} (\bibinfo {year} {2006})}\BibitemShut {NoStop}%
\bibitem [{\citenamefont {Ta~Phuoc}\ \emph {et~al.}(2012)\citenamefont
  {Ta~Phuoc}, \citenamefont {Corde}, \citenamefont {Thaury}, \citenamefont
  {Malka}, \citenamefont {Tafzi}, \citenamefont {Goddet}, \citenamefont {Shah},
  \citenamefont {Sebban},\ and\ \citenamefont {Rousse}}]{ta2012all}%
  \BibitemOpen
  \bibfield  {author} {\bibinfo {author} {\bibfnamefont {K.}~\bibnamefont
  {Ta~Phuoc}}, \bibinfo {author} {\bibfnamefont {S.}~\bibnamefont {Corde}},
  \bibinfo {author} {\bibfnamefont {C.}~\bibnamefont {Thaury}}, \bibinfo
  {author} {\bibfnamefont {V.}~\bibnamefont {Malka}}, \bibinfo {author}
  {\bibfnamefont {A.}~\bibnamefont {Tafzi}}, \bibinfo {author} {\bibfnamefont
  {J.-P.}\ \bibnamefont {Goddet}}, \bibinfo {author} {\bibfnamefont
  {R.}~\bibnamefont {Shah}}, \bibinfo {author} {\bibfnamefont {S.}~\bibnamefont
  {Sebban}},\ and\ \bibinfo {author} {\bibfnamefont {A.}~\bibnamefont
  {Rousse}},\ }\bibfield  {title} {\bibinfo {title} {All-optical compton
  gamma-ray source},\ }\href@noop {} {\bibfield  {journal} {\bibinfo  {journal}
  {Nature Photonics}\ }\textbf {\bibinfo {volume} {6}},\ \bibinfo {pages} {308}
  (\bibinfo {year} {2012})}\BibitemShut {NoStop}%
\bibitem [{\citenamefont {Chen}\ \emph {et~al.}(2013)\citenamefont {Chen},
  \citenamefont {Powers}, \citenamefont {Ghebregziabher}, \citenamefont
  {Maharjan}, \citenamefont {Liu}, \citenamefont {Golovin}, \citenamefont
  {Banerjee}, \citenamefont {Zhang}, \citenamefont {Cunningham}, \citenamefont
  {Moorti} \emph {et~al.}}]{chen2013mev}%
  \BibitemOpen
  \bibfield  {author} {\bibinfo {author} {\bibfnamefont {S.}~\bibnamefont
  {Chen}}, \bibinfo {author} {\bibfnamefont {N.}~\bibnamefont {Powers}},
  \bibinfo {author} {\bibfnamefont {I.}~\bibnamefont {Ghebregziabher}},
  \bibinfo {author} {\bibfnamefont {C.}~\bibnamefont {Maharjan}}, \bibinfo
  {author} {\bibfnamefont {C.}~\bibnamefont {Liu}}, \bibinfo {author}
  {\bibfnamefont {G.}~\bibnamefont {Golovin}}, \bibinfo {author} {\bibfnamefont
  {S.}~\bibnamefont {Banerjee}}, \bibinfo {author} {\bibfnamefont
  {J.}~\bibnamefont {Zhang}}, \bibinfo {author} {\bibfnamefont
  {N.}~\bibnamefont {Cunningham}}, \bibinfo {author} {\bibfnamefont
  {A.}~\bibnamefont {Moorti}}, \emph {et~al.},\ }\bibfield  {title} {\bibinfo
  {title} {Mev-energy x rays from inverse compton scattering with
  laser-wakefield accelerated electrons},\ }\href@noop {} {\bibfield  {journal}
  {\bibinfo  {journal} {Physical review letters}\ }\textbf {\bibinfo {volume}
  {110}},\ \bibinfo {pages} {155003} (\bibinfo {year} {2013})}\BibitemShut
  {NoStop}%
\bibitem [{\citenamefont {Powers}\ \emph {et~al.}(2014)\citenamefont {Powers},
  \citenamefont {Ghebregziabher}, \citenamefont {Golovin}, \citenamefont {Liu},
  \citenamefont {Chen}, \citenamefont {Banerjee}, \citenamefont {Zhang},\ and\
  \citenamefont {Umstadter}}]{powers2014quasi}%
  \BibitemOpen
  \bibfield  {author} {\bibinfo {author} {\bibfnamefont {N.~D.}\ \bibnamefont
  {Powers}}, \bibinfo {author} {\bibfnamefont {I.}~\bibnamefont
  {Ghebregziabher}}, \bibinfo {author} {\bibfnamefont {G.}~\bibnamefont
  {Golovin}}, \bibinfo {author} {\bibfnamefont {C.}~\bibnamefont {Liu}},
  \bibinfo {author} {\bibfnamefont {S.}~\bibnamefont {Chen}}, \bibinfo {author}
  {\bibfnamefont {S.}~\bibnamefont {Banerjee}}, \bibinfo {author}
  {\bibfnamefont {J.}~\bibnamefont {Zhang}},\ and\ \bibinfo {author}
  {\bibfnamefont {D.~P.}\ \bibnamefont {Umstadter}},\ }\bibfield  {title}
  {\bibinfo {title} {Quasi-monoenergetic and tunable x-rays from a laser-driven
  compton light source},\ }\href@noop {} {\bibfield  {journal} {\bibinfo
  {journal} {Nature Photonics}\ }\textbf {\bibinfo {volume} {8}},\ \bibinfo
  {pages} {28} (\bibinfo {year} {2014})}\BibitemShut {NoStop}%
\bibitem [{\citenamefont {Sarri}\ \emph {et~al.}(2014)\citenamefont {Sarri},
  \citenamefont {Corvan}, \citenamefont {Schumaker}, \citenamefont {Cole},
  \citenamefont {Di~Piazza}, \citenamefont {Ahmed}, \citenamefont {Harvey},
  \citenamefont {Keitel}, \citenamefont {Krushelnick}, \citenamefont {Mangles}
  \emph {et~al.}}]{sarri2014ultrahigh}%
  \BibitemOpen
  \bibfield  {author} {\bibinfo {author} {\bibfnamefont {G.}~\bibnamefont
  {Sarri}}, \bibinfo {author} {\bibfnamefont {D.}~\bibnamefont {Corvan}},
  \bibinfo {author} {\bibfnamefont {W.}~\bibnamefont {Schumaker}}, \bibinfo
  {author} {\bibfnamefont {J.}~\bibnamefont {Cole}}, \bibinfo {author}
  {\bibfnamefont {A.}~\bibnamefont {Di~Piazza}}, \bibinfo {author}
  {\bibfnamefont {H.}~\bibnamefont {Ahmed}}, \bibinfo {author} {\bibfnamefont
  {C.}~\bibnamefont {Harvey}}, \bibinfo {author} {\bibfnamefont {C.~H.}\
  \bibnamefont {Keitel}}, \bibinfo {author} {\bibfnamefont {K.}~\bibnamefont
  {Krushelnick}}, \bibinfo {author} {\bibfnamefont {S.}~\bibnamefont
  {Mangles}}, \emph {et~al.},\ }\bibfield  {title} {\bibinfo {title} {Ultrahigh
  brilliance multi-mev $\gamma$-ray beams from nonlinear relativistic thomson
  scattering},\ }\href@noop {} {\bibfield  {journal} {\bibinfo  {journal}
  {Physical review letters}\ }\textbf {\bibinfo {volume} {113}},\ \bibinfo
  {pages} {224801} (\bibinfo {year} {2014})}\BibitemShut {NoStop}%
\bibitem [{\citenamefont {Khrennikov}\ \emph {et~al.}(2015)\citenamefont
  {Khrennikov}, \citenamefont {Wenz}, \citenamefont {Buck}, \citenamefont {Xu},
  \citenamefont {Heigoldt}, \citenamefont {Veisz},\ and\ \citenamefont
  {Karsch}}]{khrennikov2015tunable}%
  \BibitemOpen
  \bibfield  {author} {\bibinfo {author} {\bibfnamefont {K.}~\bibnamefont
  {Khrennikov}}, \bibinfo {author} {\bibfnamefont {J.}~\bibnamefont {Wenz}},
  \bibinfo {author} {\bibfnamefont {A.}~\bibnamefont {Buck}}, \bibinfo {author}
  {\bibfnamefont {J.}~\bibnamefont {Xu}}, \bibinfo {author} {\bibfnamefont
  {M.}~\bibnamefont {Heigoldt}}, \bibinfo {author} {\bibfnamefont
  {L.}~\bibnamefont {Veisz}},\ and\ \bibinfo {author} {\bibfnamefont
  {S.}~\bibnamefont {Karsch}},\ }\bibfield  {title} {\bibinfo {title} {Tunable
  all-optical quasimonochromatic thomson x-ray source in the nonlinear
  regime},\ }\href@noop {} {\bibfield  {journal} {\bibinfo  {journal} {Physical
  review letters}\ }\textbf {\bibinfo {volume} {114}},\ \bibinfo {pages}
  {195003} (\bibinfo {year} {2015})}\BibitemShut {NoStop}%
\bibitem [{\citenamefont {Yu}\ \emph {et~al.}(2016)\citenamefont {Yu},
  \citenamefont {Qi}, \citenamefont {Wang}, \citenamefont {Liu}, \citenamefont
  {Li}, \citenamefont {Wang}, \citenamefont {Zhang}, \citenamefont {Liu},
  \citenamefont {Qin}, \citenamefont {Fang} \emph {et~al.}}]{yu2016ultrahigh}%
  \BibitemOpen
  \bibfield  {author} {\bibinfo {author} {\bibfnamefont {C.}~\bibnamefont
  {Yu}}, \bibinfo {author} {\bibfnamefont {R.}~\bibnamefont {Qi}}, \bibinfo
  {author} {\bibfnamefont {W.}~\bibnamefont {Wang}}, \bibinfo {author}
  {\bibfnamefont {J.}~\bibnamefont {Liu}}, \bibinfo {author} {\bibfnamefont
  {W.}~\bibnamefont {Li}}, \bibinfo {author} {\bibfnamefont {C.}~\bibnamefont
  {Wang}}, \bibinfo {author} {\bibfnamefont {Z.}~\bibnamefont {Zhang}},
  \bibinfo {author} {\bibfnamefont {J.}~\bibnamefont {Liu}}, \bibinfo {author}
  {\bibfnamefont {Z.}~\bibnamefont {Qin}}, \bibinfo {author} {\bibfnamefont
  {M.}~\bibnamefont {Fang}}, \emph {et~al.},\ }\bibfield  {title} {\bibinfo
  {title} {Ultrahigh brilliance quasi-monochromatic mev $\gamma$-rays based on
  self-synchronized all-optical compton scattering},\ }\href@noop {} {\bibfield
   {journal} {\bibinfo  {journal} {Scientific Reports}\ }\textbf {\bibinfo
  {volume} {6}},\ \bibinfo {pages} {1} (\bibinfo {year} {2016})}\BibitemShut
  {NoStop}%
\bibitem [{\citenamefont {Yan}\ \emph {et~al.}(2017)\citenamefont {Yan},
  \citenamefont {Fruhling}, \citenamefont {Golovin}, \citenamefont {Haden},
  \citenamefont {Luo}, \citenamefont {Zhang}, \citenamefont {Zhao},
  \citenamefont {Zhang}, \citenamefont {Liu}, \citenamefont {Chen} \emph
  {et~al.}}]{yan2017high}%
  \BibitemOpen
  \bibfield  {author} {\bibinfo {author} {\bibfnamefont {W.}~\bibnamefont
  {Yan}}, \bibinfo {author} {\bibfnamefont {C.}~\bibnamefont {Fruhling}},
  \bibinfo {author} {\bibfnamefont {G.}~\bibnamefont {Golovin}}, \bibinfo
  {author} {\bibfnamefont {D.}~\bibnamefont {Haden}}, \bibinfo {author}
  {\bibfnamefont {J.}~\bibnamefont {Luo}}, \bibinfo {author} {\bibfnamefont
  {P.}~\bibnamefont {Zhang}}, \bibinfo {author} {\bibfnamefont
  {B.}~\bibnamefont {Zhao}}, \bibinfo {author} {\bibfnamefont {J.}~\bibnamefont
  {Zhang}}, \bibinfo {author} {\bibfnamefont {C.}~\bibnamefont {Liu}}, \bibinfo
  {author} {\bibfnamefont {M.}~\bibnamefont {Chen}}, \emph {et~al.},\
  }\bibfield  {title} {\bibinfo {title} {High-order multiphoton thomson
  scattering},\ }\href@noop {} {\bibfield  {journal} {\bibinfo  {journal}
  {Nature Photonics}\ }\textbf {\bibinfo {volume} {11}},\ \bibinfo {pages}
  {514} (\bibinfo {year} {2017})}\BibitemShut {NoStop}%
\bibitem [{\citenamefont {Cole}\ \emph {et~al.}(2018)\citenamefont {Cole},
  \citenamefont {Behm}, \citenamefont {Gerstmayr}, \citenamefont {Blackburn},
  \citenamefont {Wood}, \citenamefont {Baird}, \citenamefont {Duff},
  \citenamefont {Harvey}, \citenamefont {Ilderton}, \citenamefont {Joglekar}
  \emph {et~al.}}]{cole2018experimental}%
  \BibitemOpen
  \bibfield  {author} {\bibinfo {author} {\bibfnamefont {J.}~\bibnamefont
  {Cole}}, \bibinfo {author} {\bibfnamefont {K.}~\bibnamefont {Behm}}, \bibinfo
  {author} {\bibfnamefont {E.}~\bibnamefont {Gerstmayr}}, \bibinfo {author}
  {\bibfnamefont {T.}~\bibnamefont {Blackburn}}, \bibinfo {author}
  {\bibfnamefont {J.}~\bibnamefont {Wood}}, \bibinfo {author} {\bibfnamefont
  {C.}~\bibnamefont {Baird}}, \bibinfo {author} {\bibfnamefont {M.~J.}\
  \bibnamefont {Duff}}, \bibinfo {author} {\bibfnamefont {C.}~\bibnamefont
  {Harvey}}, \bibinfo {author} {\bibfnamefont {A.}~\bibnamefont {Ilderton}},
  \bibinfo {author} {\bibfnamefont {A.}~\bibnamefont {Joglekar}}, \emph
  {et~al.},\ }\bibfield  {title} {\bibinfo {title} {Experimental evidence of
  radiation reaction in the collision of a high-intensity laser pulse with a
  laser-wakefield accelerated electron beam},\ }\href@noop {} {\bibfield
  {journal} {\bibinfo  {journal} {Physical Review X}\ }\textbf {\bibinfo
  {volume} {8}},\ \bibinfo {pages} {011020} (\bibinfo {year}
  {2018})}\BibitemShut {NoStop}%
\bibitem [{\citenamefont {Poder}\ \emph {et~al.}(2018)\citenamefont {Poder},
  \citenamefont {Tamburini}, \citenamefont {Sarri}, \citenamefont {Di~Piazza},
  \citenamefont {Kuschel}, \citenamefont {Baird}, \citenamefont {Behm},
  \citenamefont {Bohlen}, \citenamefont {Cole}, \citenamefont {Corvan} \emph
  {et~al.}}]{poder2018experimental}%
  \BibitemOpen
  \bibfield  {author} {\bibinfo {author} {\bibfnamefont {K.}~\bibnamefont
  {Poder}}, \bibinfo {author} {\bibfnamefont {M.}~\bibnamefont {Tamburini}},
  \bibinfo {author} {\bibfnamefont {G.}~\bibnamefont {Sarri}}, \bibinfo
  {author} {\bibfnamefont {A.}~\bibnamefont {Di~Piazza}}, \bibinfo {author}
  {\bibfnamefont {S.}~\bibnamefont {Kuschel}}, \bibinfo {author} {\bibfnamefont
  {C.}~\bibnamefont {Baird}}, \bibinfo {author} {\bibfnamefont
  {K.}~\bibnamefont {Behm}}, \bibinfo {author} {\bibfnamefont {S.}~\bibnamefont
  {Bohlen}}, \bibinfo {author} {\bibfnamefont {J.}~\bibnamefont {Cole}},
  \bibinfo {author} {\bibfnamefont {D.}~\bibnamefont {Corvan}}, \emph
  {et~al.},\ }\bibfield  {title} {\bibinfo {title} {Experimental signatures of
  the quantum nature of radiation reaction in the field of an ultraintense
  laser},\ }\href@noop {} {\bibfield  {journal} {\bibinfo  {journal} {Physical
  Review X}\ }\textbf {\bibinfo {volume} {8}},\ \bibinfo {pages} {031004}
  (\bibinfo {year} {2018})}\BibitemShut {NoStop}%
\bibitem [{\citenamefont {Abramowicz}\ \emph {et~al.}(2019)\citenamefont
  {Abramowicz}, \citenamefont {Altarelli}, \citenamefont {A{\ss}mann},
  \citenamefont {Behnke}, \citenamefont {Benhammou}, \citenamefont {Borysov},
  \citenamefont {Borysova}, \citenamefont {Brinkmann}, \citenamefont {Burkart},
  \citenamefont {B{\"u}{\ss}er} \emph {et~al.}}]{abramowicz2019letter}%
  \BibitemOpen
  \bibfield  {author} {\bibinfo {author} {\bibfnamefont {H.}~\bibnamefont
  {Abramowicz}}, \bibinfo {author} {\bibfnamefont {M.}~\bibnamefont
  {Altarelli}}, \bibinfo {author} {\bibfnamefont {R.}~\bibnamefont
  {A{\ss}mann}}, \bibinfo {author} {\bibfnamefont {T.}~\bibnamefont {Behnke}},
  \bibinfo {author} {\bibfnamefont {Y.}~\bibnamefont {Benhammou}}, \bibinfo
  {author} {\bibfnamefont {O.}~\bibnamefont {Borysov}}, \bibinfo {author}
  {\bibfnamefont {M.}~\bibnamefont {Borysova}}, \bibinfo {author}
  {\bibfnamefont {R.}~\bibnamefont {Brinkmann}}, \bibinfo {author}
  {\bibfnamefont {F.}~\bibnamefont {Burkart}}, \bibinfo {author} {\bibfnamefont
  {K.}~\bibnamefont {B{\"u}{\ss}er}}, \emph {et~al.},\ }\bibfield  {title}
  {\bibinfo {title} {Letter of intent for the luxe experiment},\ }\href@noop {}
  {\bibfield  {journal} {\bibinfo  {journal} {arXiv preprint arXiv:1909.00860}\
  } (\bibinfo {year} {2019})}\BibitemShut {NoStop}%
\bibitem [{\citenamefont {Di~Piazza}\ \emph {et~al.}(2012)\citenamefont
  {Di~Piazza}, \citenamefont {M{\"u}ller}, \citenamefont {Hatsagortsyan},\ and\
  \citenamefont {Keitel}}]{di2012extremely}%
  \BibitemOpen
  \bibfield  {author} {\bibinfo {author} {\bibfnamefont {A.}~\bibnamefont
  {Di~Piazza}}, \bibinfo {author} {\bibfnamefont {C.}~\bibnamefont
  {M{\"u}ller}}, \bibinfo {author} {\bibfnamefont {K.}~\bibnamefont
  {Hatsagortsyan}},\ and\ \bibinfo {author} {\bibfnamefont {C.~H.}\
  \bibnamefont {Keitel}},\ }\bibfield  {title} {\bibinfo {title} {Extremely
  high-intensity laser interactions with fundamental quantum systems},\
  }\href@noop {} {\bibfield  {journal} {\bibinfo  {journal} {Reviews of Modern
  Physics}\ }\textbf {\bibinfo {volume} {84}},\ \bibinfo {pages} {1177}
  (\bibinfo {year} {2012})}\BibitemShut {NoStop}%
\bibitem [{\citenamefont {Blackburn}(2020)}]{blackburn2020review}%
  \BibitemOpen
  \bibfield  {author} {\bibinfo {author} {\bibfnamefont {T.}~\bibnamefont
  {Blackburn}},\ }\bibfield  {title} {\bibinfo {title} {Radiation reaction in
  electron--beam interactions with high-intensity lasers},\ }\href@noop {}
  {\bibfield  {journal} {\bibinfo  {journal} {Reviews of Modern Plasma
  Physics}\ }\textbf {\bibinfo {volume} {4}},\ \bibinfo {pages} {1} (\bibinfo
  {year} {2020})}\BibitemShut {NoStop}%
\bibitem [{\citenamefont {Gonoskov}\ \emph {et~al.}(2022)\citenamefont
  {Gonoskov}, \citenamefont {Blackburn}, \citenamefont {Marklund},\ and\
  \citenamefont {Bulanov}}]{gonoskov2022charged}%
  \BibitemOpen
  \bibfield  {author} {\bibinfo {author} {\bibfnamefont {A.}~\bibnamefont
  {Gonoskov}}, \bibinfo {author} {\bibfnamefont {T.}~\bibnamefont {Blackburn}},
  \bibinfo {author} {\bibfnamefont {M.}~\bibnamefont {Marklund}},\ and\
  \bibinfo {author} {\bibfnamefont {S.}~\bibnamefont {Bulanov}},\ }\bibfield
  {title} {\bibinfo {title} {Charged particle motion and radiation in strong
  electromagnetic fields},\ }\href@noop {} {\bibfield  {journal} {\bibinfo
  {journal} {Reviews of Modern Physics}\ }\textbf {\bibinfo {volume} {94}},\
  \bibinfo {pages} {045001} (\bibinfo {year} {2022})}\BibitemShut {NoStop}%
\bibitem [{\citenamefont {Fedotov}\ \emph {et~al.}(2023)\citenamefont
  {Fedotov}, \citenamefont {Ilderton}, \citenamefont {Karbstein}, \citenamefont
  {King}, \citenamefont {Seipt}, \citenamefont {Taya},\ and\ \citenamefont
  {Torgrimsson}}]{fedotov2022advances}%
  \BibitemOpen
  \bibfield  {author} {\bibinfo {author} {\bibfnamefont {A.}~\bibnamefont
  {Fedotov}}, \bibinfo {author} {\bibfnamefont {A.}~\bibnamefont {Ilderton}},
  \bibinfo {author} {\bibfnamefont {F.}~\bibnamefont {Karbstein}}, \bibinfo
  {author} {\bibfnamefont {B.}~\bibnamefont {King}}, \bibinfo {author}
  {\bibfnamefont {D.}~\bibnamefont {Seipt}}, \bibinfo {author} {\bibfnamefont
  {H.}~\bibnamefont {Taya}},\ and\ \bibinfo {author} {\bibfnamefont
  {G.}~\bibnamefont {Torgrimsson}},\ }\bibfield  {title} {\bibinfo {title}
  {{Advances in QED with intense background fields}},\ }\href
  {https://doi.org/10.1016/j.physrep.2023.01.003} {\bibfield  {journal}
  {\bibinfo  {journal} {Phys. Rept.}\ }\textbf {\bibinfo {volume} {1010}},\
  \bibinfo {pages} {1} (\bibinfo {year} {2023})},\ \Eprint
  {https://arxiv.org/abs/2203.00019} {arXiv:2203.00019 [hep-ph]} \BibitemShut
  {NoStop}%
\bibitem [{\citenamefont {Schwinger}(1951)}]{schwinger1951gauge}%
  \BibitemOpen
  \bibfield  {author} {\bibinfo {author} {\bibfnamefont {J.}~\bibnamefont
  {Schwinger}},\ }\bibfield  {title} {\bibinfo {title} {On gauge invariance and
  vacuum polarization},\ }\href@noop {} {\bibfield  {journal} {\bibinfo
  {journal} {Physical Review}\ }\textbf {\bibinfo {volume} {82}},\ \bibinfo
  {pages} {664} (\bibinfo {year} {1951})}\BibitemShut {NoStop}%
\bibitem [{\citenamefont {Labun}\ and\ \citenamefont
  {Rafelski}(2009)}]{labun2009vacuum}%
  \BibitemOpen
  \bibfield  {author} {\bibinfo {author} {\bibfnamefont {L.}~\bibnamefont
  {Labun}}\ and\ \bibinfo {author} {\bibfnamefont {J.}~\bibnamefont
  {Rafelski}},\ }\bibfield  {title} {\bibinfo {title} {Vacuum-decay time in
  strong external fields},\ }\href@noop {} {\bibfield  {journal} {\bibinfo
  {journal} {Physical Review D}\ }\textbf {\bibinfo {volume} {79}},\ \bibinfo
  {pages} {057901} (\bibinfo {year} {2009})}\BibitemShut {NoStop}%
\bibitem [{\citenamefont {Harvey}\ \emph {et~al.}(2009)\citenamefont {Harvey},
  \citenamefont {Heinzl},\ and\ \citenamefont
  {Ilderton}}]{harvey2009signatures}%
  \BibitemOpen
  \bibfield  {author} {\bibinfo {author} {\bibfnamefont {C.}~\bibnamefont
  {Harvey}}, \bibinfo {author} {\bibfnamefont {T.}~\bibnamefont {Heinzl}},\
  and\ \bibinfo {author} {\bibfnamefont {A.}~\bibnamefont {Ilderton}},\
  }\bibfield  {title} {\bibinfo {title} {Signatures of high-intensity compton
  scattering},\ }\href@noop {} {\bibfield  {journal} {\bibinfo  {journal}
  {Physical Review A}\ }\textbf {\bibinfo {volume} {79}},\ \bibinfo {pages}
  {063407} (\bibinfo {year} {2009})}\BibitemShut {NoStop}%
\bibitem [{\citenamefont {Di~Piazza}\ \emph {et~al.}(2009)\citenamefont
  {Di~Piazza}, \citenamefont {Hatsagortsyan},\ and\ \citenamefont
  {Keitel}}]{di2009strong}%
  \BibitemOpen
  \bibfield  {author} {\bibinfo {author} {\bibfnamefont {A.}~\bibnamefont
  {Di~Piazza}}, \bibinfo {author} {\bibfnamefont {K.}~\bibnamefont
  {Hatsagortsyan}},\ and\ \bibinfo {author} {\bibfnamefont {C.}~\bibnamefont
  {Keitel}},\ }\bibfield  {title} {\bibinfo {title} {Strong signatures of
  radiation reaction below the radiation-dominated regime},\ }\href@noop {}
  {\bibfield  {journal} {\bibinfo  {journal} {Physical review letters}\
  }\textbf {\bibinfo {volume} {102}},\ \bibinfo {pages} {254802} (\bibinfo
  {year} {2009})}\BibitemShut {NoStop}%
\bibitem [{\citenamefont {Heinzl}\ \emph {et~al.}(2010)\citenamefont {Heinzl},
  \citenamefont {Seipt},\ and\ \citenamefont {Kampfer}}]{Heinzl:2009nd}%
  \BibitemOpen
  \bibfield  {author} {\bibinfo {author} {\bibfnamefont {T.}~\bibnamefont
  {Heinzl}}, \bibinfo {author} {\bibfnamefont {D.}~\bibnamefont {Seipt}},\ and\
  \bibinfo {author} {\bibfnamefont {B.}~\bibnamefont {Kampfer}},\ }\bibfield
  {title} {\bibinfo {title} {{Beam-Shape Effects in Nonlinear Compton and
  Thomson Scattering}},\ }\href {https://doi.org/10.1103/PhysRevA.81.022125}
  {\bibfield  {journal} {\bibinfo  {journal} {Phys. Rev.}\ }\textbf {\bibinfo
  {volume} {A81}},\ \bibinfo {pages} {022125} (\bibinfo {year} {2010})},\
  \Eprint {https://arxiv.org/abs/0911.1622} {arXiv:0911.1622 [hep-ph]}
  \BibitemShut {NoStop}%
\bibitem [{\citenamefont {Di~Piazza}\ \emph {et~al.}(2010)\citenamefont
  {Di~Piazza}, \citenamefont {Hatsagortsyan},\ and\ \citenamefont
  {Keitel}}]{di2010quantum}%
  \BibitemOpen
  \bibfield  {author} {\bibinfo {author} {\bibfnamefont {A.}~\bibnamefont
  {Di~Piazza}}, \bibinfo {author} {\bibfnamefont {K.}~\bibnamefont
  {Hatsagortsyan}},\ and\ \bibinfo {author} {\bibfnamefont {C.~H.}\
  \bibnamefont {Keitel}},\ }\bibfield  {title} {\bibinfo {title} {Quantum
  radiation reaction effects in multiphoton compton scattering},\ }\href@noop
  {} {\bibfield  {journal} {\bibinfo  {journal} {Physical review letters}\
  }\textbf {\bibinfo {volume} {105}},\ \bibinfo {pages} {220403} (\bibinfo
  {year} {2010})}\BibitemShut {NoStop}%
\bibitem [{\citenamefont {Seipt}\ and\ \citenamefont
  {Kampfer}(2011)}]{Seipt:2010ya}%
  \BibitemOpen
  \bibfield  {author} {\bibinfo {author} {\bibfnamefont {D.}~\bibnamefont
  {Seipt}}\ and\ \bibinfo {author} {\bibfnamefont {B.}~\bibnamefont
  {Kampfer}},\ }\bibfield  {title} {\bibinfo {title} {{Non-Linear Compton
  Scattering of Ultrashort and Ultraintense Laser Pulses}},\ }\href
  {https://doi.org/10.1103/PhysRevA.83.022101} {\bibfield  {journal} {\bibinfo
  {journal} {Phys. Rev.}\ }\textbf {\bibinfo {volume} {A83}},\ \bibinfo {pages}
  {022101} (\bibinfo {year} {2011})},\ \Eprint
  {https://arxiv.org/abs/1010.3301} {arXiv:1010.3301 [hep-ph]} \BibitemShut
  {NoStop}%
\bibitem [{\citenamefont {Seipt}\ and\ \citenamefont
  {K{\"a}mpfer}(2011)}]{seipt2011nonlinear}%
  \BibitemOpen
  \bibfield  {author} {\bibinfo {author} {\bibfnamefont {D.}~\bibnamefont
  {Seipt}}\ and\ \bibinfo {author} {\bibfnamefont {B.}~\bibnamefont
  {K{\"a}mpfer}},\ }\bibfield  {title} {\bibinfo {title} {Nonlinear compton
  scattering of ultrashort intense laser pulses},\ }\href@noop {} {\bibfield
  {journal} {\bibinfo  {journal} {Physical Review A}\ }\textbf {\bibinfo
  {volume} {83}},\ \bibinfo {pages} {022101} (\bibinfo {year}
  {2011})}\BibitemShut {NoStop}%
\bibitem [{\citenamefont {Harvey}\ \emph {et~al.}(2011)\citenamefont {Harvey},
  \citenamefont {Heinzl},\ and\ \citenamefont {Marklund}}]{harvey2011symmetry}%
  \BibitemOpen
  \bibfield  {author} {\bibinfo {author} {\bibfnamefont {C.}~\bibnamefont
  {Harvey}}, \bibinfo {author} {\bibfnamefont {T.}~\bibnamefont {Heinzl}},\
  and\ \bibinfo {author} {\bibfnamefont {M.}~\bibnamefont {Marklund}},\
  }\bibfield  {title} {\bibinfo {title} {Symmetry breaking from radiation
  reaction in ultra-intense laser fields},\ }\href@noop {} {\bibfield
  {journal} {\bibinfo  {journal} {Physical Review D}\ }\textbf {\bibinfo
  {volume} {84}},\ \bibinfo {pages} {116005} (\bibinfo {year}
  {2011})}\BibitemShut {NoStop}%
\bibitem [{\citenamefont {Seipt}\ and\ \citenamefont
  {Kampfer}(2012)}]{Seipt:2012tn}%
  \BibitemOpen
  \bibfield  {author} {\bibinfo {author} {\bibfnamefont {D.}~\bibnamefont
  {Seipt}}\ and\ \bibinfo {author} {\bibfnamefont {B.}~\bibnamefont
  {Kampfer}},\ }\bibfield  {title} {\bibinfo {title} {{Two-photon Compton
  process in pulsed intense laser fields}},\ }\href
  {https://doi.org/10.1103/PhysRevD.85.101701} {\bibfield  {journal} {\bibinfo
  {journal} {Phys. Rev.}\ }\textbf {\bibinfo {volume} {D85}},\ \bibinfo {pages}
  {101701} (\bibinfo {year} {2012})},\ \Eprint
  {https://arxiv.org/abs/1201.4045} {arXiv:1201.4045 [hep-ph]} \BibitemShut
  {NoStop}%
\bibitem [{\citenamefont {Thomas}\ \emph {et~al.}(2012)\citenamefont {Thomas},
  \citenamefont {Ridgers}, \citenamefont {Bulanov}, \citenamefont {Griffin},\
  and\ \citenamefont {Mangles}}]{thomas2012strong}%
  \BibitemOpen
  \bibfield  {author} {\bibinfo {author} {\bibfnamefont {A.}~\bibnamefont
  {Thomas}}, \bibinfo {author} {\bibfnamefont {C.}~\bibnamefont {Ridgers}},
  \bibinfo {author} {\bibfnamefont {S.}~\bibnamefont {Bulanov}}, \bibinfo
  {author} {\bibfnamefont {B.}~\bibnamefont {Griffin}},\ and\ \bibinfo {author}
  {\bibfnamefont {S.}~\bibnamefont {Mangles}},\ }\bibfield  {title} {\bibinfo
  {title} {Strong radiation-damping effects in a gamma-ray source generated by
  the interaction of a high-intensity laser with a wakefield-accelerated
  electron beam},\ }\href@noop {} {\bibfield  {journal} {\bibinfo  {journal}
  {Physical Review X}\ }\textbf {\bibinfo {volume} {2}},\ \bibinfo {pages}
  {041004} (\bibinfo {year} {2012})}\BibitemShut {NoStop}%
\bibitem [{\citenamefont {Blackburn}\ \emph {et~al.}(2014)\citenamefont
  {Blackburn}, \citenamefont {Ridgers}, \citenamefont {Kirk},\ and\
  \citenamefont {Bell}}]{blackburn2014quantum}%
  \BibitemOpen
  \bibfield  {author} {\bibinfo {author} {\bibfnamefont {T.}~\bibnamefont
  {Blackburn}}, \bibinfo {author} {\bibfnamefont {C.~P.}\ \bibnamefont
  {Ridgers}}, \bibinfo {author} {\bibfnamefont {J.~G.}\ \bibnamefont {Kirk}},\
  and\ \bibinfo {author} {\bibfnamefont {A.}~\bibnamefont {Bell}},\ }\bibfield
  {title} {\bibinfo {title} {Quantum radiation reaction in laser--electron-beam
  collisions},\ }\href@noop {} {\bibfield  {journal} {\bibinfo  {journal}
  {Physical review letters}\ }\textbf {\bibinfo {volume} {112}},\ \bibinfo
  {pages} {015001} (\bibinfo {year} {2014})}\BibitemShut {NoStop}%
\bibitem [{\citenamefont {Neitz}\ and\ \citenamefont
  {Di~Piazza}(2014)}]{neitz2014electron}%
  \BibitemOpen
  \bibfield  {author} {\bibinfo {author} {\bibfnamefont {N.}~\bibnamefont
  {Neitz}}\ and\ \bibinfo {author} {\bibfnamefont {A.}~\bibnamefont
  {Di~Piazza}},\ }\bibfield  {title} {\bibinfo {title} {Electron-beam dynamics
  in a strong laser field including quantum radiation reaction},\ }\href@noop
  {} {\bibfield  {journal} {\bibinfo  {journal} {Physical Review A}\ }\textbf
  {\bibinfo {volume} {90}},\ \bibinfo {pages} {022102} (\bibinfo {year}
  {2014})}\BibitemShut {NoStop}%
\bibitem [{\citenamefont {Vranic}\ \emph {et~al.}(2014)\citenamefont {Vranic},
  \citenamefont {Martins}, \citenamefont {Vieira}, \citenamefont {Fonseca},\
  and\ \citenamefont {Silva}}]{vranic2014all}%
  \BibitemOpen
  \bibfield  {author} {\bibinfo {author} {\bibfnamefont {M.}~\bibnamefont
  {Vranic}}, \bibinfo {author} {\bibfnamefont {J.~L.}\ \bibnamefont {Martins}},
  \bibinfo {author} {\bibfnamefont {J.}~\bibnamefont {Vieira}}, \bibinfo
  {author} {\bibfnamefont {R.~A.}\ \bibnamefont {Fonseca}},\ and\ \bibinfo
  {author} {\bibfnamefont {L.~O.}\ \bibnamefont {Silva}},\ }\bibfield  {title}
  {\bibinfo {title} {All-optical radiation reaction at 1 0 21 w/cm 2},\
  }\href@noop {} {\bibfield  {journal} {\bibinfo  {journal} {Physical review
  letters}\ }\textbf {\bibinfo {volume} {113}},\ \bibinfo {pages} {134801}
  (\bibinfo {year} {2014})}\BibitemShut {NoStop}%
\bibitem [{\citenamefont {Blackburn}(2015)}]{blackburn2015measuring}%
  \BibitemOpen
  \bibfield  {author} {\bibinfo {author} {\bibfnamefont {T.}~\bibnamefont
  {Blackburn}},\ }\bibfield  {title} {\bibinfo {title} {Measuring quantum
  radiation reaction in laser--electron-beam collisions},\ }\href@noop {}
  {\bibfield  {journal} {\bibinfo  {journal} {Plasma Physics and Controlled
  Fusion}\ }\textbf {\bibinfo {volume} {57}},\ \bibinfo {pages} {075012}
  (\bibinfo {year} {2015})}\BibitemShut {NoStop}%
\bibitem [{\citenamefont {Heinzl}\ \emph {et~al.}(2015)\citenamefont {Heinzl},
  \citenamefont {Harvey}, \citenamefont {Ilderton}, \citenamefont {Marklund},
  \citenamefont {Bulanov}, \citenamefont {Rykovanov}, \citenamefont
  {Schroeder}, \citenamefont {Esarey},\ and\ \citenamefont
  {Leemans}}]{heinzl2015detecting}%
  \BibitemOpen
  \bibfield  {author} {\bibinfo {author} {\bibfnamefont {T.}~\bibnamefont
  {Heinzl}}, \bibinfo {author} {\bibfnamefont {C.}~\bibnamefont {Harvey}},
  \bibinfo {author} {\bibfnamefont {A.}~\bibnamefont {Ilderton}}, \bibinfo
  {author} {\bibfnamefont {M.}~\bibnamefont {Marklund}}, \bibinfo {author}
  {\bibfnamefont {S.~S.}\ \bibnamefont {Bulanov}}, \bibinfo {author}
  {\bibfnamefont {S.}~\bibnamefont {Rykovanov}}, \bibinfo {author}
  {\bibfnamefont {C.~B.}\ \bibnamefont {Schroeder}}, \bibinfo {author}
  {\bibfnamefont {E.}~\bibnamefont {Esarey}},\ and\ \bibinfo {author}
  {\bibfnamefont {W.~P.}\ \bibnamefont {Leemans}},\ }\bibfield  {title}
  {\bibinfo {title} {Detecting radiation reaction at moderate laser
  intensities},\ }\href@noop {} {\bibfield  {journal} {\bibinfo  {journal}
  {Physical Review E}\ }\textbf {\bibinfo {volume} {91}},\ \bibinfo {pages}
  {023207} (\bibinfo {year} {2015})}\BibitemShut {NoStop}%
\bibitem [{\citenamefont {Lobet}\ \emph {et~al.}(2017)\citenamefont {Lobet},
  \citenamefont {Davoine}, \citenamefont {d’Humi{\`e}res},\ and\
  \citenamefont {Gremillet}}]{lobet2017generation}%
  \BibitemOpen
  \bibfield  {author} {\bibinfo {author} {\bibfnamefont {M.}~\bibnamefont
  {Lobet}}, \bibinfo {author} {\bibfnamefont {X.}~\bibnamefont {Davoine}},
  \bibinfo {author} {\bibfnamefont {E.}~\bibnamefont {d’Humi{\`e}res}},\ and\
  \bibinfo {author} {\bibfnamefont {L.}~\bibnamefont {Gremillet}},\ }\bibfield
  {title} {\bibinfo {title} {Generation of high-energy electron-positron pairs
  in the collision of a laser-accelerated electron beam with a multipetawatt
  laser},\ }\href@noop {} {\bibfield  {journal} {\bibinfo  {journal} {Physical
  Review Accelerators and Beams}\ }\textbf {\bibinfo {volume} {20}},\ \bibinfo
  {pages} {043401} (\bibinfo {year} {2017})}\BibitemShut {NoStop}%
\bibitem [{\citenamefont {Ridgers}\ \emph {et~al.}(2017)\citenamefont
  {Ridgers}, \citenamefont {Blackburn}, \citenamefont {Del~Sorbo},
  \citenamefont {Bradley}, \citenamefont {Slade-Lowther}, \citenamefont
  {Baird}, \citenamefont {Mangles}, \citenamefont {McKenna}, \citenamefont
  {Marklund}, \citenamefont {Murphy} \emph {et~al.}}]{ridgers2017signatures}%
  \BibitemOpen
  \bibfield  {author} {\bibinfo {author} {\bibfnamefont {C.}~\bibnamefont
  {Ridgers}}, \bibinfo {author} {\bibfnamefont {T.}~\bibnamefont {Blackburn}},
  \bibinfo {author} {\bibfnamefont {D.}~\bibnamefont {Del~Sorbo}}, \bibinfo
  {author} {\bibfnamefont {L.}~\bibnamefont {Bradley}}, \bibinfo {author}
  {\bibfnamefont {C.}~\bibnamefont {Slade-Lowther}}, \bibinfo {author}
  {\bibfnamefont {C.}~\bibnamefont {Baird}}, \bibinfo {author} {\bibfnamefont
  {S.}~\bibnamefont {Mangles}}, \bibinfo {author} {\bibfnamefont
  {P.}~\bibnamefont {McKenna}}, \bibinfo {author} {\bibfnamefont
  {M.}~\bibnamefont {Marklund}}, \bibinfo {author} {\bibfnamefont
  {C.}~\bibnamefont {Murphy}}, \emph {et~al.},\ }\bibfield  {title} {\bibinfo
  {title} {Signatures of quantum effects on radiation reaction in
  laser--electron-beam collisions},\ }\href@noop {} {\bibfield  {journal}
  {\bibinfo  {journal} {Journal of Plasma Physics}\ }\textbf {\bibinfo {volume}
  {83}} (\bibinfo {year} {2017})}\BibitemShut {NoStop}%
\bibitem [{\citenamefont {Blackburn}\ \emph {et~al.}(2020)\citenamefont
  {Blackburn}, \citenamefont {Seipt}, \citenamefont {Bulanov},\ and\
  \citenamefont {Marklund}}]{blackburn2020radiation}%
  \BibitemOpen
  \bibfield  {author} {\bibinfo {author} {\bibfnamefont {T.}~\bibnamefont
  {Blackburn}}, \bibinfo {author} {\bibfnamefont {D.}~\bibnamefont {Seipt}},
  \bibinfo {author} {\bibfnamefont {S.}~\bibnamefont {Bulanov}},\ and\ \bibinfo
  {author} {\bibfnamefont {M.}~\bibnamefont {Marklund}},\ }\bibfield  {title}
  {\bibinfo {title} {Radiation beaming in the quantum regime},\ }\href@noop {}
  {\bibfield  {journal} {\bibinfo  {journal} {Physical Review A}\ }\textbf
  {\bibinfo {volume} {101}},\ \bibinfo {pages} {012505} (\bibinfo {year}
  {2020})}\BibitemShut {NoStop}%
\bibitem [{\citenamefont {Holkundkar}\ and\ \citenamefont
  {Mackenroth}(2020)}]{holkundkar2020complete}%
  \BibitemOpen
  \bibfield  {author} {\bibinfo {author} {\bibfnamefont {A.~R.}\ \bibnamefont
  {Holkundkar}}\ and\ \bibinfo {author} {\bibfnamefont {F.}~\bibnamefont
  {Mackenroth}},\ }\bibfield  {title} {\bibinfo {title} {Complete
  characterization of ultra-intense laser pulses in radiation damping regime},\
  }\href@noop {} {\bibfield  {journal} {\bibinfo  {journal} {arXiv preprint
  arXiv:2010.09344}\ } (\bibinfo {year} {2020})}\BibitemShut {NoStop}%
\bibitem [{\citenamefont {Tamburini}\ and\ \citenamefont
  {Meuren}(2021)}]{tamburini2021efficient}%
  \BibitemOpen
  \bibfield  {author} {\bibinfo {author} {\bibfnamefont {M.}~\bibnamefont
  {Tamburini}}\ and\ \bibinfo {author} {\bibfnamefont {S.}~\bibnamefont
  {Meuren}},\ }\bibfield  {title} {\bibinfo {title} {Efficient high-energy
  photon production in the supercritical qed regime},\ }\href@noop {}
  {\bibfield  {journal} {\bibinfo  {journal} {Physical Review D}\ }\textbf
  {\bibinfo {volume} {104}},\ \bibinfo {pages} {L091903} (\bibinfo {year}
  {2021})}\BibitemShut {NoStop}%
\bibitem [{\citenamefont {Golub}\ \emph {et~al.}(2022)\citenamefont {Golub},
  \citenamefont {Villalba-Ch{\'a}vez},\ and\ \citenamefont
  {M{\"u}ller}}]{golub2022non}%
  \BibitemOpen
  \bibfield  {author} {\bibinfo {author} {\bibfnamefont {A.}~\bibnamefont
  {Golub}}, \bibinfo {author} {\bibfnamefont {S.}~\bibnamefont
  {Villalba-Ch{\'a}vez}},\ and\ \bibinfo {author} {\bibfnamefont
  {C.}~\bibnamefont {M{\"u}ller}},\ }\bibfield  {title} {\bibinfo {title}
  {Non-linear breit-wheeler pair production in collisions of bremsstrahlung
  $gamma-$ quanta and a tightly focussed laser pulse},\ }\href@noop {}
  {\bibfield  {journal} {\bibinfo  {journal} {arXiv preprint arXiv:2203.14776}\
  } (\bibinfo {year} {2022})}\BibitemShut {NoStop}%
\bibitem [{\citenamefont {Ritus}(1985)}]{Ritus:1985}%
  \BibitemOpen
  \bibfield  {author} {\bibinfo {author} {\bibfnamefont {V.~I.}\ \bibnamefont
  {Ritus}},\ }\bibfield  {title} {\bibinfo {title} {Quantum effects of the
  interaction of elementary particles with an intense electromagnetic field},\
  }\href {https://doi.org/10.1007/BF01120220} {\bibfield  {journal} {\bibinfo
  {journal} {Journal of Soviet Laser Research}\ }\textbf {\bibinfo {volume}
  {6}},\ \bibinfo {pages} {497} (\bibinfo {year} {1985})}\BibitemShut {NoStop}%
\bibitem [{\citenamefont {Baier}\ \emph {et~al.}(1989)\citenamefont {Baier},
  \citenamefont {Katkov},\ and\ \citenamefont
  {Strakhovenko}}]{baier1989quantum}%
  \BibitemOpen
  \bibfield  {author} {\bibinfo {author} {\bibfnamefont {V.}~\bibnamefont
  {Baier}}, \bibinfo {author} {\bibfnamefont {V.~M.}\ \bibnamefont {Katkov}},\
  and\ \bibinfo {author} {\bibfnamefont {V.~M.}\ \bibnamefont {Strakhovenko}},\
  }\bibfield  {title} {\bibinfo {title} {Quantum radiation theory in
  inhomogeneous external fields},\ }\href@noop {} {\bibfield  {journal}
  {\bibinfo  {journal} {Nuclear Physics B}\ }\textbf {\bibinfo {volume}
  {328}},\ \bibinfo {pages} {387} (\bibinfo {year} {1989})}\BibitemShut
  {NoStop}%
\bibitem [{\citenamefont {Baier}\ and\ \citenamefont
  {Katkov}(2005)}]{baier2005concept}%
  \BibitemOpen
  \bibfield  {author} {\bibinfo {author} {\bibfnamefont {V.~N.}\ \bibnamefont
  {Baier}}\ and\ \bibinfo {author} {\bibfnamefont {V.~M.}\ \bibnamefont
  {Katkov}},\ }\bibfield  {title} {\bibinfo {title} {Concept of formation
  length in radiation theory},\ }\href@noop {} {\bibfield  {journal} {\bibinfo
  {journal} {Physics reports}\ }\textbf {\bibinfo {volume} {409}},\ \bibinfo
  {pages} {261} (\bibinfo {year} {2005})}\BibitemShut {NoStop}%
\bibitem [{\citenamefont {Esarey}\ \emph {et~al.}(1993)\citenamefont {Esarey},
  \citenamefont {Ride},\ and\ \citenamefont {Sprangle}}]{Esarey:1993zz}%
  \BibitemOpen
  \bibfield  {author} {\bibinfo {author} {\bibfnamefont {E.}~\bibnamefont
  {Esarey}}, \bibinfo {author} {\bibfnamefont {S.~K.}\ \bibnamefont {Ride}},\
  and\ \bibinfo {author} {\bibfnamefont {P.}~\bibnamefont {Sprangle}},\
  }\bibfield  {title} {\bibinfo {title} {{Nonlinear Thomson scattering of
  intense laser pulses from beams and plasmas}},\ }\href
  {https://doi.org/10.1103/PhysRevE.48.3003} {\bibfield  {journal} {\bibinfo
  {journal} {Phys. Rev.}\ }\textbf {\bibinfo {volume} {E48}},\ \bibinfo {pages}
  {3003} (\bibinfo {year} {1993})}\BibitemShut {NoStop}%
\bibitem [{\citenamefont {Ride}\ \emph {et~al.}(1995)\citenamefont {Ride},
  \citenamefont {Esarey},\ and\ \citenamefont {Baine}}]{Ride:1995zz}%
  \BibitemOpen
  \bibfield  {author} {\bibinfo {author} {\bibfnamefont {S.~K.}\ \bibnamefont
  {Ride}}, \bibinfo {author} {\bibfnamefont {E.}~\bibnamefont {Esarey}},\ and\
  \bibinfo {author} {\bibfnamefont {M.}~\bibnamefont {Baine}},\ }\bibfield
  {title} {\bibinfo {title} {{Thomson scattering of intense lasers from
  electron beams at arbitrary interaction angles}},\ }\href
  {https://doi.org/10.1103/PhysRevE.52.5425} {\bibfield  {journal} {\bibinfo
  {journal} {Phys. Rev.}\ }\textbf {\bibinfo {volume} {E52}},\ \bibinfo {pages}
  {5425} (\bibinfo {year} {1995})}\BibitemShut {NoStop}%
\bibitem [{\citenamefont {Di~Piazza}\ \emph {et~al.}(2019)\citenamefont
  {Di~Piazza}, \citenamefont {Tamburini}, \citenamefont {Meuren},\ and\
  \citenamefont {Keitel}}]{di2019improved}%
  \BibitemOpen
  \bibfield  {author} {\bibinfo {author} {\bibfnamefont {A.}~\bibnamefont
  {Di~Piazza}}, \bibinfo {author} {\bibfnamefont {M.}~\bibnamefont
  {Tamburini}}, \bibinfo {author} {\bibfnamefont {S.}~\bibnamefont {Meuren}},\
  and\ \bibinfo {author} {\bibfnamefont {C.~H.}\ \bibnamefont {Keitel}},\
  }\bibfield  {title} {\bibinfo {title} {Improved local-constant-field
  approximation for strong-field qed codes},\ }\href@noop {} {\bibfield
  {journal} {\bibinfo  {journal} {Physical Review A}\ }\textbf {\bibinfo
  {volume} {99}},\ \bibinfo {pages} {022125} (\bibinfo {year}
  {2019})}\BibitemShut {NoStop}%
\bibitem [{\citenamefont {Ilderton}\ \emph {et~al.}(2019)\citenamefont
  {Ilderton}, \citenamefont {King},\ and\ \citenamefont
  {Seipt}}]{ilderton2019extended}%
  \BibitemOpen
  \bibfield  {author} {\bibinfo {author} {\bibfnamefont {A.}~\bibnamefont
  {Ilderton}}, \bibinfo {author} {\bibfnamefont {B.}~\bibnamefont {King}},\
  and\ \bibinfo {author} {\bibfnamefont {D.}~\bibnamefont {Seipt}},\ }\bibfield
   {title} {\bibinfo {title} {Extended locally constant field approximation for
  nonlinear compton scattering},\ }\href@noop {} {\bibfield  {journal}
  {\bibinfo  {journal} {Physical Review A}\ }\textbf {\bibinfo {volume} {99}},\
  \bibinfo {pages} {042121} (\bibinfo {year} {2019})}\BibitemShut {NoStop}%
\bibitem [{\citenamefont {Bai}\ \emph {et~al.}(2022)\citenamefont {Bai},
  \citenamefont {Blackburn}, \citenamefont {Borysov}, \citenamefont {Davidi},
  \citenamefont {Hartin}, \citenamefont {Heinemann}, \citenamefont {Ma},
  \citenamefont {Perez}, \citenamefont {Santra}, \citenamefont {Soreq} \emph
  {et~al.}}]{bai2022new}%
  \BibitemOpen
  \bibfield  {author} {\bibinfo {author} {\bibfnamefont {Z.}~\bibnamefont
  {Bai}}, \bibinfo {author} {\bibfnamefont {T.}~\bibnamefont {Blackburn}},
  \bibinfo {author} {\bibfnamefont {O.}~\bibnamefont {Borysov}}, \bibinfo
  {author} {\bibfnamefont {O.}~\bibnamefont {Davidi}}, \bibinfo {author}
  {\bibfnamefont {A.}~\bibnamefont {Hartin}}, \bibinfo {author} {\bibfnamefont
  {B.}~\bibnamefont {Heinemann}}, \bibinfo {author} {\bibfnamefont
  {T.}~\bibnamefont {Ma}}, \bibinfo {author} {\bibfnamefont {G.}~\bibnamefont
  {Perez}}, \bibinfo {author} {\bibfnamefont {A.}~\bibnamefont {Santra}},
  \bibinfo {author} {\bibfnamefont {Y.}~\bibnamefont {Soreq}}, \emph {et~al.},\
  }\bibfield  {title} {\bibinfo {title} {New physics searches with an optical
  dump at luxe},\ }\href@noop {} {\bibfield  {journal} {\bibinfo  {journal}
  {Physical Review D}\ }\textbf {\bibinfo {volume} {106}},\ \bibinfo {pages}
  {115034} (\bibinfo {year} {2022})}\BibitemShut {NoStop}%
\bibitem [{\citenamefont {Hadad}\ \emph {et~al.}(2010)\citenamefont {Hadad},
  \citenamefont {Labun}, \citenamefont {Rafelski}, \citenamefont {Elkina},
  \citenamefont {Klier},\ and\ \citenamefont {Ruhl}}]{Hadad:2010mt}%
  \BibitemOpen
  \bibfield  {author} {\bibinfo {author} {\bibfnamefont {Y.}~\bibnamefont
  {Hadad}}, \bibinfo {author} {\bibfnamefont {L.}~\bibnamefont {Labun}},
  \bibinfo {author} {\bibfnamefont {J.}~\bibnamefont {Rafelski}}, \bibinfo
  {author} {\bibfnamefont {N.}~\bibnamefont {Elkina}}, \bibinfo {author}
  {\bibfnamefont {C.}~\bibnamefont {Klier}},\ and\ \bibinfo {author}
  {\bibfnamefont {H.}~\bibnamefont {Ruhl}},\ }\bibfield  {title} {\bibinfo
  {title} {{Effects of Radiation-Reaction in Relativistic Laser
  Acceleration}},\ }\href {https://doi.org/10.1103/PhysRevD.82.096012}
  {\bibfield  {journal} {\bibinfo  {journal} {Phys. Rev.}\ }\textbf {\bibinfo
  {volume} {D82}},\ \bibinfo {pages} {096012} (\bibinfo {year} {2010})},\
  \Eprint {https://arxiv.org/abs/1005.3980} {arXiv:1005.3980 [hep-ph]}
  \BibitemShut {NoStop}%
\bibitem [{\citenamefont {Ilderton}\ and\ \citenamefont
  {Torgrimsson}(2013{\natexlab{a}})}]{ilderton2013radiation1}%
  \BibitemOpen
  \bibfield  {author} {\bibinfo {author} {\bibfnamefont {A.}~\bibnamefont
  {Ilderton}}\ and\ \bibinfo {author} {\bibfnamefont {G.}~\bibnamefont
  {Torgrimsson}},\ }\bibfield  {title} {\bibinfo {title} {Radiation reaction in
  strong field qed},\ }\href@noop {} {\bibfield  {journal} {\bibinfo  {journal}
  {Physics Letters B}\ }\textbf {\bibinfo {volume} {725}},\ \bibinfo {pages}
  {481} (\bibinfo {year} {2013}{\natexlab{a}})}\BibitemShut {NoStop}%
\bibitem [{\citenamefont {Ilderton}\ and\ \citenamefont
  {Torgrimsson}(2013{\natexlab{b}})}]{ilderton2013radiation2}%
  \BibitemOpen
  \bibfield  {author} {\bibinfo {author} {\bibfnamefont {A.}~\bibnamefont
  {Ilderton}}\ and\ \bibinfo {author} {\bibfnamefont {G.}~\bibnamefont
  {Torgrimsson}},\ }\bibfield  {title} {\bibinfo {title} {Radiation reaction
  from qed: lightfront perturbation theory in a plane wave background},\
  }\href@noop {} {\bibfield  {journal} {\bibinfo  {journal} {Physical Review
  D}\ }\textbf {\bibinfo {volume} {88}},\ \bibinfo {pages} {025021} (\bibinfo
  {year} {2013}{\natexlab{b}})}\BibitemShut {NoStop}%
\bibitem [{\citenamefont {Heinzl}\ \emph {et~al.}(2021)\citenamefont {Heinzl},
  \citenamefont {Ilderton},\ and\ \citenamefont {King}}]{heinzl2021classical}%
  \BibitemOpen
  \bibfield  {author} {\bibinfo {author} {\bibfnamefont {T.}~\bibnamefont
  {Heinzl}}, \bibinfo {author} {\bibfnamefont {A.}~\bibnamefont {Ilderton}},\
  and\ \bibinfo {author} {\bibfnamefont {B.}~\bibnamefont {King}},\ }\bibfield
  {title} {\bibinfo {title} {Classical resummation and breakdown of
  strong-field qed},\ }\href@noop {} {\bibfield  {journal} {\bibinfo  {journal}
  {Physical Review Letters}\ }\textbf {\bibinfo {volume} {127}},\ \bibinfo
  {pages} {061601} (\bibinfo {year} {2021})}\BibitemShut {NoStop}%
\bibitem [{\citenamefont {Edwards}\ and\ \citenamefont
  {Ilderton}(2021)}]{edwards2021resummation}%
  \BibitemOpen
  \bibfield  {author} {\bibinfo {author} {\bibfnamefont {J.~P.}\ \bibnamefont
  {Edwards}}\ and\ \bibinfo {author} {\bibfnamefont {A.}~\bibnamefont
  {Ilderton}},\ }\bibfield  {title} {\bibinfo {title} {Resummation of
  background-collinear corrections in strong field qed},\ }\href@noop {}
  {\bibfield  {journal} {\bibinfo  {journal} {Physical Review D}\ }\textbf
  {\bibinfo {volume} {103}},\ \bibinfo {pages} {016004} (\bibinfo {year}
  {2021})}\BibitemShut {NoStop}%
\bibitem [{\citenamefont {Torgrimsson}(2021)}]{torgrimsson2021resummation}%
  \BibitemOpen
  \bibfield  {author} {\bibinfo {author} {\bibfnamefont {G.}~\bibnamefont
  {Torgrimsson}},\ }\bibfield  {title} {\bibinfo {title} {Resummation of
  quantum radiation reaction in plane waves},\ }\href@noop {} {\bibfield
  {journal} {\bibinfo  {journal} {Physical review letters}\ }\textbf {\bibinfo
  {volume} {127}},\ \bibinfo {pages} {111602} (\bibinfo {year}
  {2021})}\BibitemShut {NoStop}%
\bibitem [{\citenamefont {Fedotov}(2017)}]{fedotov2017conjecture}%
  \BibitemOpen
  \bibfield  {author} {\bibinfo {author} {\bibfnamefont {A.}~\bibnamefont
  {Fedotov}},\ }\bibfield  {title} {\bibinfo {title} {Conjecture of
  perturbative qed breakdown at $\alpha$$\chi 2/3$ 1},\ }in\ \href@noop {}
  {\emph {\bibinfo {booktitle} {Journal of Physics: Conference Series}}},\
  Vol.\ \bibinfo {volume} {826}\ (\bibinfo {organization} {IOP Publishing},\
  \bibinfo {year} {2017})\ p.\ \bibinfo {pages} {012027}\BibitemShut {NoStop}%
\bibitem [{\citenamefont {Ilderton}(2019)}]{ilderton2019note}%
  \BibitemOpen
  \bibfield  {author} {\bibinfo {author} {\bibfnamefont {A.}~\bibnamefont
  {Ilderton}},\ }\bibfield  {title} {\bibinfo {title} {Note on the conjectured
  breakdown of qed perturbation theory in strong fields},\ }\href@noop {}
  {\bibfield  {journal} {\bibinfo  {journal} {Physical Review D}\ }\textbf
  {\bibinfo {volume} {99}},\ \bibinfo {pages} {085002} (\bibinfo {year}
  {2019})}\BibitemShut {NoStop}%
\bibitem [{\citenamefont {Mironov}\ \emph {et~al.}(2020)\citenamefont
  {Mironov}, \citenamefont {Meuren},\ and\ \citenamefont
  {Fedotov}}]{mironov2020resummation}%
  \BibitemOpen
  \bibfield  {author} {\bibinfo {author} {\bibfnamefont {A.}~\bibnamefont
  {Mironov}}, \bibinfo {author} {\bibfnamefont {S.}~\bibnamefont {Meuren}},\
  and\ \bibinfo {author} {\bibfnamefont {A.}~\bibnamefont {Fedotov}},\
  }\bibfield  {title} {\bibinfo {title} {Resummation of qed radiative
  corrections in a strong constant crossed field},\ }\href@noop {} {\bibfield
  {journal} {\bibinfo  {journal} {Physical Review D}\ }\textbf {\bibinfo
  {volume} {102}},\ \bibinfo {pages} {053005} (\bibinfo {year}
  {2020})}\BibitemShut {NoStop}%
\bibitem [{\citenamefont {Heintzmann}\ and\ \citenamefont
  {Grewing}(1972)}]{heintzmann1972acceleration}%
  \BibitemOpen
  \bibfield  {author} {\bibinfo {author} {\bibfnamefont {H.}~\bibnamefont
  {Heintzmann}}\ and\ \bibinfo {author} {\bibfnamefont {M.}~\bibnamefont
  {Grewing}},\ }\bibfield  {title} {\bibinfo {title} {Acceleration of charged
  particles and radiation-reaction in strong plane and spherical waves},\
  }\href@noop {} {\bibfield  {journal} {\bibinfo  {journal} {Zeitschrift
  f{\"u}r Physik A Hadrons and nuclei}\ }\textbf {\bibinfo {volume} {251}},\
  \bibinfo {pages} {77} (\bibinfo {year} {1972})}\BibitemShut {NoStop}%
\bibitem [{\citenamefont {Piazza}(2008)}]{piazza2008exact}%
  \BibitemOpen
  \bibfield  {author} {\bibinfo {author} {\bibfnamefont {A.~D.}\ \bibnamefont
  {Piazza}},\ }\bibfield  {title} {\bibinfo {title} {Exact solution of the
  landau-lifshitz equation in a plane wave},\ }\href@noop {} {\bibfield
  {journal} {\bibinfo  {journal} {Letters in Mathematical Physics}\ }\textbf
  {\bibinfo {volume} {83}},\ \bibinfo {pages} {305} (\bibinfo {year}
  {2008})}\BibitemShut {NoStop}%
\bibitem [{\citenamefont {Ekman}\ \emph {et~al.}(2021)\citenamefont {Ekman},
  \citenamefont {Heinzl},\ and\ \citenamefont {Ilderton}}]{ekman2021reduction}%
  \BibitemOpen
  \bibfield  {author} {\bibinfo {author} {\bibfnamefont {R.}~\bibnamefont
  {Ekman}}, \bibinfo {author} {\bibfnamefont {T.}~\bibnamefont {Heinzl}},\ and\
  \bibinfo {author} {\bibfnamefont {A.}~\bibnamefont {Ilderton}},\ }\bibfield
  {title} {\bibinfo {title} {Reduction of order, resummation, and radiation
  reaction},\ }\href@noop {} {\bibfield  {journal} {\bibinfo  {journal}
  {Physical Review D}\ }\textbf {\bibinfo {volume} {104}},\ \bibinfo {pages}
  {036002} (\bibinfo {year} {2021})}\BibitemShut {NoStop}%
\bibitem [{\citenamefont {Rafelski}\ and\ \citenamefont
  {Labun}(2013)}]{rafelski2013critical}%
  \BibitemOpen
  \bibfield  {author} {\bibinfo {author} {\bibfnamefont {J.}~\bibnamefont
  {Rafelski}}\ and\ \bibinfo {author} {\bibfnamefont {L.}~\bibnamefont
  {Labun}},\ }\bibfield  {title} {\bibinfo {title} {Critical acceleration and
  quantum vacuum},\ }\href@noop {} {\bibfield  {journal} {\bibinfo  {journal}
  {Modern Physics Letters A}\ }\textbf {\bibinfo {volume} {28}},\ \bibinfo
  {pages} {1340014} (\bibinfo {year} {2013})}\BibitemShut {NoStop}%
\bibitem [{\citenamefont {Landau}\ \emph {et~al.}(1989)\citenamefont {Landau},
  \citenamefont {Landau},\ and\ \citenamefont
  {Lifshitz}}]{landau1989classical}%
  \BibitemOpen
  \bibfield  {author} {\bibinfo {author} {\bibfnamefont {L.~D.}\ \bibnamefont
  {Landau}}, \bibinfo {author} {\bibfnamefont {L.}~\bibnamefont {Landau}},\
  and\ \bibinfo {author} {\bibfnamefont {E.}~\bibnamefont {Lifshitz}},\
  }\href@noop {} {\emph {\bibinfo {title} {The Classical Theory of Fields:
  Volume 2}}},\ Vol.~\bibinfo {volume} {2}\ (\bibinfo  {publisher} {Permagon},\
  \bibinfo {address} {Oxford},\ \bibinfo {year} {1989})\BibitemShut {NoStop}%
\bibitem [{\citenamefont {Beers}\ and\ \citenamefont
  {Nickle}(1972)}]{beers1972algebraic}%
  \BibitemOpen
  \bibfield  {author} {\bibinfo {author} {\bibfnamefont {B.}~\bibnamefont
  {Beers}}\ and\ \bibinfo {author} {\bibfnamefont {H.}~\bibnamefont {Nickle}},\
  }\bibfield  {title} {\bibinfo {title} {Algebraic solution for a dirac
  electron in a plane-wave electromagnetic field},\ }\href@noop {} {\bibfield
  {journal} {\bibinfo  {journal} {Journal of Mathematical Physics}\ }\textbf
  {\bibinfo {volume} {13}},\ \bibinfo {pages} {1592} (\bibinfo {year}
  {1972})}\BibitemShut {NoStop}%
\bibitem [{\citenamefont {Heinzl}\ and\ \citenamefont
  {Ilderton}(2017)}]{heinzl2017exact}%
  \BibitemOpen
  \bibfield  {author} {\bibinfo {author} {\bibfnamefont {T.}~\bibnamefont
  {Heinzl}}\ and\ \bibinfo {author} {\bibfnamefont {A.}~\bibnamefont
  {Ilderton}},\ }\bibfield  {title} {\bibinfo {title} {Exact classical and
  quantum dynamics in background electromagnetic fields},\ }\href@noop {}
  {\bibfield  {journal} {\bibinfo  {journal} {Physical Review Letters}\
  }\textbf {\bibinfo {volume} {118}},\ \bibinfo {pages} {113202} (\bibinfo
  {year} {2017})}\BibitemShut {NoStop}%
\bibitem [{\citenamefont {Dirac}(1938)}]{dirac1938classical}%
  \BibitemOpen
  \bibfield  {author} {\bibinfo {author} {\bibfnamefont {P.~A.~M.}\
  \bibnamefont {Dirac}},\ }\bibfield  {title} {\bibinfo {title} {Classical
  theory of radiating electrons},\ }\href@noop {} {\bibfield  {journal}
  {\bibinfo  {journal} {Proceedings of the Royal Society of London. Series A.
  Mathematical and Physical Sciences}\ }\textbf {\bibinfo {volume} {167}},\
  \bibinfo {pages} {148} (\bibinfo {year} {1938})}\BibitemShut {NoStop}%
\bibitem [{\citenamefont {Spohn}(2000)}]{spohn2000critical}%
  \BibitemOpen
  \bibfield  {author} {\bibinfo {author} {\bibfnamefont {H.}~\bibnamefont
  {Spohn}},\ }\bibfield  {title} {\bibinfo {title} {The critical manifold of
  the lorentz-dirac equation},\ }\href@noop {} {\bibfield  {journal} {\bibinfo
  {journal} {EPL (Europhysics Letters)}\ }\textbf {\bibinfo {volume} {50}},\
  \bibinfo {pages} {287} (\bibinfo {year} {2000})}\BibitemShut {NoStop}%
\bibitem [{\citenamefont {Herr}\ and\ \citenamefont
  {Muratori}(2006)}]{herr2006concept}%
  \BibitemOpen
  \bibfield  {author} {\bibinfo {author} {\bibfnamefont {W.}~\bibnamefont
  {Herr}}\ and\ \bibinfo {author} {\bibfnamefont {B.}~\bibnamefont
  {Muratori}},\ }\bibfield  {title} {\bibinfo {title} {Concept of luminosity},\
  }in\ \href@noop {} {\emph {\bibinfo {booktitle} {CAS - CERN Accelerator
  School: Intermediate Course on Accelerator Physics}}}\ (\bibinfo  {publisher}
  {Cern},\ \bibinfo {year} {2006})\ pp.\ \bibinfo {pages}
  {361--378}\BibitemShut {NoStop}%
\bibitem [{\citenamefont {Grafstr{\"o}m}\ and\ \citenamefont
  {Kozanecki}(2015)}]{grafstrom2015luminosity}%
  \BibitemOpen
  \bibfield  {author} {\bibinfo {author} {\bibfnamefont {P.}~\bibnamefont
  {Grafstr{\"o}m}}\ and\ \bibinfo {author} {\bibfnamefont {W.}~\bibnamefont
  {Kozanecki}},\ }\bibfield  {title} {\bibinfo {title} {Luminosity
  determination at proton colliders},\ }\href@noop {} {\bibfield  {journal}
  {\bibinfo  {journal} {Progress in Particle and Nuclear Physics}\ }\textbf
  {\bibinfo {volume} {81}},\ \bibinfo {pages} {97} (\bibinfo {year}
  {2015})}\BibitemShut {NoStop}%
\bibitem [{\citenamefont {Group}(2020)}]{pdg2020}%
  \BibitemOpen
  \bibfield  {author} {\bibinfo {author} {\bibfnamefont {P.~D.}\ \bibnamefont
  {Group}},\ }\bibfield  {title} {\bibinfo {title} {{Review of Particle
  Physics}},\ }\bibfield  {journal} {\bibinfo  {journal} {Progress of
  Theoretical and Experimental Physics}\ }\textbf {\bibinfo {volume} {2020}},\
  \href {https://doi.org/10.1093/ptep/ptaa104} {10.1093/ptep/ptaa104} (\bibinfo
  {year} {2020}),\ \bibinfo {note} {see section Accelerator Physics of
  Colliders},\ \Eprint
  {https://arxiv.org/abs/https://academic.oup.com/ptep/article-pdf/2020/8/083C01/34673722/ptaa104.pdf}
  {https://academic.oup.com/ptep/article-pdf/2020/8/083C01/34673722/ptaa104.pdf}
  \BibitemShut {NoStop}%
\bibitem [{\citenamefont {Noble}(1987)}]{noble1987beamstrahlung}%
  \BibitemOpen
  \bibfield  {author} {\bibinfo {author} {\bibfnamefont {R.~J.}\ \bibnamefont
  {Noble}},\ }\bibfield  {title} {\bibinfo {title} {Beamstrahlung from
  colliding electron-positron beams with negligible disruption},\ }\href@noop
  {} {\bibfield  {journal} {\bibinfo  {journal} {Nuclear Instruments and
  Methods in Physics Research Section A: Accelerators, Spectrometers, Detectors
  and Associated Equipment}\ }\textbf {\bibinfo {volume} {256}},\ \bibinfo
  {pages} {427} (\bibinfo {year} {1987})}\BibitemShut {NoStop}%
\bibitem [{\citenamefont {Telnov}(1990)}]{telnov1990problems}%
  \BibitemOpen
  \bibfield  {author} {\bibinfo {author} {\bibfnamefont {V.~I.}\ \bibnamefont
  {Telnov}},\ }\bibfield  {title} {\bibinfo {title} {Problems in obtaining
  $\gamma$$\gamma$ and $\gamma$e colliding beams at linear colliders},\
  }\href@noop {} {\bibfield  {journal} {\bibinfo  {journal} {Nuclear
  Instruments and Methods in Physics Research Section A: Accelerators,
  Spectrometers, Detectors and Associated Equipment}\ }\textbf {\bibinfo
  {volume} {294}},\ \bibinfo {pages} {72} (\bibinfo {year} {1990})}\BibitemShut
  {NoStop}%
\bibitem [{\citenamefont {Chen}(1992)}]{chen1992differential}%
  \BibitemOpen
  \bibfield  {author} {\bibinfo {author} {\bibfnamefont {P.}~\bibnamefont
  {Chen}},\ }\bibfield  {title} {\bibinfo {title} {Differential luminosity
  under multiphoton beamstrahlung},\ }\href@noop {} {\bibfield  {journal}
  {\bibinfo  {journal} {Physical Review D}\ }\textbf {\bibinfo {volume} {46}},\
  \bibinfo {pages} {1186} (\bibinfo {year} {1992})}\BibitemShut {NoStop}%
\bibitem [{\citenamefont {Yokoya}\ and\ \citenamefont
  {Chen}(1992)}]{yokoya1992beam}%
  \BibitemOpen
  \bibfield  {author} {\bibinfo {author} {\bibfnamefont {K.}~\bibnamefont
  {Yokoya}}\ and\ \bibinfo {author} {\bibfnamefont {P.}~\bibnamefont {Chen}},\
  }\bibfield  {title} {\bibinfo {title} {Beam-beam phenomena in linear
  colliders},\ }in\ \href@noop {} {\emph {\bibinfo {booktitle} {Frontiers of
  Particle Beams: Intensity Limitations}}}\ (\bibinfo  {publisher} {Springer},\
  \bibinfo {year} {1992})\ pp.\ \bibinfo {pages} {415--445}\BibitemShut
  {NoStop}%
\bibitem [{\citenamefont {Esberg}\ \emph {et~al.}(2014)\citenamefont {Esberg},
  \citenamefont {Uggerh{\o}j}, \citenamefont {Dalena},\ and\ \citenamefont
  {Schulte}}]{esberg2014strong}%
  \BibitemOpen
  \bibfield  {author} {\bibinfo {author} {\bibfnamefont {J.}~\bibnamefont
  {Esberg}}, \bibinfo {author} {\bibfnamefont {U.}~\bibnamefont {Uggerh{\o}j}},
  \bibinfo {author} {\bibfnamefont {B.}~\bibnamefont {Dalena}},\ and\ \bibinfo
  {author} {\bibfnamefont {D.}~\bibnamefont {Schulte}},\ }\bibfield  {title}
  {\bibinfo {title} {Strong field processes in beam-beam interactions at the
  compact linear collider},\ }\href@noop {} {\bibfield  {journal} {\bibinfo
  {journal} {Physical Review Special Topics-Accelerators and Beams}\ }\textbf
  {\bibinfo {volume} {17}},\ \bibinfo {pages} {051003} (\bibinfo {year}
  {2014})}\BibitemShut {NoStop}%
\bibitem [{\citenamefont {Bambade}\ \emph {et~al.}(2019)\citenamefont
  {Bambade}, \citenamefont {Barklow}, \citenamefont {Behnke}, \citenamefont
  {Berggren}, \citenamefont {Brau}, \citenamefont {Burrows}, \citenamefont
  {Denisov}, \citenamefont {Faus-Golfe}, \citenamefont {Foster}, \citenamefont
  {Fujii} \emph {et~al.}}]{bambade2019international}%
  \BibitemOpen
  \bibfield  {author} {\bibinfo {author} {\bibfnamefont {P.}~\bibnamefont
  {Bambade}}, \bibinfo {author} {\bibfnamefont {T.}~\bibnamefont {Barklow}},
  \bibinfo {author} {\bibfnamefont {T.}~\bibnamefont {Behnke}}, \bibinfo
  {author} {\bibfnamefont {M.}~\bibnamefont {Berggren}}, \bibinfo {author}
  {\bibfnamefont {J.}~\bibnamefont {Brau}}, \bibinfo {author} {\bibfnamefont
  {P.}~\bibnamefont {Burrows}}, \bibinfo {author} {\bibfnamefont
  {D.}~\bibnamefont {Denisov}}, \bibinfo {author} {\bibfnamefont
  {A.}~\bibnamefont {Faus-Golfe}}, \bibinfo {author} {\bibfnamefont
  {B.}~\bibnamefont {Foster}}, \bibinfo {author} {\bibfnamefont
  {K.}~\bibnamefont {Fujii}}, \emph {et~al.},\ }\bibfield  {title} {\bibinfo
  {title} {The international linear collider: a global project},\ }\href@noop
  {} {\bibfield  {journal} {\bibinfo  {journal} {arXiv preprint
  arXiv:1903.01629}\ } (\bibinfo {year} {2019})}\BibitemShut {NoStop}%
\bibitem [{\citenamefont {Yakimenko}\ \emph {et~al.}(2019)\citenamefont
  {Yakimenko}, \citenamefont {Meuren}, \citenamefont {Del~Gaudio},
  \citenamefont {Baumann}, \citenamefont {Fedotov}, \citenamefont {Fiuza},
  \citenamefont {Grismayer}, \citenamefont {Hogan}, \citenamefont {Pukhov},
  \citenamefont {Silva} \emph {et~al.}}]{yakimenko2019prospect}%
  \BibitemOpen
  \bibfield  {author} {\bibinfo {author} {\bibfnamefont {V.}~\bibnamefont
  {Yakimenko}}, \bibinfo {author} {\bibfnamefont {S.}~\bibnamefont {Meuren}},
  \bibinfo {author} {\bibfnamefont {F.}~\bibnamefont {Del~Gaudio}}, \bibinfo
  {author} {\bibfnamefont {C.}~\bibnamefont {Baumann}}, \bibinfo {author}
  {\bibfnamefont {A.}~\bibnamefont {Fedotov}}, \bibinfo {author} {\bibfnamefont
  {F.}~\bibnamefont {Fiuza}}, \bibinfo {author} {\bibfnamefont
  {T.}~\bibnamefont {Grismayer}}, \bibinfo {author} {\bibfnamefont
  {M.}~\bibnamefont {Hogan}}, \bibinfo {author} {\bibfnamefont
  {A.}~\bibnamefont {Pukhov}}, \bibinfo {author} {\bibfnamefont
  {L.}~\bibnamefont {Silva}}, \emph {et~al.},\ }\bibfield  {title} {\bibinfo
  {title} {Prospect of studying nonperturbative qed with beam-beam
  collisions},\ }\href@noop {} {\bibfield  {journal} {\bibinfo  {journal}
  {Physical review letters}\ }\textbf {\bibinfo {volume} {122}},\ \bibinfo
  {pages} {190404} (\bibinfo {year} {2019})}\BibitemShut {NoStop}%
\bibitem [{\citenamefont {Ginzburg}\ \emph {et~al.}(1981)\citenamefont
  {Ginzburg}, \citenamefont {Kotkin}, \citenamefont {Serbo},\ and\
  \citenamefont {Tel'Nov}}]{ginzburg1981production}%
  \BibitemOpen
  \bibfield  {author} {\bibinfo {author} {\bibfnamefont {I.}~\bibnamefont
  {Ginzburg}}, \bibinfo {author} {\bibfnamefont {G.}~\bibnamefont {Kotkin}},
  \bibinfo {author} {\bibfnamefont {V.}~\bibnamefont {Serbo}},\ and\ \bibinfo
  {author} {\bibfnamefont {V.}~\bibnamefont {Tel'Nov}},\ }\bibfield  {title}
  {\bibinfo {title} {Production of high-energy colliding. gamma gamma. and.
  gamma. e beams with a high luminosity at vlepp accelerators},\ }\href@noop {}
  {\bibfield  {journal} {\bibinfo  {journal} {JETP Lett.(Engl. Transl.);(United
  States)}\ }\textbf {\bibinfo {volume} {34}} (\bibinfo {year}
  {1981})}\BibitemShut {NoStop}%
\bibitem [{\citenamefont {Telnov}(1995)}]{telnov1995principles}%
  \BibitemOpen
  \bibfield  {author} {\bibinfo {author} {\bibfnamefont {V.}~\bibnamefont
  {Telnov}},\ }\bibfield  {title} {\bibinfo {title} {Principles of photon
  colliders},\ }\href@noop {} {\bibfield  {journal} {\bibinfo  {journal}
  {Nuclear Instruments and Methods in Physics Research Section A: Accelerators,
  Spectrometers, Detectors and Associated Equipment}\ }\textbf {\bibinfo
  {volume} {355}},\ \bibinfo {pages} {3} (\bibinfo {year} {1995})}\BibitemShut
  {NoStop}%
\bibitem [{\citenamefont {Gronberg}(2014)}]{gronberg2014photon}%
  \BibitemOpen
  \bibfield  {author} {\bibinfo {author} {\bibfnamefont {J.}~\bibnamefont
  {Gronberg}},\ }\bibfield  {title} {\bibinfo {title} {The photon collider},\
  }\href@noop {} {\bibfield  {journal} {\bibinfo  {journal} {Reviews of
  Accelerator Science and Technology}\ }\textbf {\bibinfo {volume} {7}},\
  \bibinfo {pages} {161} (\bibinfo {year} {2014})}\BibitemShut {NoStop}%
\bibitem [{\citenamefont {Takahashi}(2019)}]{takahashi2019future}%
  \BibitemOpen
  \bibfield  {author} {\bibinfo {author} {\bibfnamefont {T.}~\bibnamefont
  {Takahashi}},\ }\bibfield  {title} {\bibinfo {title} {Future prospects of
  gamma--gamma collider},\ }\href@noop {} {\bibfield  {journal} {\bibinfo
  {journal} {Reviews of Accelerator Science and Technology}\ }\textbf {\bibinfo
  {volume} {10}},\ \bibinfo {pages} {215} (\bibinfo {year} {2019})}\BibitemShut
  {NoStop}%
\bibitem [{\citenamefont {Yandow}\ \emph {et~al.}(2019)\citenamefont {Yandow},
  \citenamefont {Toncian},\ and\ \citenamefont {Ditmire}}]{yandow2019direct}%
  \BibitemOpen
  \bibfield  {author} {\bibinfo {author} {\bibfnamefont {A.}~\bibnamefont
  {Yandow}}, \bibinfo {author} {\bibfnamefont {T.}~\bibnamefont {Toncian}},\
  and\ \bibinfo {author} {\bibfnamefont {T.}~\bibnamefont {Ditmire}},\
  }\bibfield  {title} {\bibinfo {title} {Direct laser ion acceleration and
  above-threshold ionization at intensities from 10 21 w/cm 2 to 3$\times$ 10
  23 w/cm 2},\ }\href@noop {} {\bibfield  {journal} {\bibinfo  {journal}
  {Physical Review A}\ }\textbf {\bibinfo {volume} {100}},\ \bibinfo {pages}
  {053406} (\bibinfo {year} {2019})}\BibitemShut {NoStop}%
\bibitem [{\citenamefont {Hegelich}\ \emph {et~al.}(2020)\citenamefont
  {Hegelich}, \citenamefont {Labun},\ and\ \citenamefont
  {Labun}}]{hegelich2020reconciling}%
  \BibitemOpen
  \bibfield  {author} {\bibinfo {author} {\bibfnamefont {B.~M.}\ \bibnamefont
  {Hegelich}}, \bibinfo {author} {\bibfnamefont {L.}~\bibnamefont {Labun}},\
  and\ \bibinfo {author} {\bibfnamefont {O.~Z.}\ \bibnamefont {Labun}},\
  }\bibfield  {title} {\bibinfo {title} {Reconciling vacuum laser acceleration
  theory and experiment},\ }\href@noop {} {\bibfield  {journal} {\bibinfo
  {journal} {arXiv preprint arXiv:2009.00659}\ } (\bibinfo {year}
  {2020})}\BibitemShut {NoStop}%
\bibitem [{\citenamefont {Wang}\ \emph {et~al.}(2021)\citenamefont {Wang},
  \citenamefont {Feng}, \citenamefont {Ke}, \citenamefont {Yu}, \citenamefont
  {Xu}, \citenamefont {Qi}, \citenamefont {Chen}, \citenamefont {Qin},
  \citenamefont {Zhang}, \citenamefont {Fang} \emph {et~al.}}]{wang2021free}%
  \BibitemOpen
  \bibfield  {author} {\bibinfo {author} {\bibfnamefont {W.}~\bibnamefont
  {Wang}}, \bibinfo {author} {\bibfnamefont {K.}~\bibnamefont {Feng}}, \bibinfo
  {author} {\bibfnamefont {L.}~\bibnamefont {Ke}}, \bibinfo {author}
  {\bibfnamefont {C.}~\bibnamefont {Yu}}, \bibinfo {author} {\bibfnamefont
  {Y.}~\bibnamefont {Xu}}, \bibinfo {author} {\bibfnamefont {R.}~\bibnamefont
  {Qi}}, \bibinfo {author} {\bibfnamefont {Y.}~\bibnamefont {Chen}}, \bibinfo
  {author} {\bibfnamefont {Z.}~\bibnamefont {Qin}}, \bibinfo {author}
  {\bibfnamefont {Z.}~\bibnamefont {Zhang}}, \bibinfo {author} {\bibfnamefont
  {M.}~\bibnamefont {Fang}}, \emph {et~al.},\ }\bibfield  {title} {\bibinfo
  {title} {Free-electron lasing at 27 nanometres based on a laser wakefield
  accelerator},\ }\href@noop {} {\bibfield  {journal} {\bibinfo  {journal}
  {Nature}\ }\textbf {\bibinfo {volume} {595}},\ \bibinfo {pages} {516}
  (\bibinfo {year} {2021})}\BibitemShut {NoStop}%
\bibitem [{\citenamefont {Ke}\ \emph {et~al.}(2021)\citenamefont {Ke},
  \citenamefont {Feng}, \citenamefont {Wang}, \citenamefont {Qin},
  \citenamefont {Yu}, \citenamefont {Wu}, \citenamefont {Chen}, \citenamefont
  {Qi}, \citenamefont {Zhang}, \citenamefont {Xu} \emph {et~al.}}]{ke2021near}%
  \BibitemOpen
  \bibfield  {author} {\bibinfo {author} {\bibfnamefont {L.}~\bibnamefont
  {Ke}}, \bibinfo {author} {\bibfnamefont {K.}~\bibnamefont {Feng}}, \bibinfo
  {author} {\bibfnamefont {W.}~\bibnamefont {Wang}}, \bibinfo {author}
  {\bibfnamefont {Z.}~\bibnamefont {Qin}}, \bibinfo {author} {\bibfnamefont
  {C.}~\bibnamefont {Yu}}, \bibinfo {author} {\bibfnamefont {Y.}~\bibnamefont
  {Wu}}, \bibinfo {author} {\bibfnamefont {Y.}~\bibnamefont {Chen}}, \bibinfo
  {author} {\bibfnamefont {R.}~\bibnamefont {Qi}}, \bibinfo {author}
  {\bibfnamefont {Z.}~\bibnamefont {Zhang}}, \bibinfo {author} {\bibfnamefont
  {Y.}~\bibnamefont {Xu}}, \emph {et~al.},\ }\bibfield  {title} {\bibinfo
  {title} {Near-gev electron beams at a few per-mille level from a laser
  wakefield accelerator via density-tailored plasma},\ }\href@noop {}
  {\bibfield  {journal} {\bibinfo  {journal} {Physical review letters}\
  }\textbf {\bibinfo {volume} {126}},\ \bibinfo {pages} {214801} (\bibinfo
  {year} {2021})}\BibitemShut {NoStop}%
\bibitem [{\citenamefont {Weingartner}\ \emph {et~al.}(2012)\citenamefont
  {Weingartner}, \citenamefont {Raith}, \citenamefont {Popp}, \citenamefont
  {Chou}, \citenamefont {Wenz}, \citenamefont {Khrennikov}, \citenamefont
  {Heigoldt}, \citenamefont {Maier}, \citenamefont {Kajumba}, \citenamefont
  {Fuchs} \emph {et~al.}}]{weingartner2012ultralow}%
  \BibitemOpen
  \bibfield  {author} {\bibinfo {author} {\bibfnamefont {R.}~\bibnamefont
  {Weingartner}}, \bibinfo {author} {\bibfnamefont {S.}~\bibnamefont {Raith}},
  \bibinfo {author} {\bibfnamefont {A.}~\bibnamefont {Popp}}, \bibinfo {author}
  {\bibfnamefont {S.}~\bibnamefont {Chou}}, \bibinfo {author} {\bibfnamefont
  {J.}~\bibnamefont {Wenz}}, \bibinfo {author} {\bibfnamefont {K.}~\bibnamefont
  {Khrennikov}}, \bibinfo {author} {\bibfnamefont {M.}~\bibnamefont
  {Heigoldt}}, \bibinfo {author} {\bibfnamefont {A.~R.}\ \bibnamefont {Maier}},
  \bibinfo {author} {\bibfnamefont {N.}~\bibnamefont {Kajumba}}, \bibinfo
  {author} {\bibfnamefont {M.}~\bibnamefont {Fuchs}}, \emph {et~al.},\
  }\bibfield  {title} {\bibinfo {title} {Ultralow emittance electron beams from
  a laser-wakefield accelerator},\ }\href@noop {} {\bibfield  {journal}
  {\bibinfo  {journal} {Physical Review Special Topics-Accelerators and Beams}\
  }\textbf {\bibinfo {volume} {15}},\ \bibinfo {pages} {111302} (\bibinfo
  {year} {2012})}\BibitemShut {NoStop}%
\bibitem [{\citenamefont {Plateau}\ \emph {et~al.}(2012)\citenamefont
  {Plateau}, \citenamefont {Geddes}, \citenamefont {Thorn}, \citenamefont
  {Chen}, \citenamefont {Benedetti}, \citenamefont {Esarey}, \citenamefont
  {Gonsalves}, \citenamefont {Matlis}, \citenamefont {Nakamura}, \citenamefont
  {Schroeder} \emph {et~al.}}]{plateau2012low}%
  \BibitemOpen
  \bibfield  {author} {\bibinfo {author} {\bibfnamefont {G.}~\bibnamefont
  {Plateau}}, \bibinfo {author} {\bibfnamefont {C.}~\bibnamefont {Geddes}},
  \bibinfo {author} {\bibfnamefont {D.}~\bibnamefont {Thorn}}, \bibinfo
  {author} {\bibfnamefont {M.}~\bibnamefont {Chen}}, \bibinfo {author}
  {\bibfnamefont {C.}~\bibnamefont {Benedetti}}, \bibinfo {author}
  {\bibfnamefont {E.}~\bibnamefont {Esarey}}, \bibinfo {author} {\bibfnamefont
  {A.}~\bibnamefont {Gonsalves}}, \bibinfo {author} {\bibfnamefont
  {N.}~\bibnamefont {Matlis}}, \bibinfo {author} {\bibfnamefont
  {K.}~\bibnamefont {Nakamura}}, \bibinfo {author} {\bibfnamefont
  {C.}~\bibnamefont {Schroeder}}, \emph {et~al.},\ }\bibfield  {title}
  {\bibinfo {title} {Low-emittance electron bunches from a laser-plasma
  accelerator measured using single-shot x-ray spectroscopy},\ }\href@noop {}
  {\bibfield  {journal} {\bibinfo  {journal} {Physical review letters}\
  }\textbf {\bibinfo {volume} {109}},\ \bibinfo {pages} {064802} (\bibinfo
  {year} {2012})}\BibitemShut {NoStop}%
\bibitem [{\citenamefont {Couperus}\ \emph {et~al.}(2017)\citenamefont
  {Couperus}, \citenamefont {Pausch}, \citenamefont {K{\"o}hler}, \citenamefont
  {Zarini}, \citenamefont {Kr{\"a}mer}, \citenamefont {Garten}, \citenamefont
  {Huebl}, \citenamefont {Gebhardt}, \citenamefont {Helbig}, \citenamefont
  {Bock} \emph {et~al.}}]{couperus2017demonstration}%
  \BibitemOpen
  \bibfield  {author} {\bibinfo {author} {\bibfnamefont {J.}~\bibnamefont
  {Couperus}}, \bibinfo {author} {\bibfnamefont {R.}~\bibnamefont {Pausch}},
  \bibinfo {author} {\bibfnamefont {A.}~\bibnamefont {K{\"o}hler}}, \bibinfo
  {author} {\bibfnamefont {O.}~\bibnamefont {Zarini}}, \bibinfo {author}
  {\bibfnamefont {J.}~\bibnamefont {Kr{\"a}mer}}, \bibinfo {author}
  {\bibfnamefont {M.}~\bibnamefont {Garten}}, \bibinfo {author} {\bibfnamefont
  {A.}~\bibnamefont {Huebl}}, \bibinfo {author} {\bibfnamefont
  {R.}~\bibnamefont {Gebhardt}}, \bibinfo {author} {\bibfnamefont
  {U.}~\bibnamefont {Helbig}}, \bibinfo {author} {\bibfnamefont
  {S.}~\bibnamefont {Bock}}, \emph {et~al.},\ }\bibfield  {title} {\bibinfo
  {title} {Demonstration of a beam loaded nanocoulomb-class laser wakefield
  accelerator},\ }\href@noop {} {\bibfield  {journal} {\bibinfo  {journal}
  {Nature communications}\ }\textbf {\bibinfo {volume} {8}},\ \bibinfo {pages}
  {1} (\bibinfo {year} {2017})}\BibitemShut {NoStop}%
\bibitem [{\citenamefont {G{\"o}tzfried}\ \emph {et~al.}(2020)\citenamefont
  {G{\"o}tzfried}, \citenamefont {D{\"o}pp}, \citenamefont {Gilljohann},
  \citenamefont {Foerster}, \citenamefont {Ding}, \citenamefont {Schindler},
  \citenamefont {Schilling}, \citenamefont {Buck}, \citenamefont {Veisz},\ and\
  \citenamefont {Karsch}}]{gotzfried2020physics}%
  \BibitemOpen
  \bibfield  {author} {\bibinfo {author} {\bibfnamefont {J.}~\bibnamefont
  {G{\"o}tzfried}}, \bibinfo {author} {\bibfnamefont {A.}~\bibnamefont
  {D{\"o}pp}}, \bibinfo {author} {\bibfnamefont {M.}~\bibnamefont
  {Gilljohann}}, \bibinfo {author} {\bibfnamefont {F.}~\bibnamefont
  {Foerster}}, \bibinfo {author} {\bibfnamefont {H.}~\bibnamefont {Ding}},
  \bibinfo {author} {\bibfnamefont {S.}~\bibnamefont {Schindler}}, \bibinfo
  {author} {\bibfnamefont {G.}~\bibnamefont {Schilling}}, \bibinfo {author}
  {\bibfnamefont {A.}~\bibnamefont {Buck}}, \bibinfo {author} {\bibfnamefont
  {L.}~\bibnamefont {Veisz}},\ and\ \bibinfo {author} {\bibfnamefont
  {S.}~\bibnamefont {Karsch}},\ }\bibfield  {title} {\bibinfo {title} {Physics
  of high-charge electron beams in laser-plasma wakefields},\ }\href@noop {}
  {\bibfield  {journal} {\bibinfo  {journal} {Physical Review X}\ }\textbf
  {\bibinfo {volume} {10}},\ \bibinfo {pages} {041015} (\bibinfo {year}
  {2020})}\BibitemShut {NoStop}%
\bibitem [{\citenamefont {Aniculaesei}\ \emph {et~al.}(2022)\citenamefont
  {Aniculaesei} \emph {et~al.}}]{aniculaesei2022}%
  \BibitemOpen
  \bibfield  {author} {\bibinfo {author} {\bibfnamefont {C.}~\bibnamefont
  {Aniculaesei}} \emph {et~al.},\ }\bibfield  {title} {\bibinfo {title}
  {Acceleration of electron bunches beyond 10 gev using a nanoparticle-assisted
  hybrid wakefield accelerator}} (\bibinfo {year} {2022}),\ \bibinfo {note}
  {submitted}\BibitemShut {NoStop}%
\bibitem [{\citenamefont {Quesnel}\ and\ \citenamefont
  {Mora}(1998)}]{Quesnel:1998zz}%
  \BibitemOpen
  \bibfield  {author} {\bibinfo {author} {\bibfnamefont {B.}~\bibnamefont
  {Quesnel}}\ and\ \bibinfo {author} {\bibfnamefont {P.}~\bibnamefont {Mora}},\
  }\bibfield  {title} {\bibinfo {title} {{Theory and simulation of the
  interaction of ultraintense laser pulses with electrons in vacuum}},\ }\href
  {https://doi.org/10.1103/PhysRevE.58.3719} {\bibfield  {journal} {\bibinfo
  {journal} {Phys. Rev.}\ }\textbf {\bibinfo {volume} {E58}},\ \bibinfo {pages}
  {3719} (\bibinfo {year} {1998})}\BibitemShut {NoStop}%
\bibitem [{\citenamefont {Yoon}\ \emph {et~al.}(2021)\citenamefont {Yoon},
  \citenamefont {Kim}, \citenamefont {Choi}, \citenamefont {Sung},
  \citenamefont {Lee}, \citenamefont {Lee},\ and\ \citenamefont
  {Nam}}]{yoon2021realization}%
  \BibitemOpen
  \bibfield  {author} {\bibinfo {author} {\bibfnamefont {J.~W.}\ \bibnamefont
  {Yoon}}, \bibinfo {author} {\bibfnamefont {Y.~G.}\ \bibnamefont {Kim}},
  \bibinfo {author} {\bibfnamefont {I.~W.}\ \bibnamefont {Choi}}, \bibinfo
  {author} {\bibfnamefont {J.~H.}\ \bibnamefont {Sung}}, \bibinfo {author}
  {\bibfnamefont {H.~W.}\ \bibnamefont {Lee}}, \bibinfo {author} {\bibfnamefont
  {S.~K.}\ \bibnamefont {Lee}},\ and\ \bibinfo {author} {\bibfnamefont {C.~H.}\
  \bibnamefont {Nam}},\ }\bibfield  {title} {\bibinfo {title} {Realization of
  laser intensity over 10 23 w/cm 2},\ }\href@noop {} {\bibfield  {journal}
  {\bibinfo  {journal} {Optica}\ }\textbf {\bibinfo {volume} {8}},\ \bibinfo
  {pages} {630} (\bibinfo {year} {2021})}\BibitemShut {NoStop}%
\bibitem [{\citenamefont {Ghebregziabher}\ \emph {et~al.}(2013)\citenamefont
  {Ghebregziabher}, \citenamefont {Shadwick},\ and\ \citenamefont
  {Umstadter}}]{ghebregziabher2013spectral}%
  \BibitemOpen
  \bibfield  {author} {\bibinfo {author} {\bibfnamefont {I.}~\bibnamefont
  {Ghebregziabher}}, \bibinfo {author} {\bibfnamefont {B.~A.}\ \bibnamefont
  {Shadwick}},\ and\ \bibinfo {author} {\bibfnamefont {D.}~\bibnamefont
  {Umstadter}},\ }\bibfield  {title} {\bibinfo {title} {Spectral bandwidth
  reduction of thomson scattered light by pulse chirping},\ }\href@noop {}
  {\bibfield  {journal} {\bibinfo  {journal} {Physical Review Special
  Topics-Accelerators and Beams}\ }\textbf {\bibinfo {volume} {16}},\ \bibinfo
  {pages} {030705} (\bibinfo {year} {2013})}\BibitemShut {NoStop}%
\bibitem [{\citenamefont {Holkundkar}\ \emph {et~al.}(2015)\citenamefont
  {Holkundkar}, \citenamefont {Harvey},\ and\ \citenamefont
  {Marklund}}]{holkundkar2015thomson}%
  \BibitemOpen
  \bibfield  {author} {\bibinfo {author} {\bibfnamefont {A.~R.}\ \bibnamefont
  {Holkundkar}}, \bibinfo {author} {\bibfnamefont {C.}~\bibnamefont {Harvey}},\
  and\ \bibinfo {author} {\bibfnamefont {M.}~\bibnamefont {Marklund}},\
  }\bibfield  {title} {\bibinfo {title} {Thomson scattering in high-intensity
  chirped laser pulses},\ }\href@noop {} {\bibfield  {journal} {\bibinfo
  {journal} {Physics of Plasmas}\ }\textbf {\bibinfo {volume} {22}},\ \bibinfo
  {pages} {103103} (\bibinfo {year} {2015})}\BibitemShut {NoStop}%
\bibitem [{\citenamefont {Rykovanov}\ \emph {et~al.}(2016)\citenamefont
  {Rykovanov}, \citenamefont {Geddes}, \citenamefont {Schroeder}, \citenamefont
  {Esarey},\ and\ \citenamefont {Leemans}}]{rykovanov2016controlling}%
  \BibitemOpen
  \bibfield  {author} {\bibinfo {author} {\bibfnamefont {S.}~\bibnamefont
  {Rykovanov}}, \bibinfo {author} {\bibfnamefont {C.}~\bibnamefont {Geddes}},
  \bibinfo {author} {\bibfnamefont {C.}~\bibnamefont {Schroeder}}, \bibinfo
  {author} {\bibfnamefont {E.}~\bibnamefont {Esarey}},\ and\ \bibinfo {author}
  {\bibfnamefont {W.}~\bibnamefont {Leemans}},\ }\bibfield  {title} {\bibinfo
  {title} {Controlling the spectral shape of nonlinear thomson scattering with
  proper laser chirping},\ }\href@noop {} {\bibfield  {journal} {\bibinfo
  {journal} {Physical Review Accelerators and Beams}\ }\textbf {\bibinfo
  {volume} {19}},\ \bibinfo {pages} {030701} (\bibinfo {year}
  {2016})}\BibitemShut {NoStop}%
\bibitem [{\citenamefont {McDonald}(1998)}]{mcdonald1998comment}%
  \BibitemOpen
  \bibfield  {author} {\bibinfo {author} {\bibfnamefont {K.~T.}\ \bibnamefont
  {McDonald}},\ }\bibfield  {title} {\bibinfo {title} {Comment on
  “experimental observation of electrons accelerated in vacuum to
  relativistic energies by a high-intensity laser”},\ }\href@noop {}
  {\bibfield  {journal} {\bibinfo  {journal} {Physical review letters}\
  }\textbf {\bibinfo {volume} {80}},\ \bibinfo {pages} {1350} (\bibinfo {year}
  {1998})}\BibitemShut {NoStop}%
\bibitem [{\citenamefont {Esarey}\ \emph {et~al.}(2009)\citenamefont {Esarey},
  \citenamefont {Schroeder},\ and\ \citenamefont
  {Leemans}}]{esarey2009physics}%
  \BibitemOpen
  \bibfield  {author} {\bibinfo {author} {\bibfnamefont {E.}~\bibnamefont
  {Esarey}}, \bibinfo {author} {\bibfnamefont {C.}~\bibnamefont {Schroeder}},\
  and\ \bibinfo {author} {\bibfnamefont {W.}~\bibnamefont {Leemans}},\
  }\bibfield  {title} {\bibinfo {title} {Physics of laser-driven plasma-based
  electron accelerators},\ }\href@noop {} {\bibfield  {journal} {\bibinfo
  {journal} {Reviews of modern physics}\ }\textbf {\bibinfo {volume} {81}},\
  \bibinfo {pages} {1229} (\bibinfo {year} {2009})}\BibitemShut {NoStop}%
\bibitem [{\citenamefont {Lu}\ \emph {et~al.}(2007)\citenamefont {Lu},
  \citenamefont {Tzoufras}, \citenamefont {Joshi}, \citenamefont {Tsung},
  \citenamefont {Mori}, \citenamefont {Vieira}, \citenamefont {Fonseca},\ and\
  \citenamefont {Silva}}]{lu2007generating}%
  \BibitemOpen
  \bibfield  {author} {\bibinfo {author} {\bibfnamefont {W.}~\bibnamefont
  {Lu}}, \bibinfo {author} {\bibfnamefont {M.}~\bibnamefont {Tzoufras}},
  \bibinfo {author} {\bibfnamefont {C.}~\bibnamefont {Joshi}}, \bibinfo
  {author} {\bibfnamefont {F.}~\bibnamefont {Tsung}}, \bibinfo {author}
  {\bibfnamefont {W.}~\bibnamefont {Mori}}, \bibinfo {author} {\bibfnamefont
  {J.}~\bibnamefont {Vieira}}, \bibinfo {author} {\bibfnamefont
  {R.}~\bibnamefont {Fonseca}},\ and\ \bibinfo {author} {\bibfnamefont
  {L.}~\bibnamefont {Silva}},\ }\bibfield  {title} {\bibinfo {title}
  {Generating multi-gev electron bunches using single stage laser wakefield
  acceleration in a 3d nonlinear regime},\ }\href@noop {} {\bibfield  {journal}
  {\bibinfo  {journal} {Physical Review Special Topics-Accelerators and Beams}\
  }\textbf {\bibinfo {volume} {10}},\ \bibinfo {pages} {061301} (\bibinfo
  {year} {2007})}\BibitemShut {NoStop}%
\bibitem [{\citenamefont {Fedotov}\ \emph {et~al.}(2010)\citenamefont
  {Fedotov}, \citenamefont {Narozhny}, \citenamefont {Mourou},\ and\
  \citenamefont {Korn}}]{fedotov2010limitations}%
  \BibitemOpen
  \bibfield  {author} {\bibinfo {author} {\bibfnamefont {A.}~\bibnamefont
  {Fedotov}}, \bibinfo {author} {\bibfnamefont {N.}~\bibnamefont {Narozhny}},
  \bibinfo {author} {\bibfnamefont {G.}~\bibnamefont {Mourou}},\ and\ \bibinfo
  {author} {\bibfnamefont {G.}~\bibnamefont {Korn}},\ }\bibfield  {title}
  {\bibinfo {title} {Limitations on the attainable intensity of high power
  lasers},\ }\href@noop {} {\bibfield  {journal} {\bibinfo  {journal} {Physical
  review letters}\ }\textbf {\bibinfo {volume} {105}},\ \bibinfo {pages}
  {080402} (\bibinfo {year} {2010})}\BibitemShut {NoStop}%
\bibitem [{\citenamefont {Mironov}\ \emph {et~al.}(2014)\citenamefont
  {Mironov}, \citenamefont {Narozhny},\ and\ \citenamefont
  {Fedotov}}]{mironov2014collapse}%
  \BibitemOpen
  \bibfield  {author} {\bibinfo {author} {\bibfnamefont {A.}~\bibnamefont
  {Mironov}}, \bibinfo {author} {\bibfnamefont {N.}~\bibnamefont {Narozhny}},\
  and\ \bibinfo {author} {\bibfnamefont {A.}~\bibnamefont {Fedotov}},\
  }\bibfield  {title} {\bibinfo {title} {Collapse and revival of
  electromagnetic cascades in focused intense laser pulses},\ }\href@noop {}
  {\bibfield  {journal} {\bibinfo  {journal} {Physics Letters A}\ }\textbf
  {\bibinfo {volume} {378}},\ \bibinfo {pages} {3254} (\bibinfo {year}
  {2014})}\BibitemShut {NoStop}%
\bibitem [{\citenamefont {Vais}\ and\ \citenamefont
  {Bychenkov}(2018)}]{vais2018direct}%
  \BibitemOpen
  \bibfield  {author} {\bibinfo {author} {\bibfnamefont {O.}~\bibnamefont
  {Vais}}\ and\ \bibinfo {author} {\bibfnamefont {V.~Y.}\ \bibnamefont
  {Bychenkov}},\ }\bibfield  {title} {\bibinfo {title} {Direct electron
  acceleration for diagnostics of a laser pulse focused by an off-axis
  parabolic mirror},\ }\href@noop {} {\bibfield  {journal} {\bibinfo  {journal}
  {Applied Physics B}\ }\textbf {\bibinfo {volume} {124}},\ \bibinfo {pages}
  {1} (\bibinfo {year} {2018})}\BibitemShut {NoStop}%
\bibitem [{\citenamefont {Budker}\ \emph {et~al.}(2022)\citenamefont {Budker},
  \citenamefont {Gorchtein}, \citenamefont {Krasny}, \citenamefont {Pálffy},\
  and\ \citenamefont {Surzhykov}}]{budker2022gamma}%
  \BibitemOpen
  \bibfield  {author} {\bibinfo {author} {\bibfnamefont {D.}~\bibnamefont
  {Budker}}, \bibinfo {author} {\bibfnamefont {M.}~\bibnamefont {Gorchtein}},
  \bibinfo {author} {\bibfnamefont {M.~W.}\ \bibnamefont {Krasny}}, \bibinfo
  {author} {\bibfnamefont {A.}~\bibnamefont {Pálffy}},\ and\ \bibinfo {author}
  {\bibfnamefont {A.}~\bibnamefont {Surzhykov}},\ }\bibfield  {title} {\bibinfo
  {title} {Physics opportunities with the gamma factory},\ }\href
  {https://doi.org/https://doi.org/10.1002/andp.202200004} {\bibfield
  {journal} {\bibinfo  {journal} {Annalen der Physik}\ }\textbf {\bibinfo
  {volume} {534}},\ \bibinfo {pages} {2200004} (\bibinfo {year} {2022})},\
  \Eprint
  {https://arxiv.org/abs/https://onlinelibrary.wiley.com/doi/pdf/10.1002/andp.202200004}
  {https://onlinelibrary.wiley.com/doi/pdf/10.1002/andp.202200004} \BibitemShut
  {NoStop}%
\bibitem [{\citenamefont {Karbstein}(2022)}]{karbstein2022birefringence}%
  \BibitemOpen
  \bibfield  {author} {\bibinfo {author} {\bibfnamefont {F.}~\bibnamefont
  {Karbstein}},\ }\bibfield  {title} {\bibinfo {title} {Vacuum birefringence at
  the gamma factory},\ }\href
  {https://doi.org/https://doi.org/10.1002/andp.202100137} {\bibfield
  {journal} {\bibinfo  {journal} {Annalen der Physik}\ }\textbf {\bibinfo
  {volume} {534}},\ \bibinfo {pages} {2100137} (\bibinfo {year} {2022})},\
  \Eprint
  {https://arxiv.org/abs/https://onlinelibrary.wiley.com/doi/pdf/10.1002/andp.202100137}
  {https://onlinelibrary.wiley.com/doi/pdf/10.1002/andp.202100137} \BibitemShut
  {NoStop}%
\bibitem [{\citenamefont {Blackburn}\ and\ \citenamefont
  {Marklund}(2018)}]{blackburn2018nonlinear}%
  \BibitemOpen
  \bibfield  {author} {\bibinfo {author} {\bibfnamefont {T.}~\bibnamefont
  {Blackburn}}\ and\ \bibinfo {author} {\bibfnamefont {M.}~\bibnamefont
  {Marklund}},\ }\bibfield  {title} {\bibinfo {title} {Nonlinear breit--wheeler
  pair creation with bremsstrahlung $\gamma$ rays},\ }\href@noop {} {\bibfield
  {journal} {\bibinfo  {journal} {Plasma Physics and Controlled Fusion}\
  }\textbf {\bibinfo {volume} {60}},\ \bibinfo {pages} {054009} (\bibinfo
  {year} {2018})}\BibitemShut {NoStop}%
\bibitem [{\citenamefont {Meuren}\ \emph {et~al.}(2015)\citenamefont {Meuren},
  \citenamefont {Hatsagortsyan}, \citenamefont {Keitel},\ and\ \citenamefont
  {Di~Piazza}}]{meuren2015polarization}%
  \BibitemOpen
  \bibfield  {author} {\bibinfo {author} {\bibfnamefont {S.}~\bibnamefont
  {Meuren}}, \bibinfo {author} {\bibfnamefont {K.~Z.}\ \bibnamefont
  {Hatsagortsyan}}, \bibinfo {author} {\bibfnamefont {C.~H.}\ \bibnamefont
  {Keitel}},\ and\ \bibinfo {author} {\bibfnamefont {A.}~\bibnamefont
  {Di~Piazza}},\ }\bibfield  {title} {\bibinfo {title} {Polarization-operator
  approach to pair creation in short laser pulses},\ }\href@noop {} {\bibfield
  {journal} {\bibinfo  {journal} {Physical Review D}\ }\textbf {\bibinfo
  {volume} {91}},\ \bibinfo {pages} {013009} (\bibinfo {year}
  {2015})}\BibitemShut {NoStop}%
\bibitem [{\citenamefont {Meuren}\ \emph {et~al.}(2016)\citenamefont {Meuren},
  \citenamefont {Keitel},\ and\ \citenamefont
  {Di~Piazza}}]{meuren2016semiclassical}%
  \BibitemOpen
  \bibfield  {author} {\bibinfo {author} {\bibfnamefont {S.}~\bibnamefont
  {Meuren}}, \bibinfo {author} {\bibfnamefont {C.~H.}\ \bibnamefont {Keitel}},\
  and\ \bibinfo {author} {\bibfnamefont {A.}~\bibnamefont {Di~Piazza}},\
  }\bibfield  {title} {\bibinfo {title} {Semiclassical picture for
  electron-positron photoproduction in strong laser fields},\ }\href@noop {}
  {\bibfield  {journal} {\bibinfo  {journal} {Physical Review D}\ }\textbf
  {\bibinfo {volume} {93}},\ \bibinfo {pages} {085028} (\bibinfo {year}
  {2016})}\BibitemShut {NoStop}%
\bibitem [{\citenamefont {Pariente}\ \emph {et~al.}(2016)\citenamefont
  {Pariente}, \citenamefont {Gallet}, \citenamefont {Borot}, \citenamefont
  {Gobert},\ and\ \citenamefont {Qu{\'e}r{\'e}}}]{pariente2016space}%
  \BibitemOpen
  \bibfield  {author} {\bibinfo {author} {\bibfnamefont {G.}~\bibnamefont
  {Pariente}}, \bibinfo {author} {\bibfnamefont {V.}~\bibnamefont {Gallet}},
  \bibinfo {author} {\bibfnamefont {A.}~\bibnamefont {Borot}}, \bibinfo
  {author} {\bibfnamefont {O.}~\bibnamefont {Gobert}},\ and\ \bibinfo {author}
  {\bibfnamefont {F.}~\bibnamefont {Qu{\'e}r{\'e}}},\ }\bibfield  {title}
  {\bibinfo {title} {Space--time characterization of ultra-intense femtosecond
  laser beams},\ }\href@noop {} {\bibfield  {journal} {\bibinfo  {journal}
  {Nature Photonics}\ }\textbf {\bibinfo {volume} {10}},\ \bibinfo {pages}
  {547} (\bibinfo {year} {2016})}\BibitemShut {NoStop}%
\bibitem [{\citenamefont {Tiwari}\ \emph {et~al.}(2019)\citenamefont {Tiwari},
  \citenamefont {Gaul}, \citenamefont {Martinez}, \citenamefont {Dyer},
  \citenamefont {Gordon}, \citenamefont {Spinks}, \citenamefont {Toncian},
  \citenamefont {Bowers}, \citenamefont {Jiao}, \citenamefont {Kupfer} \emph
  {et~al.}}]{tiwari2019beam}%
  \BibitemOpen
  \bibfield  {author} {\bibinfo {author} {\bibfnamefont {G.}~\bibnamefont
  {Tiwari}}, \bibinfo {author} {\bibfnamefont {E.}~\bibnamefont {Gaul}},
  \bibinfo {author} {\bibfnamefont {M.}~\bibnamefont {Martinez}}, \bibinfo
  {author} {\bibfnamefont {G.}~\bibnamefont {Dyer}}, \bibinfo {author}
  {\bibfnamefont {J.}~\bibnamefont {Gordon}}, \bibinfo {author} {\bibfnamefont
  {M.}~\bibnamefont {Spinks}}, \bibinfo {author} {\bibfnamefont
  {T.}~\bibnamefont {Toncian}}, \bibinfo {author} {\bibfnamefont
  {B.}~\bibnamefont {Bowers}}, \bibinfo {author} {\bibfnamefont
  {X.}~\bibnamefont {Jiao}}, \bibinfo {author} {\bibfnamefont {R.}~\bibnamefont
  {Kupfer}}, \emph {et~al.},\ }\bibfield  {title} {\bibinfo {title} {Beam
  distortion effects upon focusing an ultrashort petawatt laser pulse to
  greater than 10 22 w/cm 2},\ }\href@noop {} {\bibfield  {journal} {\bibinfo
  {journal} {Optics Letters}\ }\textbf {\bibinfo {volume} {44}},\ \bibinfo
  {pages} {2764} (\bibinfo {year} {2019})}\BibitemShut {NoStop}%
\bibitem [{\citenamefont {Salamin}(2007)}]{salamin2007fields}%
  \BibitemOpen
  \bibfield  {author} {\bibinfo {author} {\bibfnamefont {Y.~I.}\ \bibnamefont
  {Salamin}},\ }\bibfield  {title} {\bibinfo {title} {Fields of a gaussian beam
  beyond the paraxial approximation},\ }\href@noop {} {\bibfield  {journal}
  {\bibinfo  {journal} {Applied Physics B}\ }\textbf {\bibinfo {volume} {86}},\
  \bibinfo {pages} {319} (\bibinfo {year} {2007})}\BibitemShut {NoStop}%
\end{thebibliography}%

\end{document}